\documentclass{WileyMSP-template}

\usepackage{cite}
\usepackage{amsmath,amsfonts,amssymb}
\usepackage{array}
\usepackage{booktabs}
\usepackage{tabularx}
\usepackage{float}
\usepackage{placeins}
\usepackage{textcomp}
\usepackage{url}
\usepackage{siunitx}
\usepackage{xcolor}

\DeclareGraphicsExtensions{.pdf,.jpeg,.jpg,.png}
\usepackage{needspace}
\renewcommand{\threesubsection}[1]{\Needspace*{4\baselineskip}\medskip\textit{#1}:\space}
\begin{document}

\pagestyle{fancy}
\fancyhf{}
\renewcommand{\headrulewidth}{0pt}
\cfoot{\thepage}

\title{Task-Adaptive Physical Reservoir Computing via Tunable Molecular Communication Dynamics}
\maketitle

\author{Saad Yousuf, Kaan Burak Ikiz, and Murat Kuscu*}

\begin{affiliations}
S. Yousuf, K. B. Ikiz, M. Kuscu\\
Nano/Bio/Physical Information and Communications Laboratory (CALICO Lab),\\
Department of Electrical and Electronics Engineering, Ko\c{c} University,\\
Istanbul 34450, Turkey\\
Email: mkuscu@ku.edu.tr
\end{affiliations}


\begin{abstract}
Diffusion-based molecular communication (MC) channels suffer from two impairments: a fading channel memory from diffusion and the saturating, nonlinear dose response of ligand receptors. These impairments have been shown to enable the channel to serve as a physical reservoir computer, with memory and nonlinearity set by channel parameters that living cells already regulate. Since computing tasks require different balances of memory and nonlinearity, this paper investigates whether these parameters can tune the MC channel to different tasks, even when only some are adjustable. Bayesian optimization of a mean-field model of the MC channel reveals that the optimal configurations of three benchmark tasks, i.e., time-series forecasting, sine-to-square transformation, and their hybrid, differ in receptor operating point, timescale, and diffusion memory. Tuning the receptor kinetics alone yields most of the error reduction of tuning all parameters, while reusing a task-specific configuration across tasks incurs a penalty despite readout retraining. A stochastic receptor model shows that receptor noise penalizes memory-demanding tasks more than threshold-like ones, and that time-averaging readouts recover much of this loss. Lastly, a synthetic glucose case study demonstrates that a single channel can be switched during operation from forecasting the glucose level to detecting threshold crossings within tens of symbols.
\end{abstract}

\section{Introduction}
\label{sec:introduction}

Molecular communication (MC) is a bio-inspired communication technique that uses molecules to encode, transmit, and receive information. It is the native communication modality of living systems and the enabling technology of the Internet of Bio-Nano Things (IoBNT), in which engineered cells, biosensors, and micro- and nanoscale devices exchange information with each other and with living tissue \cite{akyildiz2015iobnt,farsad2016survey,akyildiz2019moving,kuscu2021iobnt}. In the canonical diffusion-based MC channel, information molecules released by a transmitter propagate via diffusion to a receiver, where they reversibly bind to ligand receptors \cite{pierobon2010physical,nakano2013book,jamali2019channel}. From a communication perspective, two inherent properties of this channel limit its performance. The first one is the channel memory resulting from the slow diffusion of molecules. The molecules released for one symbol arrive at the receiver over many symbol intervals, and thus, the received signal also depends on the previously transmitted symbols, which creates inter-symbol interference (ISI) \cite{pierobon2010physical,kilinc2013receiver,noel2014enzymes}. The second one is the nonlinearity of the receiver. The receiver detects the transmitted information, which is encoded into the ligand concentration, via the number of ligand-bound receptors. Since the receiver has a finite number of receptors, the number of bound receptors is a saturating function of the ligand concentration, i.e., the receiver exhibits a nonlinear dose response \cite{pierobon2011noise,chou2015receiver,kuscu2016sinw}. Therefore, both effects have conventionally been treated as impairments to be mitigated in the design of MC receivers, e.g., through ISI-aware detection \cite{kilinc2013receiver,kuscu2019transmitter}, maximum-likelihood detection based on the statistics of receptor binding \cite{kuscu2018ml}, detection under molecular interference \cite{kuscu2022interference}, and enzymatic degradation of residual molecules \cite{noel2014enzymes}.

On the other hand, reservoir computing treats the same two properties, i.e., fading memory and nonlinearity, as computational resources. A reservoir is a fixed nonlinear dynamical system with fading memory, which is driven by an input signal, and whose response is mapped to the desired output by a trained linear readout \cite{jaeger2001echo,maass2002realtime,lukosevicius2009reservoir}. The computation performed by a reservoir relies on exactly these two resources, i.e., the memory of past inputs and their nonlinear transformation, and the balance between the two determines which tasks a given reservoir can perform well \cite{dambre2012capacity,inubushi2017beyond}. In physical reservoir computing, the reservoir is implemented by a physical system, e.g., a photonic, electronic, spintronic, mechanical, or chemical system \cite{tanaka2019recent,nakajima2020physical,liang2024emerging,baltussen2024chemical}. Moreover, even a single nonlinear node can serve as a reservoir when its transient response within each input interval is sampled into multiple virtual nodes through time multiplexing \cite{appeltant2011information}.

We have previously shown that the canonical MC channel can be harnessed as such a physical reservoir \cite{uzun2025molecular}, since its two impairments provide exactly these two resources. Specifically, the overlap of successive molecule pulses at the receiver creates fading memory, and the saturation of the receptors provides nonlinearity. The transient of the receptor occupancy, i.e., the fraction of bound receptors, is sampled within each symbol interval to obtain the virtual nodes, and a linear readout of these samples is the only trained part of the reservoir. In addition, the receptors contribute a further resource as well as a complication. First, since ligand-receptor binding is reversible, the occupancy does not follow the ligand concentration instantaneously but relaxes toward its equilibrium value with a finite time constant set by the binding rates, which constitutes a second, receptor-level memory. Second, since ligand-receptor binding is stochastic, the occupancy fluctuates around its mean, and this receptor noise, which grows as the number of receptors decreases, poses a problem for any reservoir built on this channel. In that study, we also evaluated the forecasting performance of the MC channel on chaotic time series, and showed that this performance depends non-monotonically on the receptor kinetics, the diffusion coefficient, and the transmitter-receiver distance, peaking at a particular balance of memory and nonlinearity \cite{uzun2025molecular}.

A reservoir, however, performs well on the tasks to which its balance of memory and nonlinearity is suited, and a different task calls for a different balance. Task adaptivity, i.e., the ability to shift this balance toward the task at hand, therefore determines how broadly a single reservoir can be used. This ability is of particular importance for MC systems, which are envisioned to operate inside the body for continuous health monitoring, drug delivery, and the coordination of engineered cells. In such applications, a deployed system often cannot be retrieved or replaced, and it is expected to operate for a long time \cite{akyildiz2015iobnt,kuscu2021iobnt,akyildiz2019moving}. Hence, an MC channel that also acts as a reservoir in such a system will have to serve more than one computational purpose over its lifetime. For example, an implanted biosensor monitoring a biomarker may first need to forecast the trend of the biomarker concentration in order to anticipate an episode \cite{downs2022realtime,kuscu2021iobnt,akyildiz2019moving}. It may then need to decide whether the concentration stays above a threshold, which converts the concentration waveform into a binary output, and later to recognize a nonlinear pattern in the recent concentration history. Similarly, in closed-loop drug delivery, a forecast of the biomarker concentration determines when a dose is released, whereas a threshold decision determines when the release stops \cite{akyildiz2015iobnt,kuscu2021iobnt}. In a network of engineered cells, the same intercellular channel may carry a signal at one stage, and integrate or classify it at another \cite{smith2023engineering,zhu2022modular}. These tasks impose different requirements on the reservoir: forecasting a smooth trend requires a long and faithful memory of the input, a threshold decision requires a strong nonlinearity with little memory, and recognizing a nonlinear pattern requires both. This raises the question of whether the physical configuration of the channel has to be adapted to each task, or whether retraining the readout suffices.

Most physical reservoirs, however, are fixed once deployed. Their memory and nonlinearity are determined by the material and the geometry, and adapting them to another task requires an external field, a redesign, or a new device \cite{lee2023task,liang2024emerging}. Nevertheless, task adaptivity has been demonstrated in certain physical substrates. For instance, a chiral-magnet reservoir was driven across its magnetic phases by tuning the magnetic field and temperature, which changed its memory and nonlinearity to suit different tasks \cite{lee2023task}. Adaptive or multitask operation has also been reported for biomolecular memristors and enzymatic reaction networks \cite{maraj2023sensory,mohamed2025memimpedance,ghosh2026recursive}.

The MC channel, on the other hand, offers a different route to task adaptivity, since its memory and nonlinearity are set by the physical parameters of its own components. These parameters include the amount and timing of molecule release at the transmitter, the diffusion coefficient and the transmitter-receiver distance in the channel, and the association and dissociation rates of the receptors at the receiver. Several of these parameters can be chosen at design time, and some of them can be adjusted after deployment. Moreover, living cells already regulate the same parameters in natural MC. For example, chemotactic bacteria adapt the operating point of their receptors to the ambient concentration through receptor methylation \cite{barkai1997robustness,sartori2011noise}, and cells extend their dynamic range by expressing receptors of different affinities \cite{kuscu2026adaptive}. On the engineering side, aptamer and protein libraries span orders of magnitude in ligand-receptor association and dissociation rates, and can be selected for slow dissociation \cite{ding2024kineticitc,ding2023pushing,alkhamis2024slowoff}. It was also shown that the affinity of the aptamer layer of a fabricated biosensor can be modulated by temperature or an electric field \cite{chen2022tunable,ma2017electrostatic}. Similarly, molecule release can be programmed in microfluidic transmitters, engineered sender cells, and synthetic cells \cite{bolhassan2024microfluidic,smith2023engineering,monck2024synthetic}. We have recently proposed these mechanisms as the basis of adaptive MC receivers that tune their dissociation constant to the statistics of the received signal to improve the communication performance \cite{kuscu2026adaptive}. Therefore, the MC channel is a reservoir whose memory and nonlinearity can be reconfigured through its own components; however, what this reconfigurability can offer for computation remains an open question.

In practice, not every MC system provides access to all of these parameters, and the value of this reconfigurability depends on which parameters a given system can tune. For example, a biosensor implanted in tissue acts as the receiver of an MC channel, whose transmitter is the physiological source of the biomarker and whose medium is the tissue. Hence, it can choose its receptors and its readout, whereas the source, the tissue, and the placement are imposed \cite{downs2022realtime,kuscu2021graphene}. An engineered source that signals to native cells can control only the release \cite{arjmandi2016modulator,smith2023engineering}. Two communicating engineered nodes in an IoBNT scenario can control both ends of the channel, while the medium between them remains imposed \cite{zhu2022modular,heuberger2024protocell}. Only an in vitro MC system can approach full control over all parameters \cite{bolhassan2024microfluidic,baltussen2024chemical}. Thus, a tunable parameter in the channel model does not necessarily correspond to an accessible control in a physically implemented MC system. Furthermore, all these systems are subject to a common constraint. A practical MC receiver infers the transmitted information from the number of bound receptors, and since ligand-receptor binding is stochastic, this number fluctuates around its mean \cite{pierobon2011noise,kuscu2016sinw}. A design rule derived from a mean-field model, which considers only the mean occupancy, is thus useful only if it remains valid under this receptor noise, and a receiver with few receptors may require a different configuration from one with many.

In this paper, we quantitatively investigate task-adaptive reservoir computing in the canonical MC channel under these practical constraints. In particular, we address four questions: (i) whether different temporal tasks require different physical configurations of the same channel, and what physically distinguishes these configurations; (ii) how much of the error reduction achieved by tuning all parameters remains when only the receiver, only the transmitter, or both are tunable; (iii) what it costs to reuse a configuration tuned for one task on another task when only the readout is retrained; and (iv) how these answers change when the receptor-count noise of a practical MC receiver is incorporated into the design, and how many receptors are needed for the mean-field design to remain valid.

To this end, we employ the mean-field model from our previous study \cite{uzun2025molecular} together with Bayesian optimization on three benchmark tasks, which represent the three requirements described above. These are Mackey-Glass forecasting, sine-to-square transformation, and the prediction of a cubed Mackey-Glass series, which is a hybrid task combining the memory requirement of the first task with the nonlinearity requirement of the second. We then repeat the optimization at four different receptor numbers with a stochastic receptor model, which is validated against particle-based simulations, and in which the bound receptors are counted instead of averaged. To test whether the results obtained under receptor noise hold beyond these three benchmarks, we study three additional tasks. Two of these, i.e., temporal XOR and pattern detection, operate on a bit stream encoded by on-off keying of the molecule release, which is the modulation scheme already used in most MC systems \cite{farsad2016survey}. The third one is NARMA-10, a standard reservoir computing benchmark that requires a long memory of the input. We evaluate all tasks over receptor numbers ranging from those of a cell-based receiver to those of an artificial engineered receiver. For each task, we determine its critical receptor number, $N_R^{*}$, which we define as the receptor number at which the MC channel outperforms a linear regression on its own input. We further investigate whether a readout that averages out the receptor noise, by integrating the occupancy over time windows, low-pass filtering it, and retaining more past states, can lower this receptor number. To put the errors of the MC channel into perspective, we also tune a digital reservoir of equal size, i.e., an echo state network, with the same optimizer. Finally, we apply our results to a case study in which the input is a synthetic blood glucose signal generated from a physiological model. Two targets are defined on this single input: the glucose level 30 minutes ahead (forecasting) and a binary indicator of whether the current glucose level exceeds a threshold (threshold decision). We then simulate the switching of a single channel between these two tasks during operation.

Our results reveal seven main findings. First, each task has a distinct optimal configuration, and the optimal configurations of different tasks are distinguished by the receptor operating point relative to the dissociation constant, the receptor timescale relative to the symbol interval, and the diffusion memory (Section~\ref{sec:res_optima}). Second, for the three benchmark tasks, tuning only the receptor kinetics provides most of the error reduction that tuning all parameters achieves relative to the untuned channel, whereas tuning only the transmitter yields little improvement (Section~\ref{sec:res_access}). For the two bit-stream tasks, however, tuning the receptor kinetics alone provides only a small part of this reduction (Section~\ref{sec:res_budget}). Third, a configuration tuned for one task performs worse on another task than the configuration tuned for that task, even after the readout is retrained, and this penalty is most pronounced across the boundary between memory-dominated and nonlinearity-dominated tasks (Section~\ref{sec:res_reuse}). Fourth, the receptor-count noise determines how much of the mean-field performance is retained by a receiver that counts bound receptors, and this depends on both the task and the receptor number (Section~\ref{sec:res_noise}). Fifth, when the receptor noise is incorporated into the optimization itself, the optimal configuration and the readout regularization change in a way that the mean-field model cannot anticipate (Section~\ref{sec:res_noiseaware}). Sixth, the critical receptor number divides the tasks into three groups. Smooth transformations and threshold decisions can be computed with the receptor numbers of a single cell, provided that the readout resolves the transient within a symbol interval. Nonlinear bit computations require receptor numbers at or above the upper end of what a cell expresses, whereas memory tasks require receptor numbers that only an artificial engineered receiver, such as an electrode-based biosensor \cite{kuscu2016sinw,kuscu2021graphene}, can provide. A readout that averages out the receptor noise lowers the critical receptor number of the memory benchmarks by orders of magnitude. In contrast, a readout that integrates the occupancy over the whole symbol interval, as a living cell naturally does, falls behind the linear regression on most tasks. It outperforms the linear regression on these tasks only when it resolves two half-symbol windows and is supported by more receptors (Section~\ref{sec:res_budget}). Seventh, a single channel can switch between two tasks within tens of symbols when its receptor rates are tuned to the values optimal for the new task and its readout is replaced by the one trained for that task. Retuning the receptor rates outperforms retraining only the readout when the two task-specific configurations lie far apart, as in the case of transformation and forecasting (Section~\ref{sec:res_switch}).

The remainder of the paper is organized as follows. In Section~\ref{sec:model}, we present the MC channel model, its tunable parameters, and the representative MC systems that can tune them. The results are reported in Section~\ref{sec:results}, and their implications and limitations are discussed in Section~\ref{sec:discussion}. Section~\ref{sec:conclusion} concludes the paper, and Section~\ref{sec:methods} describes the simulation and analysis methods.
\section{Molecular Communication Channel as a Tunable Physical Reservoir}
\label{sec:model}

\subsection{System Model and Reservoir Construction}
\label{sec:system}

\begin{figure*}[t!]
\centering
\includegraphics[width=\textwidth]{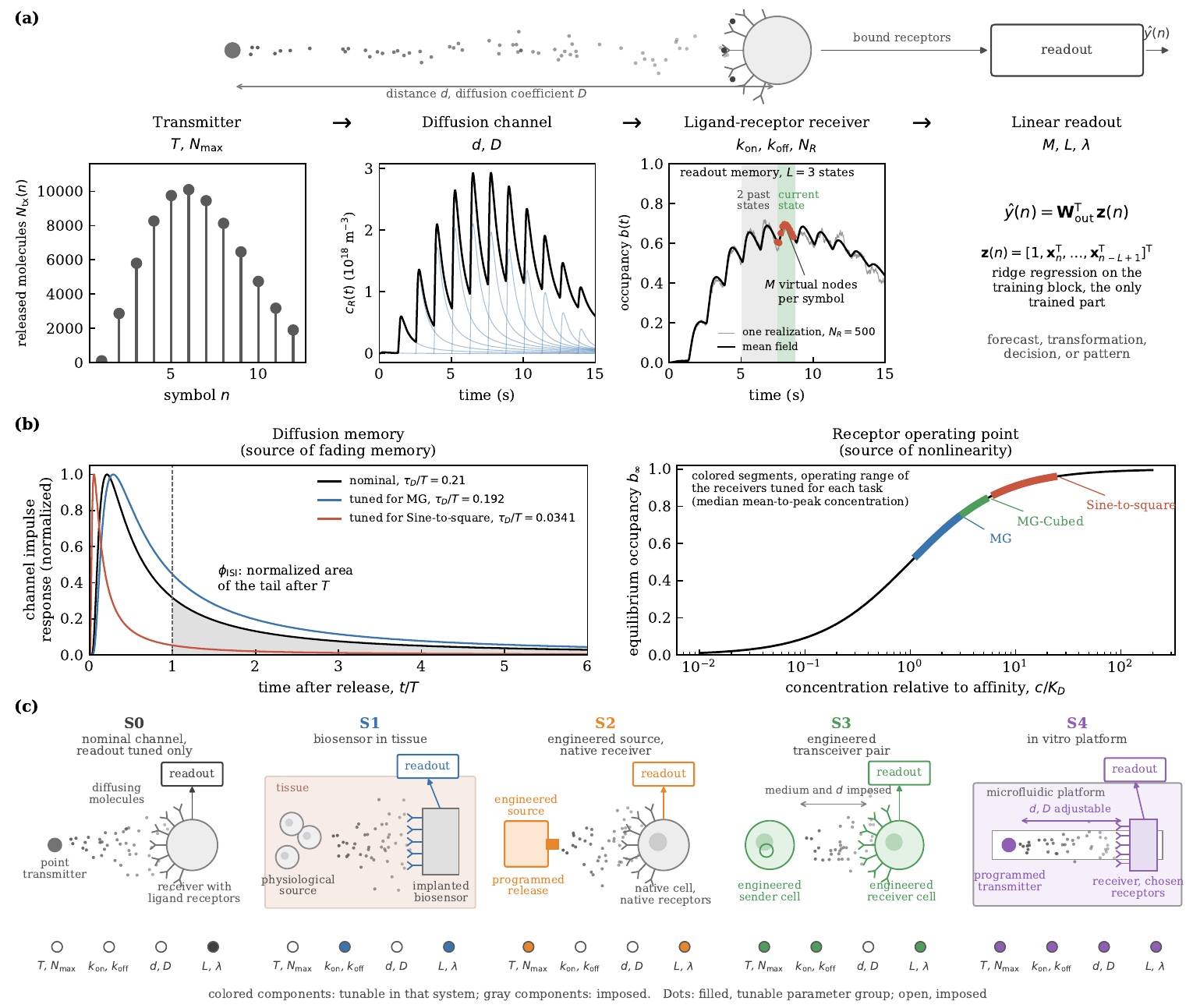}
\caption{The MC channel as a tunable physical reservoir. a) The MC channel and its signal chain for the nominal configuration and a Mackey-Glass input segment. The schematic at the top shows the channel, i.e., a point transmitter that releases molecules, which diffuse over the distance $d$ to a spherical receiver with ligand receptors, whose bound receptors drive the readout. The name of each component and the parameters that it sets are given below it, above the corresponding signal. The transmitter releases $N_{\mathrm{tx}}(n)$ molecules per symbol, and the pulses resulting from successive releases (thin lines) superpose to form the concentration at the receiver (black). The receptors convert the concentration into the occupancy $b(t)$. The mean-field occupancy (black) is the expectation of the stochastic receptor model, one realization of which is shown for $N_R=500$ receptors (gray). The occupancy transient is sampled into $M$ virtual nodes per symbol (dots, shown for $M=10$), and the readout holds the states of $L$ consecutive symbols, i.e., the current state (green) and the $L-1$ past states (gray), which together form its readout memory. A ridge-regression readout produces the output. b) The two dynamical resources. Left: normalized channel impulse response, i.e., the source of the fading memory, for the nominal channel and for the channels tuned for forecasting and transformation under S4; the kernel tail after the symbol interval, whose normalized area is $\phi_{\mathrm{ISI}}$ in Equation~\eqref{eq:isi}, is shaded for the nominal channel. Right: equilibrium binding curve, i.e., the source of the nonlinearity. The colored segments mark the part of the curve on which the receivers tuned for each task operate, extending from the median mean concentration to the median peak concentration of the tuned configurations (Section~\ref{sec:res_optima}). The ranges of the two Mackey-Glass (MG) tasks overlap near $c/K_D\approx3$. c) The readout-only reference S0 and the four representative MC systems listed in Table~\ref{tab:sets}, each drawn as an MC channel with its transmitter, medium, receiver with ligand receptors, and readout. Components drawn in the color of the set are tunable in that system, whereas gray components are imposed. The dots below each schematic mark the four parameter groups, which are filled where the set tunes them and open where they are imposed.}
\label{fig:system}
\end{figure*}

We consider the MC channel illustrated in \textbf{Figure~\ref{fig:system}}a, which we treat as a physical reservoir. It consists of a point transmitter, a diffusive medium, a receiver equipped with ligand receptors, and the readout that follows the receiver. The transmitter, located at a distance $d$ from a spherical receiver, releases $N_{\mathrm{tx}}(n)$ molecules at the beginning of symbol $n$, i.e., at time $t_n=(n-1)T$, where $T$ is the symbol interval. The task input $u(n)\in[0,1]$ is linearly encoded into the number of released molecules as $N_{\mathrm{tx}}(n)=N_{\min}+u(n)\,(N_{\max}-N_{\min})$, where $N_{\min}=100$ and $N_{\max}$ are the numbers of molecules released for $u=0$ and $u=1$, respectively, and $N_{\max}$ is one of the tunable parameters. The transmitted molecules diffuse in an unbounded three-dimensional medium with diffusion coefficient $D$, and thus, the concentration at the receiver is given by the superposition of the pulses resulting from all previous releases,
\begin{equation}
c_R(t)=\sum_{t_n<t}N_{\mathrm{tx}}(n)\,h(t-t_n),\qquad
h(t)=\frac{1}{(4\pi Dt)^{3/2}}\exp\!\left(-\frac{d^2}{4Dt}\right).
\label{eq:channel}
\end{equation}
Here $h(t)$ is the impulse response of the channel, i.e., the solution of the diffusion equation for a single molecule released at the origin at time zero and observed at distance $d$ in an unbounded medium \cite{pierobon2010physical,jamali2019channel}. The impulse response reaches its peak at $\tau_D=d^2/(6D)$ and then decays slowly. The normalized tail of this kernel after one symbol interval, i.e.,
\begin{equation}
\phi_{\mathrm{ISI}}=\frac{\int_T^{\infty}h(t)\,dt}{\int_0^{\infty}h(t)\,dt}=\operatorname{erf}\!\left(\frac{d}{\sqrt{4DT}}\right),
\label{eq:isi}
\end{equation}
measures the fraction of the exposure of the receiver to a pulse that falls after the symbol interval in which the pulse is released (Figure~\ref{fig:system}b, left) \cite{uzun2025molecular}. Since this tail decays slowly, the summation over past releases has to be truncated in any numerical evaluation. Following our previous study \cite{uzun2025molecular}, the contribution of each release is retained for $\operatorname{round}(100\,\tau_D/T)$ symbol intervals, with a minimum of one symbol interval, and neglected afterwards. This truncation is a modeling assumption, whose effect on the results is quantified in Section~\ref{sec:met_numerics}. Equation~\eqref{eq:channel} together with this truncation constitutes the MC channel model used in this paper.

The receiver is equipped with $N_R$ identical receptors, which reversibly bind the molecules with association rate $k_{\mathrm{on}}$ and dissociation rate $k_{\mathrm{off}}$. Each receptor can be either in the free or in the bound state. A free receptor binds a molecule at rate $k_{\mathrm{on}}c_R(t)$, whereas a bound receptor unbinds at rate $k_{\mathrm{off}}$; therefore, the number of bound receptors $n_b(t)$ follows a birth-death process with binomial transitions \cite{pierobon2011noise,chou2015receiver}. The fraction of bound receptors, i.e., $b(t)=n_b(t)/N_R$, is the quantity that the receiver can read, and we refer to it as the \textit{occupancy}. The expectation of the occupancy is denoted by $p(t)=\mathbb{E}[b(t)]$. For identical and independent receptors driven by the deterministic concentration $c_R(t)$, and with all receptors initially free, this expectation obeys the mean-field equation
\begin{equation}
\frac{dp(t)}{dt}=k_{\mathrm{on}}\,c_R(t)\,[1-p(t)]-k_{\mathrm{off}}\,p(t),
\label{eq:binding}
\end{equation}
and a single sample of the occupancy fluctuates around this expectation with variance $p(1-p)/N_R$, which we refer to as the \textit{receptor-count noise} \cite{pierobon2011noise,kuscu2016sinw}. Equation~\eqref{eq:binding} reveals two features that provide the receptor with its own memory and nonlinearity. First, at a constant concentration $c$, the mean occupancy does not change instantaneously but relaxes toward its new value with the time constant $\tau_R=1/(k_{\mathrm{on}}c+k_{\mathrm{off}})$, which constitutes the memory of the receptor. Second, the value to which the mean occupancy relaxes, $p_\infty=c/(c+K_D)$, where $K_D=k_{\mathrm{off}}/k_{\mathrm{on}}$ is the dissociation constant, is a saturating function of the concentration. Its sensitivity, $dp_\infty/dc=K_D/(c+K_D)^2$, decreases as $c$ increases beyond $K_D$ (Figure~\ref{fig:system}b, right), which constitutes the nonlinearity of the receptor. The receptor-count noise, on the other hand, results from the stochastic nature of ligand-receptor binding, and its variance decreases as the number of receptors increases. We refer to the evaluation based on Equation~\eqref{eq:binding}, whose state is $p(t)$, as the \textit{mean-field model}, and to the evaluation based on the birth-death process, which operates on the realizations of $b(t)$, as the \textit{stochastic receptor model}. Both models share the concentration given by Equation~\eqref{eq:channel}, and the stochastic receptor model converges to the mean-field model as $N_R$ increases.

The receptor population constitutes a single physical node, whose state at any time instant is described by a single number, i.e., the occupancy. Following the single-node approach of physical reservoir computing \cite{appeltant2011information,uzun2025molecular}, we obtain a high-dimensional reservoir state through \textit{time multiplexing}. Specifically, the occupancy transient within symbol $n$ is sampled at $M$ equidistant instants, i.e., $\mathbf x_n=[b(t_n),b(t_n+T/M),\ldots,b(t_n+(M-1)T/M)]^{\mathsf T}$, and these $M$ samples constitute the \textit{virtual nodes} of the reservoir. The readout concatenates the current state with the $L-1$ preceding states, i.e., $\mathbf z_n=[1,\mathbf x_n^{\mathsf T},\ldots,\mathbf x_{n-L+1}^{\mathsf T}]^{\mathsf T}$. Throughout the paper, we refer to $L$, which is the number of symbols whose states are explicitly held by the readout, as the \textit{readout memory}. The output is given by $\hat y(n)=\mathbf W_{\mathrm{out}}^{\mathsf T}\mathbf z_n$, where the readout weights are fitted by ridge regression on a training block as
\begin{equation}
\mathbf W_{\mathrm{out}}=(\mathbf Z^{\mathsf T}\mathbf Z+\lambda\mathbf I)^{-1}\mathbf Z^{\mathsf T}\mathbf y.
\label{eq:ridge}
\end{equation}
Here $\mathbf Z$ is the matrix whose rows are the vectors $\mathbf z_n^{\mathsf T}$ of the training block, $\mathbf y$ is the vector of the corresponding target outputs, $\lambda$ is the ridge parameter, and $\mathbf I$ is the identity matrix.

Every task sequence is divided into four blocks of consecutive symbols, each of which serves a different purpose (Section~\ref{sec:met_tasks}). The washout block allows the channel and the receptors to forget their initial state. The training block is used to fit the readout weights. The optimization block is used to evaluate the search objective and the readout settings, and every search minimizes the error on this block. The held-out block is excluded from all searches, and it serves to verify that the comparisons also hold on symbols that have not been seen by any selection. We define a \textit{configuration} as a particular choice of the six physical parameters together with $L$ and $\lambda$. For a fixed configuration, only the readout weights are fitted, whereas the physical parameters and the two readout settings, i.e., the readout memory $L$ and the ridge parameter $\lambda$, are selected by an outer search on the optimization block. In both models, $\lambda$ is chosen for every configuration from a fixed grid between $10^{-6}$ and $10^{3}$. The \textit{nominal configuration}, which serves as the reference configuration throughout the paper, consists of the parameter values listed in the caption of Table~\ref{tab:sets}, which are the same as those used in our previous study \cite{uzun2025molecular}. In Section~\ref{sec:res_budget}, we also consider three readout variants that trade the resolution of the state for the averaging of the receptor noise: integrating virtual nodes, which average the occupancy over their window instead of sampling it at a single instant; a causal low-pass filter with time constant $\tau_f$, which smooths the sequence of node values; and an extended readout memory of up to $L=20$ symbols.

The mean-field results reported in this paper are organized by four dimensionless groups. The first one is the ratio $\tau_D/T$, which compares the peak time of the impulse response with the symbol interval, and which we use as the metric for the \textit{diffusion memory}; note that $\phi_{\mathrm{ISI}}$ is a monotone function of this ratio. The second one is the product $k_{\mathrm{off}}T$, which we refer to as the \textit{scaled dissociation rate}. The relaxation time of the receptor at concentration $c$, normalized by the symbol interval, is $\tau_R/T=[k_{\mathrm{off}}T(1+c/K_D)]^{-1}$, and thus, the \textit{receptor timescale} depends jointly on the rates and the concentration. A receptor with $k_{\mathrm{off}}T\gg1$ follows the concentration within a symbol interval, whereas a receptor with $k_{\mathrm{off}}T\ll1$ integrates the concentration over several symbols at low occupancy. The third one is the ratio $c/K_D$, i.e., the \textit{operating point} of the receptor, which indicates its position on the binding curve. For $c$ well below $K_D$, the occupancy increases almost linearly with the concentration. At $c=K_D$, the occupancy is one half, and the response to relative changes in the concentration is the steepest. For $c$ well above $K_D$, the receptors approach saturation, where the sensitivity, relative to its value at low concentration, scales as $(1+c/K_D)^{-2}$, which approaches $(K_D/c)^{2}$. We describe a receiver by two coordinates, i.e., $k_{\mathrm{off}}T$ and $c_{\mathrm{peak}}/K_D$, where the latter is the operating point at the peak concentration of the task input. The fourth one is the integer $L$, which counts the symbols of explicit \textit{readout memory}. The first three groups are set by the physical parameters, whereas the last one is set by the readout. A fifth quantity, i.e., the \textit{receptor number} $N_R$, sets the size of the occupancy fluctuations, and it enters the noise results reported in Sections~\ref{sec:res_noise} to \ref{sec:res_budget}.

\subsection{Physically Tunable Parameters}
\label{sec:tunable}

In principle, seven parameters of the model are tunable. For each of them, \textbf{Table~\ref{tab:tunable}} lists the component that sets it, the mechanisms by which it can be changed, and the stage at which each mechanism is available, i.e., at design, at calibration of a built system, or during operation. The transmitter sets $T$ and $N_{\max}$. Programmable release is available wherever the release is actively driven, e.g., in hydrodynamically gated microfluidic transmitters \cite{bolhassan2024microfluidic}, engineered sender cells \cite{smith2023engineering}, synthetic cells with triggered cargo release \cite{monck2024synthetic}, and ion-channel-based modulators \cite{arjmandi2016modulator}. The receiver sets $k_{\mathrm{on}}$ and $k_{\mathrm{off}}$. At design time, the recognition element can be chosen from libraries of aptamers and engineered proteins, whose kinetic rates span orders of magnitude \cite{ding2024kineticitc,ding2023pushing,zhu2022modular}, and slow-off-rate selection directly targets the dissociation rate \cite{alkhamis2024slowoff}. The receptor density and the active fraction are set during surface preparation \cite{schoukroun2015heterogeneous}, and the kinetics of surface-bound aptamers can be measured in situ \cite{abeykoon2025interrogation,filius2025decoding}. On the receptor layer of a fabricated biosensor, it has been shown that temperature can modulate the dynamic range of electrochemical aptamer sensors in real time \cite{chen2022tunable}, and that an electric field can switch ligand binding and release at aptamer surfaces \cite{ma2017electrostatic}. Moreover, pH-responsive and allosteric aptamer switches are candidate switchable elements \cite{thompson2020ph,sun2024allosteric}. Note that these interventions change both rates jointly along a path that is specific to the receptor layer; hence, the two rates cannot be set independently. The channel sets $d$ and $D$. The distance is fixed by the placement of the transmitter and the receiver. The diffusion coefficient is determined by the carrier molecule and the medium through the Stokes-Einstein relation, i.e., $D=k_B\Theta/(6\pi\eta r_h)$, where $k_B$ is the Boltzmann constant, $\Theta$ is the absolute temperature, $\eta$ is the viscosity of the medium, and $r_h$ is the hydrodynamic radius of the carrier molecule. Therefore, both are design-time choices, although $D$ drifts with the temperature and the viscosity of the medium. The readout sets $L$, $M$, $\lambda$, and the filter time constant $\tau_f$, all of which require acquisition, storage, and computation downstream of the receptors. Lastly, the receptor number $N_R$ is set by the receptor density and the receiver area. The receptor number does not change the mean-field occupancy, and it enters the results only through the receptor-count noise. The last column of Table~\ref{tab:tunable} indicates the sets of tunable parameters, i.e., S0 to S4, which are defined in Section~\ref{sec:sets}.

\begin{table*}[t!]
\centering
\caption{Tunable parameters of the MC reservoir, together with the component that sets each parameter, the mechanisms that change it, and the tunable-parameter sets, defined in Table~\ref{tab:sets}, in which it is searched. Design, calibration, and runtime denote a choice of components before construction, a setting of a built system between runs, and an adjustment during operation through a specified actuator, respectively.}
\label{tab:tunable}
\small
\renewcommand{\arraystretch}{1.2}
\begin{tabularx}{\textwidth}{@{}p{1.45cm}p{1.75cm}XXp{3.1cm}@{}}
\toprule
\textbf{Parameter} & \textbf{Set by} & \textbf{Design-time mechanism} & \textbf{Calibration or runtime mechanism} & \textbf{Searched in} \\
\midrule
$T$, $N_{\max}$ & transmitter & release mechanism and encoding & programmed release waveform in microfluidic transmitters \cite{bolhassan2024microfluidic}, engineered sender cells \cite{smith2023engineering}, synthetic cells \cite{monck2024synthetic}, and ion-channel modulators \cite{arjmandi2016modulator}; imposed for a physiological source & S2, S3, S4 \\
$k_{\mathrm{on}}$, $k_{\mathrm{off}}$ & receiver, receptor layer & recognition element from aptamer or protein libraries \cite{ding2024kineticitc,ding2023pushing}, slow-off-rate selection \cite{alkhamis2024slowoff}, surface preparation \cite{schoukroun2015heterogeneous} & thermal \cite{chen2022tunable} or electrical \cite{ma2017electrostatic} modulation of the fabricated receptor layer, pH or allosteric switching \cite{thompson2020ph,sun2024allosteric}; both rates change together; imposed for native receptors & S1, S3, S4 \\
$d$ & geometry & placement of transmitter and receiver & adjustable between runs on a platform; fixed once implanted & S4 \\
$D$ & carrier and medium & carrier molecule and medium through the Stokes-Einstein relation & slow drift with the temperature and viscosity of the medium; imposed for a physiological biomarker in tissue & S4 \\
$L$, $M$, $\lambda$, $\tau_f$ & readout & acquisition and storage capacity & readout software; requires processing downstream of the receptor & $L$ and $\lambda$ in S0 to S4, all four in the readout designed for noise (Section~\ref{sec:res_budget}) \\
$N_R$ & receiver surface & receptor density and receiver area, surface preparation \cite{schoukroun2015heterogeneous}, expression level & fixed after preparation & noise level, 500 to $5\times10^{5}$ in the noise-aware searches, 500 to $10^{9}$ in the sweeps \\
\bottomrule
\end{tabularx}
\end{table*}

\subsection{Tunable-Parameter Sets in Representative MC Systems}
\label{sec:sets}

The parameters that a system can tune depend on which of its components play the roles of the transmitter, the medium, and the receiver. We consider four representative MC systems, and describe each of them as an MC channel (Figure~\ref{fig:system}c). In a biosensor implanted in tissue, the transmitter is the physiological source of the biomarker, the medium is the tissue, and the receiver is the sensor with its receptor layer. Such a system can choose its receptors and its readout, and it can modulate the binding kinetics of its receptor layer where such a mechanism exists; however, the release, the transport, and the placement are imposed \cite{downs2022realtime,kuscu2021graphene}. In an engineered source that signals to native cells, e.g., a drug-delivery implant or an engineered sender cell, the transmitter is the engineered part, and the receiver is an unmodified cell whose response is read externally. Such a system can control only the release \cite{arjmandi2016modulator,smith2023engineering}. In an engineered transceiver pair, e.g., sender and receiver cells with an orthogonal ligand and a synthetic receptor, or protocells with engineered signal transduction, both ends are engineered, whereas the medium and the geometry between the stationary nodes are imposed \cite{smith2023engineering,zhu2022modular,heuberger2024protocell,mashima2022shuttling}. Lastly, an in vitro microfluidic or chemical platform can select the ligand, the medium, the geometry, the recognition chemistry, the release waveform, and the readout, and it can change most of them between runs \cite{bolhassan2024microfluidic,baltussen2024chemical}.

\textbf{Table~\ref{tab:sets}} maps these systems onto five sets of tunable parameters, which we label as S0 to S4 and use throughout the paper. S0 tunes only the readout, i.e., $L$ and $\lambda$, at the nominal configuration, and it serves as the reference that is attainable by every system. S1 tunes the receptor kinetics and the readout (biosensor in tissue), S2 tunes the release parameters and the readout (engineered source), S3 tunes both ends and the readout (engineered transceiver pair), and S4 tunes all six physical parameters and the readout (in vitro platform). The sets are nested, i.e., S0 is contained in S1 and S2, both of which are contained in S3, which is in turn contained in S4; therefore, the best attainable error cannot increase as the set grows. However, the searches reported in this paper are finite, and a larger set is harder to search within the same budget. Since S1 and S2 have the same dimension, their comparison isolates the effect of where the tunable parameters are located in the channel from the effect of how many there are. Every parameter that a set does not tune is held at its nominal value. The transmitter-receiver distance is searched down to 6~$\mu$m, which keeps the point transmitter at least one receiver radius away from the surface of the receiver of radius 3~$\mu$m used in the particle model (Section~\ref{sec:met_bo}).

\begin{table}[H]
\centering
\caption{Tunable-parameter sets and the MC systems they represent. The nominal values, i.e., $k_{\mathrm{on}}=10^{-18}$~m$^3$~s$^{-1}$, $k_{\mathrm{off}}=1$~s$^{-1}$, $T=1.25$~s, $d=7.07~\mu$m, $N_{\max}=10100$, $D=3.16\times10^{-11}$~m$^2$~s$^{-1}$, are used for every parameter that a set does not tune, whereas the readout parameters $L$ and $\lambda$ are tuned in every set.}
\label{tab:sets}
\small
\begin{tabularx}{\textwidth}{@{}p{0.7cm}p{4.0cm}p{3.6cm}X@{}}
\toprule
\textbf{Set} & \textbf{Tuned} & \textbf{Held at nominal} & \textbf{Representative MC system} \\
\midrule
S0 & $L$, $\lambda$ & all six physical parameters & readout only, any system \\
S1 & $k_{\mathrm{on}}$, $k_{\mathrm{off}}$, $L$, $\lambda$ & $T$, $N_{\max}$, $d$, $D$ & biosensor in tissue reading a physiological source \\
S2 & $T$, $N_{\max}$, $L$, $\lambda$ & $k_{\mathrm{on}}$, $k_{\mathrm{off}}$, $d$, $D$ & engineered source signaling to native receptors \\
S3 & $k_{\mathrm{on}}$, $k_{\mathrm{off}}$, $T$, $N_{\max}$, $L$, $\lambda$ & $d$, $D$ & engineered transceiver pair, stationary nodes \\
S4 & all six physical parameters, $L$, $\lambda$ & none & in vitro platform \\
\bottomrule
\end{tabularx}
\end{table}
\section{Results}
\label{sec:results}

\subsection{Task-Specific Optimal Configurations}
\label{sec:res_optima}

\begin{figure*}[t!]
\centering
\includegraphics[width=\textwidth]{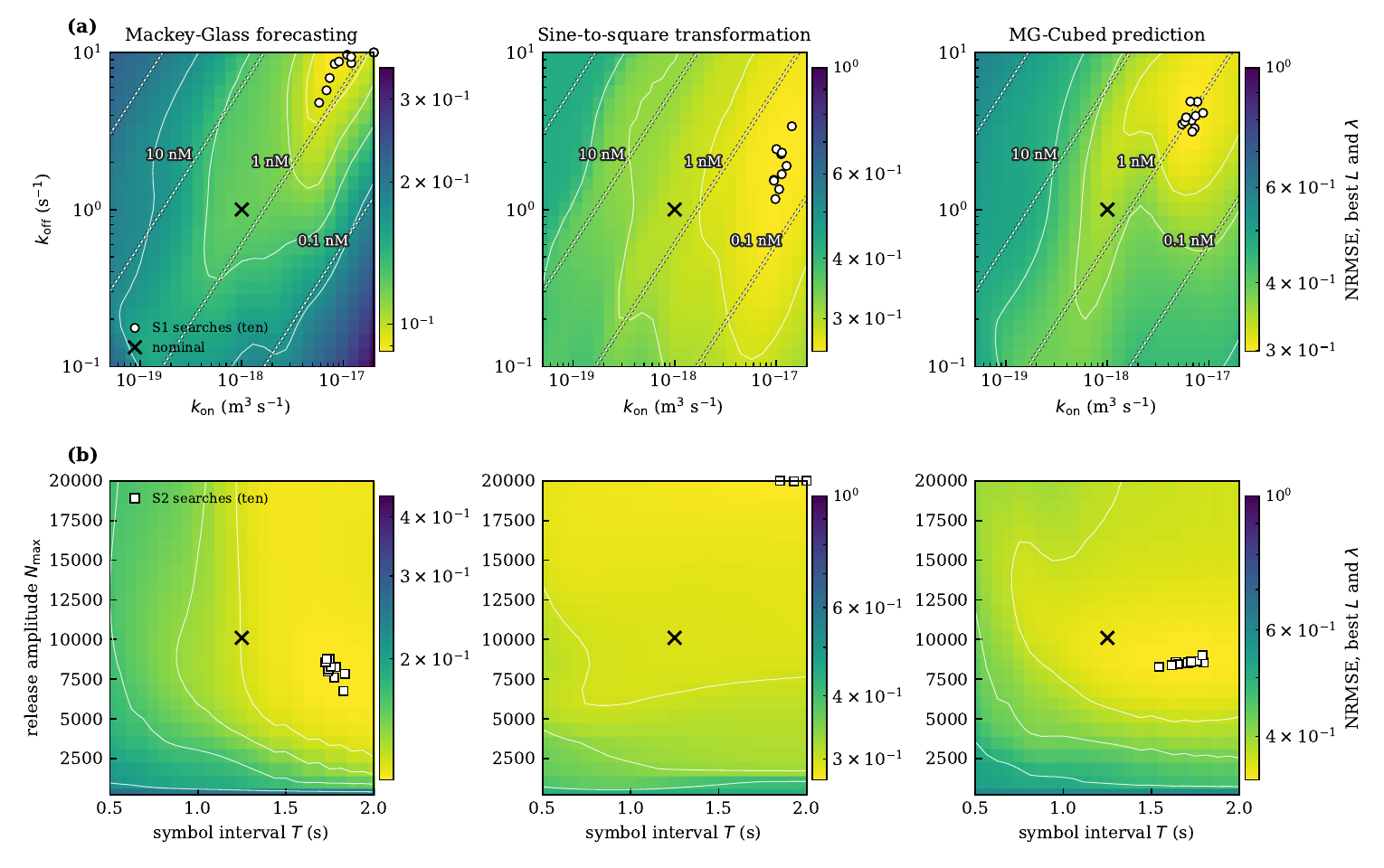}
\caption{Error surfaces over the receiver and the transmitter parameters for the three tasks (mean-field model). a) NRMSE over the association and dissociation rates at the nominal transmitter and channel settings, with the best $L$ and ridge value at every grid point. Dotted lines are contours of constant $K_D$, white circles mark the optima of the ten receiver-only searches (S1), and the cross marks the nominal receptor. b) NRMSE over the symbol interval and the release amplitude at the nominal receiver and channel settings, with the optima of the ten transmitter-only searches (S2) shown as white squares. Solid white contours enclose the regions in which the NRMSE lies within 10, 25, 50, and 100 percent of the minimum of each panel. The color scale is logarithmic and specific to each panel.}
\label{fig:landscapes}
\end{figure*}

We first consider the case in which every parameter is tunable, i.e., the set S4. For each task, Bayesian optimization over all six physical parameters and the two readout settings was run ten times, with 100 evaluations per run (Section~\ref{sec:met_bo}). The best configuration found for a task, which we refer to as its \textit{S4 optimum}, achieves a normalized root-mean-square error (NRMSE) of 0.087 for Mackey-Glass (MG) forecasting, 0.231 for sine-to-square transformation, and 0.285 for MG-Cubed prediction on the optimization block (\textbf{Table~\ref{tab:optima}}). The three optima differ in several physical parameters and in the dimensionless groups defined in Section~\ref{sec:system}, whereas the association rate of all three optima and the dissociation rate of the optima for the two MG tasks lie at the boundaries of the search domain. To reveal the structure underlying these optima, we computed the error over two regular grids (\textbf{Figure~\ref{fig:landscapes}}, Section~\ref{sec:met_landscapes}). The first grid spans the two receiver parameters at the nominal transmitter and channel settings, and the second grid spans the two transmitter parameters at the nominal receiver and channel settings. In both grids, the best readout memory $L$ and ridge value are selected at every grid point.

As shown in Figure~\ref{fig:landscapes}a, the receiver error surfaces of the three tasks, i.e., their error surfaces over the two receptor rates, differ in both position and shape. For forecasting, the minimum error of 0.088 is attained at $K_D\approx1.8$~nM and $k_{\mathrm{off}}T\approx12.5$, in a narrow valley along a line of constant $K_D$, which ends on the upper boundary of the dissociation-rate domain. Only 4 percent of the grid lies within 10 percent of this minimum. For sine-to-square, the minimum error of 0.256 is attained at $K_D\approx0.29$~nM and $k_{\mathrm{off}}T\approx2.7$, in a broad basin covering 29 percent of the grid. MG-Cubed lies between the two, with a minimum error of 0.299 at $K_D\approx0.91$~nM and $k_{\mathrm{off}}T\approx4.0$. The ten receiver-only searches (S1) end in these minima. For sine-to-square and MG-Cubed, their mean errors, 0.257 and 0.300, agree with the grid minima to within 0.002. For forecasting, 8 of the ten searches end within 0.005 of the grid minimum, whereas the remaining ones end in the corner of the domain at 0.099, since the valley is narrow and the search budget is small. On the other hand, the transmitter error surfaces are shallow (Figure~\ref{fig:landscapes}b). Their minima, 0.111, 0.273, and 0.339, lie only 9, 7, and 2 percent below the error of the nominal channel, respectively, and 41 to 62 percent of each grid lies within 10 percent of the minimum. For the two MG tasks, the low-error region extends from large release amplitudes at short symbol intervals to smaller amplitudes at long symbol intervals. Along this region, the receptor operating point remains approximately constant, whereas the diffusion memory per symbol changes.

\begin{table*}[t!]
\centering
\caption{Best configuration found by the ten S4 searches for each task (mean-field model), together with the NRMSE on the optimization block and the dimensionless groups defined in Section~\ref{sec:system}. Here, $c_{\mathrm{peak}}$ is the maximum concentration at the receiver under the task input, and $b_{\max}$ is the maximum occupancy on the optimization block.}
\label{tab:optima}
\small
\setlength{\tabcolsep}{3.5pt}
\begin{tabular}{@{}lcccccccccccc@{}}
\toprule
\textbf{Task} & $k_{\mathrm{on}}$ (m$^3$ s$^{-1}$) & $k_{\mathrm{off}}$ (s$^{-1}$) & $T$ (s) & $d$ ($\mu$m) & $D$ (m$^2$ s$^{-1}$) & $N_{\max}$ & $L$ & NRMSE & $k_{\mathrm{off}}T$ & $c_{\mathrm{peak}}/K_D$ & $\tau_D/T$ & $b_{\max}$ \\
\midrule
MG & $2.00\times10^{-17}$ & 10.00 & 1.87 & 11.1 & $5.75\times10^{-11}$ & 15597 & 5 & 0.087 & 18.7 & 2.6 & 0.19 & 0.72 \\
Sine-to-square & $2.00\times10^{-17}$ & 2.37 & 2.00 & 6.0 & $8.80\times10^{-11}$ & 20000 & 5 & 0.231 & 4.7 & 59.5 & 0.03 & 0.98 \\
MG-Cubed & $2.00\times10^{-17}$ & 10.00 & 1.52 & 7.6 & $6.68\times10^{-11}$ & 20000 & 5 & 0.285 & 15.2 & 8.0 & 0.10 & 0.89 \\
\bottomrule
\end{tabular}
\end{table*}

\begin{figure*}[t!]
\centering
\includegraphics[width=\textwidth]{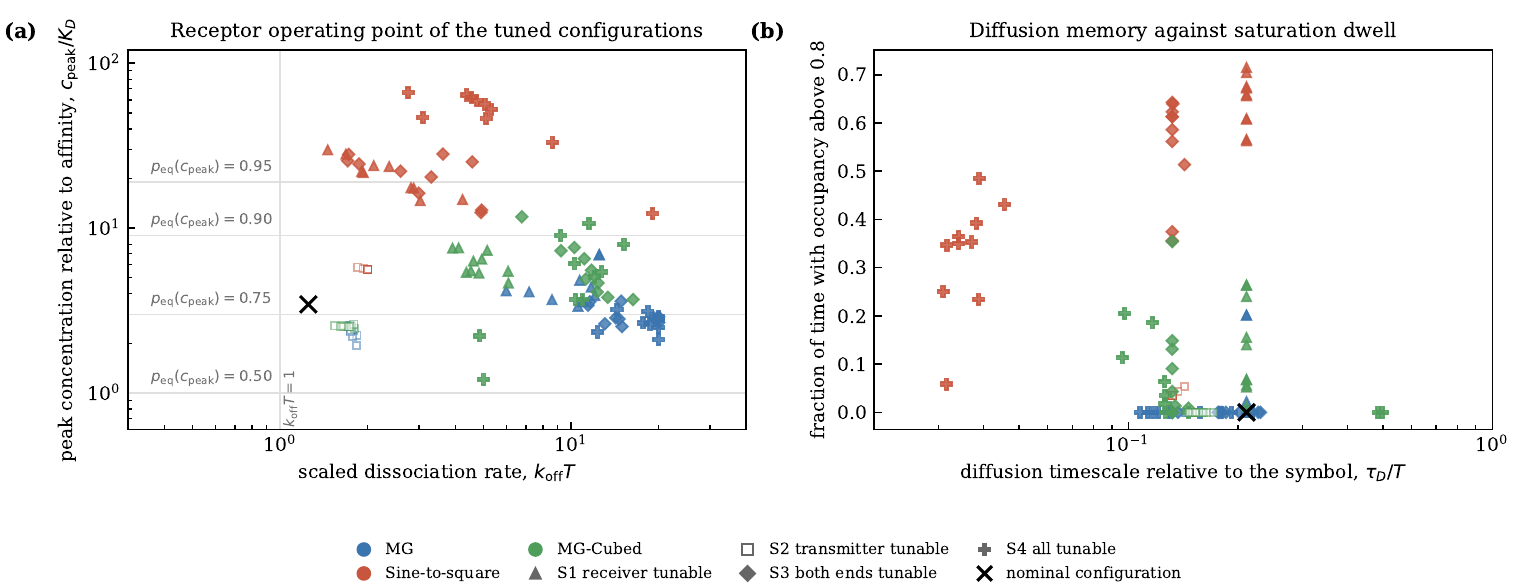}
\caption{Physical characteristics of the tuned configurations (mean-field model). Each marker represents the best configuration of one search, colored by task, with triangles, squares, diamonds, and plus signs for the sets S1 to S4, respectively. The cross indicates the nominal configuration under the MG input, since the concentration metrics depend on the task input. a) Peak concentration at the receiver relative to the dissociation constant against the scaled dissociation rate. Horizontal reference lines indicate the equilibrium occupancy at the peak concentration, $p_{\mathrm{eq}}(c_{\mathrm{peak}})$. Under S2, the receptor is fixed, and thus, those configurations (open squares) vary only through the release amplitude and the symbol interval. b) Diffusion timescale relative to the symbol interval against the fraction of the optimization block spent above an occupancy of 0.8.}
\label{fig:operating}
\end{figure*}

\textbf{Figure~\ref{fig:operating}} shows the best configurations of all 120 mean-field searches in the plane spanned by the two receptor coordinates defined in Section~\ref{sec:system}, i.e., the scaled dissociation rate and the operating point at the peak concentration. The three tasks occupy separate regions of this plane. Whenever the receptor is tunable, the configurations tuned for forecasting have a fast receptor, with $k_{\mathrm{off}}T$ between 6.0 and 20.0 and a median of 14.3, and a peak concentration of about three times $K_D$ (median $c_{\mathrm{peak}}/K_D=3.0$). Such a receptor follows the concentration within a symbol interval and operates in the responsive part of its binding curve, with a maximum occupancy of about 0.75 and no time spent above an occupancy of 0.8. The dissociation rate lies on the upper boundary of the domain in 17 of the 30 S4 searches, including every S4 search for forecasting; hence, the selected forecasting configurations favor the fastest receptor that the domain allows. In contrast, the configurations tuned for the transformation task have a slower receptor, with $k_{\mathrm{off}}T$ between 1.5 and 19.1 and a median of 3.1, and a peak concentration of 12.3 to 66.5 times $K_D$ (median 24.9). This drives the receptors into saturation for about half of the time, with a median fraction of 0.56 above an occupancy of 0.8, such that the occupancy approaches a threshold function of the recent concentration, which is the operation required by the task. The MG-Cubed configurations lie between the two (median $k_{\mathrm{off}}T=9.7$, $c_{\mathrm{peak}}/K_D=5.5$, and fraction above 0.8 occupancy 0.06). In terms of affinity, the forecasting receivers have $K_D$ values near 1.4~nM and the transformation receivers near 0.27~nM, both of which lie within the range of practical aptamers \cite{ding2024kineticitc,ding2023pushing}. The nominal receptor, with $k_{\mathrm{off}}T=1.25$ and $c_{\mathrm{peak}}/K_D=3.4$, is slower than forecasting requires and less saturated than transformation requires, which explains why the nominal channel suits neither task and why tuning improves both. The tasks thus differ in the position of the operating point on the binding curve and in the receptor timescale, and none of them requires an unusual receptor.

The diffusion memory of the tuned configurations is shown in Figure~\ref{fig:operating}b. When the channel itself is tunable (S4), the optimizer selects the shortest diffusion memory that the domain allows for the transformation task, i.e., $\tau_D/T$ between 0.03 and 0.05, with the transmitter at the 6~$\mu$m boundary and a fast diffusion coefficient in every search. For forecasting, on the other hand, it keeps $\tau_D/T$ between 0.11 and 0.22. For all three tasks, the optimizer selects the maximum readout memory $L=5$ in 114 of the 120 searches, since no storage cost enters the objective, and it selects the smallest ridge value of the grid in every search. Note that the readout memory plays an important role. Here, the best configurations found when the receptor kinetics are the only tunable physical parameters are referred to as the \textit{S1 optima}. When their readout is restricted to the current state ($L=1$) and the channel is kept unchanged, the S1 optima yield errors of 0.301, 0.273, and 0.531 for forecasting, transformation, and MG-Cubed, respectively (Supporting Information Table~S22). Hence, the channel alone provides only part of the memory that the tasks require.

Several optima lie on the boundaries of the search domain, i.e., at the fastest dissociation rate for forecasting, and at the shortest distance and the longest symbol interval for transformation. To check the effect of these boundaries, we ran additional searches in a wider domain, which extends the dissociation rate up to 100~s$^{-1}$, the symbol interval up to 5~s, the release up to $10^{5}$ molecules, and the distance down to 3.5~$\mu$m. In the mean-field model, the wider domain changes the best S1 and S4 errors by at most 2 and 3 percent, respectively, and the tasks remain separated in the receptor plane (Supporting Information Section~S7). Under noise at 50000 receptors, the wider domain reduces the errors by 0 to 9 percent, with the largest reduction for forecasting, whose noise-aware optimum under S1 shifts to a dissociation rate of 79~s$^{-1}$. Consequently, the boundaries of the domain limit the numerical values, but they do not change the ordering of the tasks.

\FloatBarrier
\subsection{Performance under Partial Tunability}
\label{sec:res_access}

\begin{figure*}[t!]
\centering
\includegraphics[width=\textwidth]{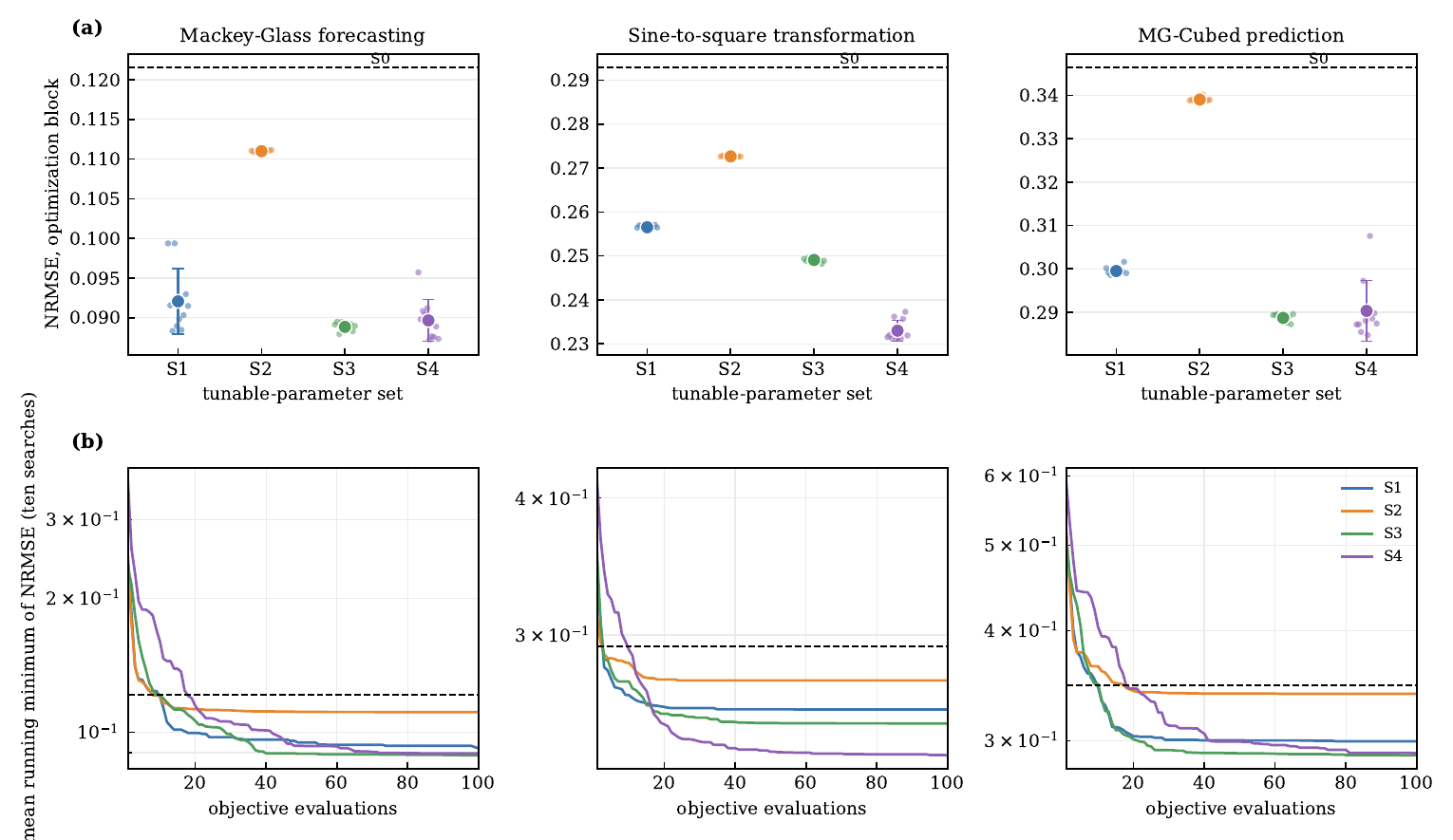}
\caption{Performance under the tunable-parameter sets (mean-field model). a) NRMSE on the optimization block for the best configuration of each of ten independent searches (small points), together with their mean and sample standard deviation (large markers), for the receiver-only (S1), transmitter-only (S2), both-ends (S3), and all-parameter (S4) sets, each of which also tunes $L$ and $\lambda$. The dashed line indicates the nominal channel with a tuned readout (S0). b) Mean over the ten searches of the lowest NRMSE found within the first $n$ objective evaluations, as a function of $n$.}
\label{fig:sets}
\end{figure*}

We now investigate how much of this gain is retained when only some of the parameters are tunable. \textbf{Figure~\ref{fig:sets}}a shows the errors achieved by ten independent searches under each of the tunable-parameter sets listed in Table~\ref{tab:sets}, with the nominal channel and a tuned readout (S0) as the reference (the values on both evaluation blocks are listed in Supporting Information Table~S6). We make three observations. First, tuning only the receptor kinetics (S1) lowers the error on every task, i.e., by 24, 12, and 14 percent relative to S0 for forecasting, transformation, and MG-Cubed, respectively. We refer to this reduction of the error relative to S0 as the \textit{gain} of a set. Second, tuning only the release parameters (S2) yields gains of 9, 7, and 2 percent, and the ten S2 searches agree to within 0.0004, which indicates that they all find the same shallow minimum, and that this minimum offers little gain. The reason can be seen in Figure~\ref{fig:operating}a. With the receptor fixed, $k_{\mathrm{off}}T$ can be changed only through $T$, i.e., between 1.5 and 2.0. As a result, the receptor remains slow, and $N_{\max}$ alone shifts the operating point along the binding curve. Release tuning changes the concentration waveform, the operating point, and the concentration-dependent relaxation time; however, it cannot change the rate constants of the receptor. Within the tested domain, this restricted access yields smaller gains than tuning the receptor kinetics together with the readout. Third, joint tuning of both ends (S3) and tuning of all parameters (S4) improve every task over S0, by 15 to 27 percent and by 16 to 26 percent, respectively. On the two MG tasks, S3 and S4 achieve the same error within the variability across searches, and both gain only about 3 percent over S1. Thus, the receptor kinetics alone recover 92 and 84 percent of the S4 gain on these tasks. On sine-to-square, S4 gains 9 percent over S1, and S1 recovers 61 percent of the S4 gain, since a short diffusion memory is an advantage that only the channel parameters can provide (Figure~\ref{fig:operating}b).

The convergence curves in Figure~\ref{fig:sets}b reveal the cost of the larger sets. After 20 evaluations, the receiver-only search for forecasting reaches a mean running minimum of 0.100, against 0.115 for the seven-parameter search, and the two curves meet only near the end of the budget, at 0.092 and 0.090, respectively. The same ordering holds on the held-out block, which was not used by any search. These results have direct implications for MC systems. A biosensor whose only tunable part is its receptor layer can recover most of the attainable gain, whereas an engineered transceiver pair should be co-designed at both ends, since release tuning alone yields little gain.

\FloatBarrier
\subsection{Cost of Reusing a Configuration across Tasks}
\label{sec:res_reuse}

The simplest response to a new task is to keep the channel unchanged and retrain only the readout. The cost of this approach in the mean-field model is quantified in Supporting Information Figure~S2. For each tunable-parameter set and each ordered pair of tasks, the best configuration of each of the ten searches for the source task is retained. Each of these configurations is then evaluated on the target task after the readout is retrained on the training block of the target task, with $L$ and the ridge value re-selected for the target (Section~\ref{sec:met_reuse}). The median error over these ten configurations is compared with the median error of the ten searches tuned for the target task within the same set. In 24 of the 24 combinations of set and task pair, the reused configuration has the higher median error, by 0.1 to 58 percent (Supporting Information Table~S12). The penalty is small between the two MG tasks, i.e., 0 to 7 percent, since they share an input series and a requirement for memory, whereas the largest penalties involve the transformation task. A receiver tuned for sine-to-square and reused for forecasting incurs a penalty of 2 to 58 percent, and in 6 of the 24 combinations, the reused configuration performs even worse than the untuned nominal channel with a tuned readout. The reason lies in the receptor operating point shown in Figure~\ref{fig:operating}a. A receptor that is driven into saturation for half of the time compresses the amplitude differences that forecasting requires, and the linear readout refitted for forecasting does not restore them. The mean penalty over the six task pairs is 5 percent for S2, 11 percent for S1, 11 percent for S3, and 22 percent for S4. The same trend is observed on the held-out block, where 24 of the 24 combinations incur a penalty (Supporting Information Table~S13). Therefore, for these task-specific configurations, retraining the readout leaves a task-dependent penalty, and it cannot substitute for retuning, except between closely related tasks.

Retuning, however, is not the only alternative. A system that has to serve all three tasks without retuning can instead be tuned for the worst case. To this end, a minimax search minimizes the largest, i.e., worst-case, ratio between the error on a task and the task-specific optimum of the same set (Section~\ref{sec:met_compromise}). In the mean-field model, this search finds a single S1 configuration that remains within 4 percent of the task-specific optimum on every task, and a single S4 configuration that remains within 8 percent (Supporting Information Table~S24). Under noise at 50000 receptors, the corresponding values are 5 and 9 percent. The minimax configuration lies at the forecasting end of the receptor plane ($k_{\mathrm{off}}T=10.2$ and $c_{\mathrm{peak}}/K_D=4.7$ under S1), since the transformation task tolerates an unsaturated receptor better than forecasting tolerates a saturated one.

\FloatBarrier
\subsection{Receptor-Count Noise and the Validity of the Mean-Field Design}
\label{sec:res_noise}

\begin{figure*}[t!]
\centering
\includegraphics[width=\textwidth]{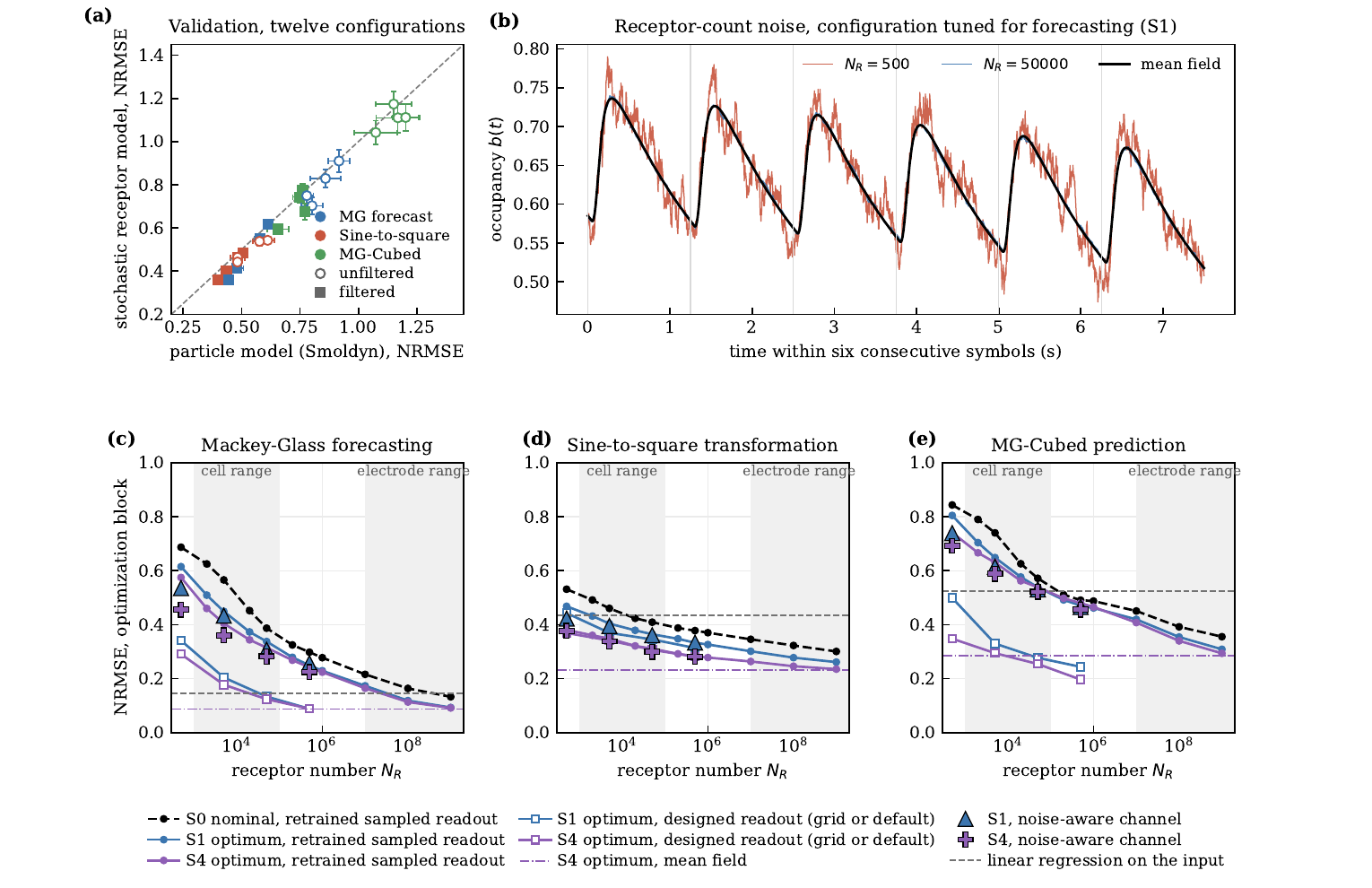}
\caption{Receptor-count noise. a) Task errors of the stochastic receptor model against those of the particle-based simulations for the twelve validation configurations, before (open) and after (filled) the trailing moving-average filter, with the mean and standard deviation over five particle realizations and 50 receptor-model realizations. The dashed line indicates equality. b) Mean-field occupancy and one realization of the stochastic receptor model with 500 and with 50000 receptors for the S1 configuration tuned for forecasting, over six consecutive symbols of the optimization block. c) to e) NRMSE as a function of the receptor number, from 500 to $10^{9}$, for the mean-field optima of S0, S1, and S4, re-evaluated with independent realizations, with $L$ and the ridge value retrained under noise (filled circles). Open squares indicate the designed readout, i.e., the lower optimization-block error of the integrating-node grid and its paired sampled default on the same realizations, where the grid admits a low-pass filter and up to 20 symbols of readout memory (Section~\ref{sec:res_budget}); the designed readout was evaluated up to $5\times10^{5}$ receptors. Large markers indicate the noise-aware optima of S1 and S4 at their design receptor numbers. The dash-dotted line indicates the mean-field value of the S4 optimum, and the dashed gray line indicates the linear regression on the transmitted input. Shaded bands mark illustrative receptor-count ranges for cells and for electrochemical aptamer electrodes.}
\label{fig:noise}
\end{figure*}

The mean-field model treats the occupancy as a deterministic fraction, whereas a practical receiver counts the bound receptors, and this count fluctuates. We evaluate the effect of these fluctuations with the stochastic receptor model described in Section~\ref{sec:system}, in which the same concentration drives a birth-death process of $N_R$ receptors, and each realization is read out with its own trained readout. \textbf{Figure~\ref{fig:noise}}a validates this model against particle-based simulations of twelve configurations with 500 receptors and five realizations each, where the outputs of the stochastic model were processed with the same filter and ridge value as those of the particle simulations (Section~\ref{sec:met_particle}). These configurations were selected in a preliminary set of searches, whose domain admitted transmitter placements closer to the receiver than the domain used for the results above. The stochastic receptor model reproduces the task errors obtained from the particle simulations with a mean difference of -0.037 after filtering and -0.041 before filtering, a largest difference of 0.095 and 0.098, respectively, and a Pearson correlation of 0.991 over the raw and filtered values together (Supporting Information Table~S14). The largest deviations belong to the S4 configurations of this set, whose transmitter was located within 1.2~$\mu$m of the receiver surface, where the point-receiver concentration given by Equation~\eqref{eq:channel} is a likely contributor to the difference. Note that the domain of the main searches excludes such placements (Section~\ref{sec:met_bo}). Hence, for these twelve configurations at 500 receptors, the stochastic receptor model closely reproduces the task errors obtained from the particle simulations, which supports its use as an efficient approximation for the receptor-count effects studied below. However, two limitations of this validation should be noted. First, the comparison does not separate the contributions of molecule-number fluctuations, receiver geometry, depletion, and rebinding. Second, since it covers twelve configurations at a single receptor number, it does not validate every configuration of the subsequent sweeps. The practical value of the stochastic receptor model lies in its speed, as it evaluates a configuration in seconds, whereas a particle simulation takes hours.

Figure~\ref{fig:noise}b illustrates the effect of the noise on the reservoir state. With 500 receptors, a single sample of the occupancy fluctuates by $\sqrt{p(1-p)/N_R}\approx0.02$, which is of the order of the variation of the occupancy across the $M$ virtual nodes within one symbol. As a result, the within-symbol structure that distinguishes the virtual nodes is lost in the noise, and only the symbol-to-symbol variation persists. With 50000 receptors, on the other hand, the fluctuation is ten times smaller, and the transient is resolved. A capacity analysis reveals where the loss occurs (Supporting Information Section~S10). For the S1 forecasting optimum, the linear memory capacity decreases from 11.3 in the mean-field model to 4.8 at 500 receptors, and the second-order capacity decreases from 11.3 to 0.8. The memory that persists is concentrated at the delays that the readout retains in its $L$ past states.

Figure~\ref{fig:noise}c to e show the consequences for the \textit{mean-field optima}, i.e., the configurations found in Sections~\ref{sec:res_optima} and \ref{sec:res_access} with the noiseless model. With the fixed readout, i.e., with $L$ and $\lambda=10^{-6}$ of the mean-field optimum, the errors of the S1 optima at 500 receptors are 0.849, 0.541, and 1.108 for forecasting, transformation, and MG-Cubed, respectively, whereas the corresponding mean-field values are 0.088, 0.256, and 0.299. This is the failure exposed by the particle simulations: a readout fitted with a very small ridge value to hundreds of strongly correlated features amplifies the noise, which it cannot separate from the signal. When $L$ and $\lambda$ are re-selected on the noisy realizations, the same channels yield 0.615, 0.468, and 0.805, with the ridge value raised to $1$, $1$, and $1$, respectively. Retraining thus removes most of the failure. Moreover, the ridge value, which the mean-field rule leaves at the smallest grid value for the benchmark optima, becomes the first parameter to adjust when a receiver counts receptors.

The remaining error decreases slowly with the receptor number. From 500 to $5\times10^{5}$ receptors, the retrained S1 optimum improves by a factor of 2.5, 1.4, and 1.7 on the three tasks, whereas the binomial standard deviation of the occupancy at a fixed $p$ decreases by a factor of 32 over the same range. The task error decreases more slowly than this count-noise scaling, since the readout that averages the noise also discards the within-symbol structure, and this loss of resolution limits the error at large receptor numbers. For this reason, we continued the sweep to the receptor numbers of artificial engineered receivers, such as electrochemical aptamer electrodes. Electrochemical aptamer sensors carry probe densities of $10^{11}$ to $10^{12}$ per cm$^2$ \cite{schoukroun2015heterogeneous,downs2022realtime}, which corresponds to $1.1\times10^{5}$ to $1.1\times10^{6}$ receptors on a receiver of the size simulated here, i.e., with a radius of 3~$\mu$m, and to $10^{7}$ to $10^{10}$ receptors on an electrode of 0.01 to 1~mm$^2$. For comparison, a cell carries on the order of $10^{3}$ to $10^{5}$ receptors of one type \cite{lauffenburger1993receptors,li2004cellular}. With the same readout, the forecasting error reaches 0.173 at $10^{7}$ receptors, 0.118 at $10^{8}$, and 0.093 at $10^{9}$, against 0.088 in the mean-field model. The transformation and MG-Cubed errors approach their mean-field values in the same way, reaching 0.235 and 0.294 at $10^{9}$ receptors, against 0.231 and 0.285, respectively. Hence, for these selected configurations and this readout scheme, the stochastic errors approach the corresponding mean-field errors at $10^{8}$ to $10^{9}$ receptors in the model.

To assess these values, we compare them with a linear regression on the transmitted input, which we refer to as the \textit{linear reference} of a task (Section~\ref{sec:met_tasks}). This is a digital reference fitted to the noise-free transmitted symbols with the same blocks and ridge rule, using the best of the tested tap lengths. Note that it is not a lower bound for a readout of nonlinear receptor features, and the mean-field channel lies below it on most tasks. An error below this reference indicates an advantage over a linear filter of the noise-free input. An error above it, on the other hand, indicates that the tested occupancy-based pipeline, with its acquisition, readout, and selection procedure, performs worse than this filter; however, it does not by itself isolate an information loss in transport or reception. The reference achieves 0.147 for forecasting and 0.526 for MG-Cubed. For sine-to-square, a fixed-coefficient linear filter of the sinusoidal input cannot reproduce the harmonics of the square wave, and its ideal periodic error is $\sqrt{1-8/\pi^{2}}\approx0.435$, which is reproduced by the computed baseline. In the mean-field model, the tuned channel lies below the reference on all three benchmark tasks, by a wide margin for forecasting (0.087 against 0.147). Under noise, however, the outcome depends on the task. The retrained forecasting optima remain above the reference up to $5\times10^{5}$ receptors, where the S4 optimum reaches 0.242, and they fall below it at about $2.2\times10^{7}$ receptors. MG-Cubed falls below its reference at about $7.8\times10^{4}$ receptors, whereas the transformation task remains below the linear error floor down to 500 receptors under S4 and 2000 receptors under S1. When a task requires fine amplitude differences held in memory, the count noise removes this information before any readout can access it, whereas when a task requires a threshold, the saturating receptor keeps supplying it. In Section~\ref{sec:res_budget}, we turn this observation into a critical receptor number for every task, i.e., the receptor number that a task requires before the channel outperforms its linear reference.

\FloatBarrier
\subsection{Noise-Aware Optimization}
\label{sec:res_noiseaware}

\begin{figure*}[t!]
\centering
\includegraphics[width=\textwidth]{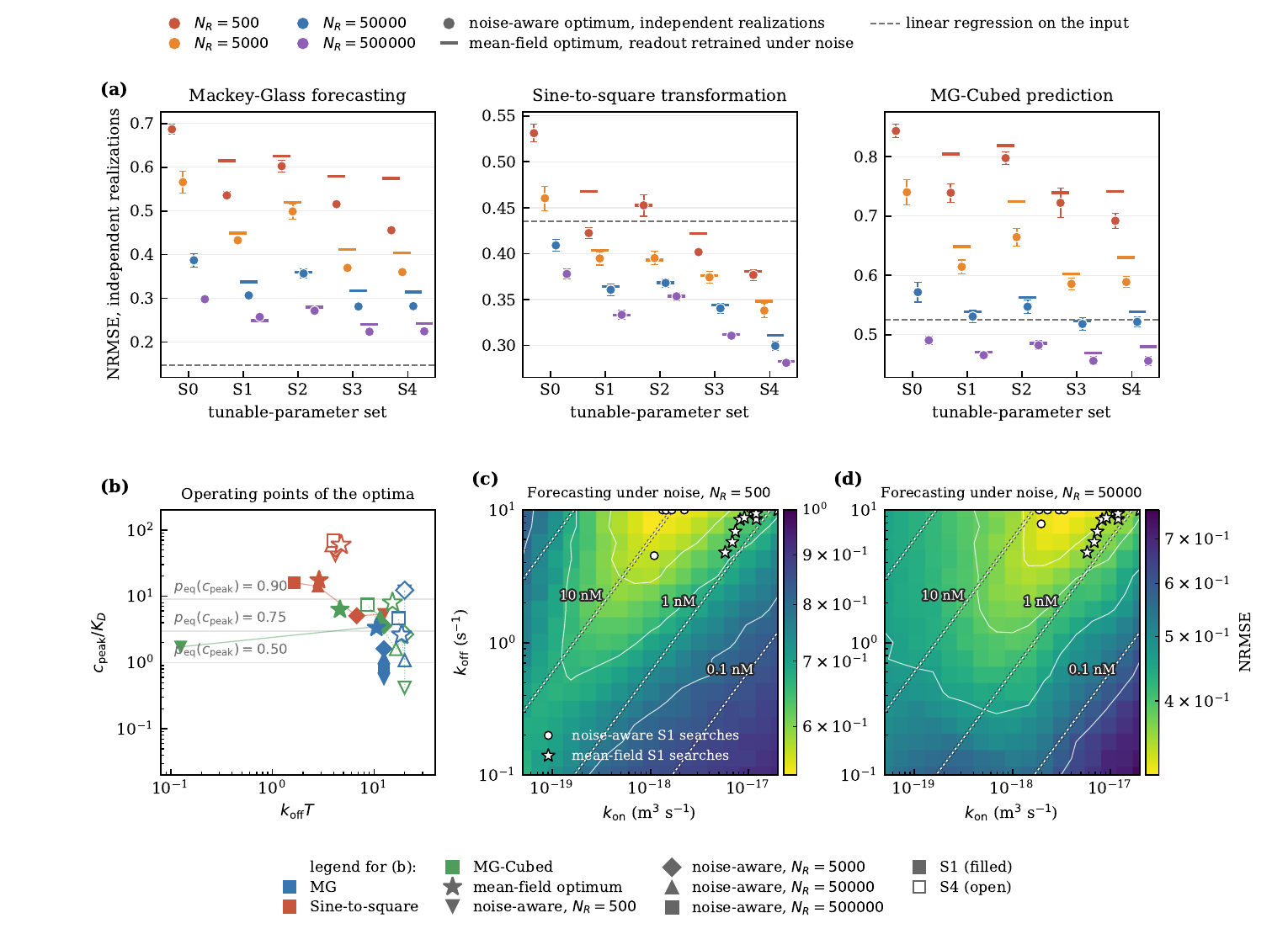}
\caption{Noise-aware design. a) NRMSE of the noise-aware optima of each tunable-parameter set at four receptor numbers (circles, mean and standard deviation over 50 independent realizations), compared with the mean-field optima of the same sets with the readout retrained under the same noise (horizontal bars) and with the nominal channel with a retrained readout (S0). The dashed line indicates the best linear regression on the transmitted input. b) Receptor operating points of the noise-aware optima of S1 (filled) and S4 (open) at each receptor number (marker shape), colored by task and connected in the order of receptor number, with the mean-field optima shown as stars. Horizontal reference lines indicate the equilibrium occupancy at the peak concentration. c) and d) Receiver error surfaces for forecasting under noise at 500 and at 50000 receptors, with the best $L$ and ridge value at every grid point and ten realizations per point, together with the optima of the noise-aware S1 searches (circles) and those of the mean-field S1 searches (stars).}
\label{fig:noiseaware}
\end{figure*}

The mean-field optima were designed for a noiseless receiver. Therefore, we repeated the searches under S1 to S4 with the stochastic receptor model as the objective, at $N_R=500$, 5000, 50000, and $5\times10^{5}$, and we refer to the resulting configurations as the \textit{noise-aware optima} (Section~\ref{sec:met_noise}). \textbf{Figure~\ref{fig:noiseaware}}a compares the noise-aware optima, re-evaluated with 50 independent realizations of the receptor process that were not seen by any search, with the mean-field optima of the same sets after their readout is retrained under the same noise (Supporting Information Table~S18). Noise-aware tuning improves on the retrained mean-field optimum most for forecasting, i.e., by 13 percent under S1 and 21 percent under S4 at 500 receptors, and by 9 and 10 percent, respectively, at 50000 receptors. For the transformation task, the gain is 10 percent under S1 at 500 receptors and within a few percent otherwise, and for MG-Cubed, it is at most 8 percent. Hence, for these two tasks, the mean-field channel with a retrained readout is already close to the best that the domain offers under noise, whereas forecasting benefits from retuning the channel. The gain decreases with the receptor number, down to 7 percent for forecasting at $5\times10^{5}$ receptors under S4, where the noise-aware error is still 2.6 times the mean-field value. As in the mean-field model, S3 and S4 generally yield the lower errors, and several within-task differences are small relative to the variability across realizations.

The noise-aware optima differ from the mean-field optima in more than their ridge value. Figure~\ref{fig:noiseaware}b traces the receptor operating points of the S1 and S4 optima from 500 receptors to the mean-field limit. For forecasting, the searches select fast receptors at every receptor number, at the upper boundary of the dissociation rate in all but one case (8.69~s$^{-1}$ under S4 at $5\times10^{5}$ receptors); however, the operating point changes. Under S1, the optimum at 500 receptors has a peak concentration of 0.57 $K_D$, well inside the linear part of the binding curve, against 3.0 $K_D$ for the mean-field configurations tuned for forecasting. Across the four receptor numbers, the selected S1 values span 0.57 to 1.62 $K_D$. A low occupancy keeps the receptor response linear, and a linear response can be averaged by the readout without distortion. The ridge value of these optima is $1$ at 500 receptors, $10^{-1}$ at 5000, $10^{-2}$ at 50000, and $10^{-3}$ at $5\times10^{5}$, i.e., one decade lower for every decade of receptor number. The ridge spreads the weights over the $ML$ correlated nodes, and thus, acts as the averaging filter that we applied manually to the particle-simulation outputs (Section~\ref{sec:met_particle}). Under S4, the forecasting optimum at 500 receptors lies at the other end of the binding curve, in saturation, with $c_{\mathrm{peak}}/K_D=12$, and it has the shortest diffusion memory of the domain, i.e., $\tau_D/T=0.02$, against 0.19 in the mean-field model. This shifts the memory from the channel to the $L$ retained states of the readout, where the model stores it without additional readout noise. For the transformation task, the optima remain in saturation at every receptor number, since the threshold-like response required by the task is also the response least affected by count noise near full occupancy. For MG-Cubed, the S1 optimum at 500 receptors is the slowest receptor of the domain, with $k_{\mathrm{off}}T=0.1$, which integrates the concentration over several symbols, and thereby trades temporal resolution for averaging within the receptor. From 5000 receptors onward, the optimum is again a fast receptor. The receiver error surfaces in Figure~\ref{fig:noiseaware}c and d show the same shift across the whole receptor plane. At 500 receptors, the narrow valley of the mean-field error surface (Figure~\ref{fig:landscapes}a) is replaced by a broad region along the fast edge of the domain, with its minimum at low affinity, whereas at 50000 receptors, the valley reappears.

We make two further observations (Supporting Information Section~S6). First, the readout memory $L=5$ remains the choice of nearly every noise-aware search, and sampled and integrating virtual nodes yield the same errors within the variability across realizations for the fast receptors selected here, since the receptor fluctuations are correlated over the node window. Second, reusing a configuration across tasks incurs a similar cost under noise as in the mean-field model. At 500 receptors, the noise-aware optimum of one task, when reused on another task with a retrained readout, performs worse than the noise-aware optimum of the target task in 23 of the 24 combinations, by up to 51 percent. As in the mean-field model, the penalty is concentrated on the pairs involving the transformation task.

\FloatBarrier
\subsection{Critical Receptor Numbers and Readouts Designed for Noise across Tasks}
\label{sec:res_budget}

\begin{figure*}[t!]
\centering
\includegraphics[width=\textwidth]{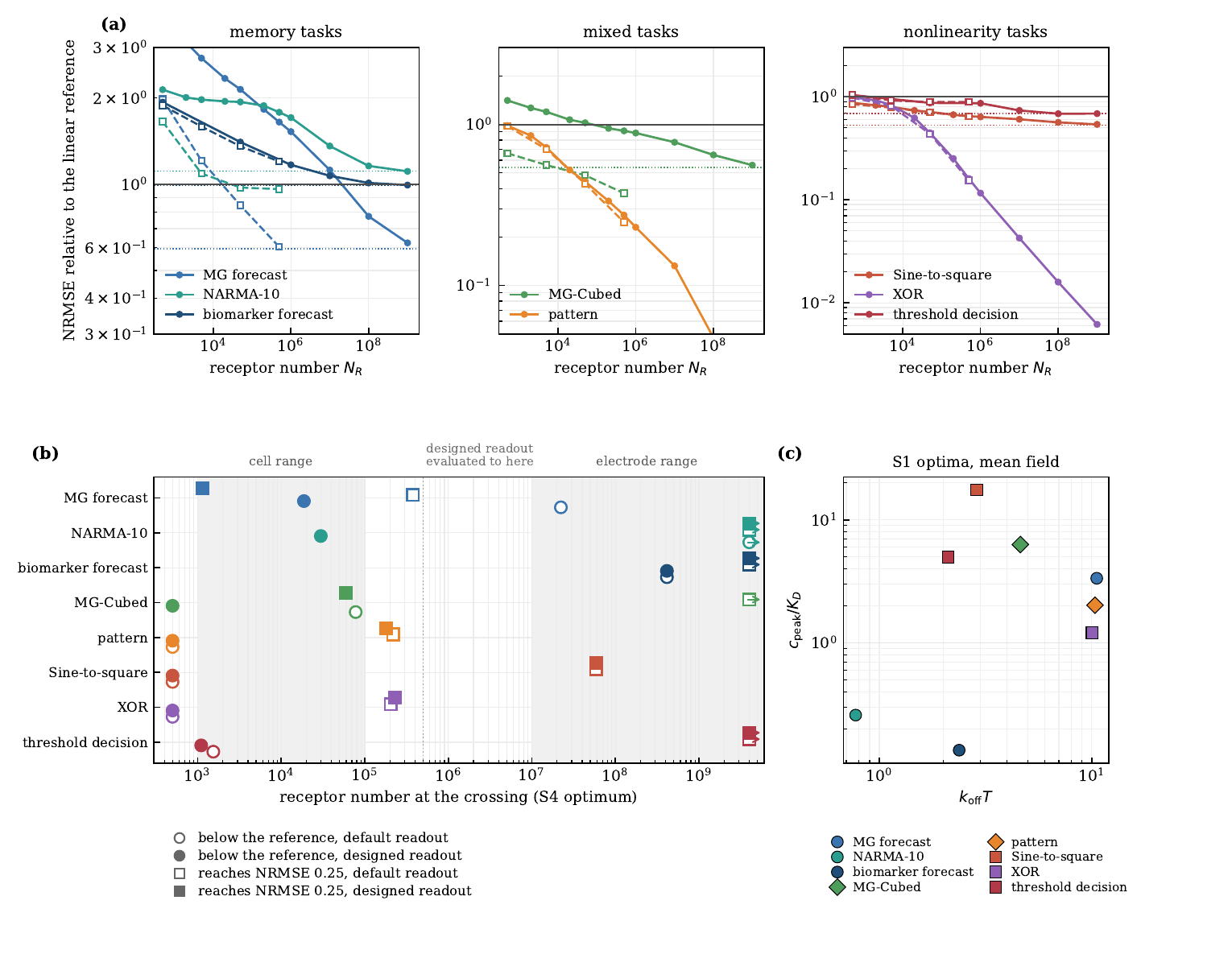}
\caption{Tasks that a receiver counting bound receptors can compute. a) NRMSE of the S4 mean-field optimum of each task, normalized by the linear reference of the task, as a function of the receptor number, grouped into memory, mixed, and nonlinearity tasks. Solid lines indicate the sampled readout retrained under noise ($L$ re-selected from 1 to 5), and dashed lines indicate the readout designed for noise (best of the integrating-node grid and the sampled default, evaluated up to $5\times10^{5}$ receptors). Dotted lines indicate the mean-field values, and the horizontal line marks the reference. b) Critical receptor numbers, i.e., the receptor number at which the channel crosses its linear reference (circles), and the receptor number at which it reaches an NRMSE of 0.25 (squares), with the default readout (open) and with the readout designed for noise (filled). Arrows mark criteria that are not reached for $N_R\le10^{9}$, markers at 500 mark crossings at or below the smallest evaluated receptor number, and the dotted vertical line marks the largest receptor number at which the designed readout was evaluated. Shaded bands mark illustrative receptor-count ranges for cells and for electrochemical aptamer electrodes. c) Operating points of the S1 mean-field optima of the eight tasks, i.e., the scaled dissociation rate against the peak concentration relative to affinity.}
\label{fig:taskmap}
\end{figure*}

The three benchmark tasks represent three requirements, i.e., memory, nonlinearity, and both. To test whether the results obtained under noise hold beyond these benchmarks, we considered three additional tasks, which use the same data blocks and the same readout (Section~\ref{sec:met_tasks}). The first one is the temporal XOR of the last two bits of an on-off keyed bit stream, where on-off keying is the modulation scheme already used in most MC systems. This task requires a strong nonlinearity and one symbol of memory, and no linear map of the input can compute it (linear reference 1.002). The second one is the detection of a three-bit pattern in the same stream, which requires nonlinearity and three symbols of memory (linear reference 0.761). The third one is NARMA-10, i.e., the standard reservoir computing benchmark, which requires ten symbols of memory and a mild nonlinearity (linear reference 0.387 with twenty input taps). In the mean-field model, the channel computes the two bit tasks almost exactly under S4, with errors of 0.003 and 0.006 (\textbf{Table~\ref{tab:budget}} and Supporting Information Section~S11), which is the clearest demonstration of its nonlinearity. However, it remains above the linear reference on NARMA-10 (0.431), since this memory task is not covered by the five-symbol readout and the diffusion memory together. On these tasks, the receptor kinetics alone recover only a small part of the gain of tuning all parameters. Under S1, the errors are 0.159 for XOR, 0.467 for the pattern, and 0.609 for NARMA-10, against 0.254, 0.511, and 0.671 for the nominal channel, which corresponds to reductions of 38, 9, and 9 percent, respectively, whereas S4 achieves reductions of 99, 99, and 36 percent. The reason can be seen in the operating points of the new tasks, which fall into the regimes identified in Section~\ref{sec:res_optima} (Figure~\ref{fig:taskmap}c and Supporting Information Table~S30). Under S4, the bit tasks combine fast receptor kinetics with a peak concentration far above the dissociation constant ($c_{\mathrm{peak}}/K_D=14$ for XOR and 33 for the pattern) and with a diffusion memory near the short end of the domain ($\tau_D/T=0.02$ and 0.02). The nominal channel of S1 offers neither of these, and the S1 optima shown in Figure~\ref{fig:taskmap}c retain lower peak ratios of 1.2 and 2. NARMA-10, on the other hand, selects a slow receptor at a low operating point, i.e., with a low $c_{\mathrm{peak}}/K_D$.

To put these mean-field errors into perspective, we also consider a digital reservoir of the same size. Specifically, a feature-count-matched digital reference, i.e., an echo state network with as many units as the number of features of the default channel readout, was tuned with the same optimizer and budget as the channel searches (Section~\ref{sec:met_esn} and Supporting Information Section~S14). The echo state network achieves 0.003 on forecasting, 0.006 on MG-Cubed, and 0.179 on NARMA-10, against 0.087, 0.285, and 0.431 for the S4 channel optima, respectively. On the transformation task, it achieves 0.103 against 0.231, and on the two bit tasks, it achieves errors below 0.001, against 0.003 and 0.006 for the S4 channel, and both compute these tasks accurately in the noise-free evaluation. On the biomarker forecast of the case study presented in Section~\ref{sec:res_switch}, it matches the channel and the linear reference (0.332), and on the threshold decision, it achieves 0.293 against 0.344. As a result, the channel cannot compete with a digital reservoir on memory tasks, where it lacks the recurrent state provided by a network of 250 units, and its value lies in a different application domain. The MC channel computes inside the communication channel itself, i.e., in tissue or between cells, where no digital reservoir can be placed, with a nonlinearity and a memory supplied by its own components. Hence, the comparisons made in this paper remain comparisons between configurations of this channel.

\textbf{Figure~\ref{fig:taskmap}}a shows the error from 500 to $10^{9}$ receptors, normalized by the linear reference of each task, for all eight tasks, i.e., the six benchmarks and the two targets of the case study presented in Section~\ref{sec:res_switch}, which are a biomarker forecast and a threshold decision. Each curve is summarized by two quantities for the selected S4 configuration and readout scheme (Figure~\ref{fig:taskmap}b, Table~\ref{tab:budget}, and Section~\ref{sec:met_remedies}). The first one is the \textit{critical receptor number}, $N_R^{*}$, i.e., the interpolated receptor number at which the channel falls below its linear reference, and the second one is the receptor number at which the channel reaches an illustrative NRMSE of 0.25. Note that both quantities are criterion crossings for one configuration and one scheme, and neither of them is a physical minimum. With the readout used in the preceding sections, which samples $M$ nodes and re-selects $L$ from 1 to 5 at each receptor number, and which we refer to as the \textit{default readout}, the tasks fall into three groups. The first group consists of sine-to-square and the threshold decision, which require a saturating nonlinearity and little memory. The former outperforms its reference at every receptor number studied ($\le 500$), and the latter from 1500 receptors, which is within the range of a cell. The second group consists of the bit tasks, which cross their reference early, since a linear map of the input cannot compute XOR at all and computes the pattern only in part; for XOR, whose reference is near one, this criterion is uninformative. These tasks reach an NRMSE of 0.25 only with $2.0\times10^{5}$ and $2.2\times10^{5}$ receptors, respectively, since the noise obscures the within-symbol structure that encodes the bit combination. The third group consists of the memory tasks, which require the largest receptor numbers. Forecasting outperforms its reference at $2.2\times10^{7}$ receptors, and MG-Cubed at $7.8\times10^{4}$. The biomarker forecast crosses its reference only at about $4.1\times10^{8}$ receptors, and NARMA-10 does not cross it for $N_R\le10^{9}$, since its mean-field error with this readout lies above the reference.

\begin{table*}[t!]
\centering
\caption{Critical receptor numbers of the eight tasks for the S4 mean-field optimum, together with the linear reference, the mean-field error, and the receptor numbers at which the channel outperforms the reference and reaches an NRMSE of 0.25 under two readout schemes. The default readout samples the nodes and re-selects $L$ and the ridge value under noise at each receptor number, whereas the designed readout is the best of the integrating-node grid and the sampled default, evaluated up to $5\times10^{5}$ receptors. Critical receptor numbers are interpolated criterion crossings for one configuration and scheme, and the two schemes use different receptor grids and realizations (Section~\ref{sec:met_remedies}; paired crossings are given in Supporting Information Table~S34). The entry ${\le}500$ marks a crossing at or below the smallest evaluated receptor number, and n.r. marks no crossing for $N_R\le10^{9}$. For temporal XOR, the linear reference is near one, and crossing it is not evidence of accurate bit computation.}
\label{tab:budget}
\scriptsize
\setlength{\tabcolsep}{4pt}
\begin{tabular}{@{}llcccccc@{}}
\toprule
 & & & & \multicolumn{2}{c}{\textbf{Below the reference}} & \multicolumn{2}{c}{\textbf{Reaches NRMSE 0.25}} \\
\cmidrule(lr){5-6}\cmidrule(lr){7-8}
\textbf{Task} & \textbf{Class} & \textbf{Linear reference} & \textbf{Mean field} & \textbf{default} & \textbf{designed} & \textbf{default} & \textbf{designed} \\
\midrule
MG forecasting & memory & 0.147 & 0.087 & $2.2\times10^{7}$ & $1.9\times10^{4}$ & $3.8\times10^{5}$ & 1100 \\
Sine-to-square & nonlinearity & 0.435 & 0.231 & $\le 500$ & $\le 500$ & $5.9\times10^{7}$ & $5.9\times10^{7}$ \\
MG-Cubed & mixed & 0.526 & 0.285 & $7.8\times10^{4}$ & $\le 500$ & n.r. & $5.9\times10^{4}$ \\
Temporal XOR & nonlinearity & 1.002 & 0.003 & $\le 500$ & $\le 500$ & $2.0\times10^{5}$ & $2.3\times10^{5}$ \\
Bit-pattern detection & mixed & 0.761 & 0.006 & $\le 500$ & $\le 500$ & $2.2\times10^{5}$ & $1.8\times10^{5}$ \\
NARMA-10 & memory & 0.387 & 0.431 & n.r. & $3.0\times10^{4}$ & n.r. & n.r. \\
Biomarker forecast & memory & 0.332 & 0.329 & $4.1\times10^{8}$ & $4.1\times10^{8}$ & n.r. & n.r. \\
Threshold decision & nonlinearity & 0.499 & 0.344 & 1500 & 1100 & n.r. & n.r. \\
\bottomrule
\end{tabular}
\end{table*}

The readout is the part of the system that can be changed without modifying the channel, and three of its parameters trade resolution for averaging. First, the virtual nodes can integrate the occupancy over their windows instead of sampling it, which is the time averaging by which a cell improves the precision of its receptor readings \cite{berg1977physics}, and the number of nodes per symbol can be reduced to lengthen these windows. Second, a causal low-pass filter with time constant $\tau_f$ can smooth the node sequence. Third, the readout memory can be extended beyond five symbols. We evaluated every combination of these settings on the mean-field optima at four receptor numbers (Section~\ref{sec:met_remedies} and Supporting Information Section~S12). We refer to the best of this grid and the default readout on the same realizations as the \textit{readout designed for noise}, or the \textit{designed readout} for short. For the memory tasks, the effect is large. At 50000 receptors, the error of the S1 forecasting optimum decreases from 0.338 to 0.133 with a low-pass filter of 2 symbol intervals and 20 symbols of readout memory, which is below the linear reference. At $5\times10^{5}$ receptors, it further decreases to 0.088, i.e., the mean-field value of the five-symbol design. The critical receptor number of forecasting decreases from $2.2\times10^{7}$ to $1.9\times10^{4}$ receptors, and the receptor number required for an NRMSE of 0.25 decreases from $3.8\times10^{5}$ to 1100. MG-Cubed outperforms its reference at every receptor number studied ($\le 500$), and NARMA-10 from $3.0\times10^{4}$ receptors. Most of this gain results from the readout memory: the longer readout stores twenty consecutive noisy reservoir-state vectors and averages over them, and the low-pass filter adds a further averaging within the symbol. The same readout also improves the noiseless model, from 0.088 to 0.027 for forecasting; hence, with this readout, the noisy receiver approaches the mean-field error of the five-symbol design, whereas the lower mean-field error of the twenty-symbol design remains out of reach. For the bit tasks and the threshold decision, on the other hand, the designed readout changes little (from 0.448 to 0.439 for XOR at 50000 receptors), since their performance is limited by the noise on the within-symbol structure, which no averaging over symbols can restore.

A joint search over the receptor rates and the readout settings confirms these findings (Supporting Information Table~S35). Under S1, this search achieves 0.189 for forecasting, 0.313 for MG-Cubed, and 0.372 for NARMA-10 at 5000 receptors, and 0.127, 0.271, and 0.367 at 50000 receptors. For S3, in which the release is tunable as well, the symbol interval was allowed up to 5~s, and the feature count was not capped (Section~\ref{sec:met_remedies}). In this case, the joint design achieves 0.083, 0.228, and 0.366 at 50000 receptors, which for forecasting matches the mean-field error of the five-symbol S4 design (0.087). The biomarker forecast is the exception among the memory tasks, as its mean-field optimum only matches the linear reference. The grid lowers its error only slightly, i.e., from 0.463 to 0.451 at 50000 receptors, and within the sweep up to $5\times10^{5}$ receptors, it does not cross the reference, which the sampled default reaches at about $4.1\times10^{8}$ receptors. Note that the two readout schemes listed in Table~\ref{tab:budget} were evaluated on different receptor grids and realizations (Section~\ref{sec:met_remedies}). Therefore, the differences between their interpolated crossings, e.g., for the two bit tasks at the 0.25 level, cannot be attributed to the readout alone, and the interpolation intervals, which are set by the spacing of the evaluated receptor numbers, should not be interpreted as error bounds. On the same grid and realizations, the designed readout crosses at or below the default readout for every task (Supporting Information Table~S34), e.g., for XOR at $2.3\times10^{5}$ against $2.5\times10^{5}$ receptors.

To examine coarser acquisition, we restricted the readout to one integrating window per symbol, i.e., the occupancy averaged over the symbol interval, and to two windows, i.e., the averages over the two halves of the symbol interval. We refer to these one- and two-window readouts as the \textit{cell-like readout}. These readouts are biologically motivated abstractions that test the loss of within-symbol information; however, they do not specify a cellular circuit that implements the windows, the stored history, the filter, and the trained weights. Both readouts keep the low-pass filter and the readout memory of the grid (Section~\ref{sec:met_remedies} and Supporting Information Tables~S36 and S37). With one window per symbol, the channel loses the within-symbol structure on which most tasks rely. In the mean-field model, the S1 forecasting optimum yields 0.399, which lies above the linear reference, against 0.088 with the default readout, and the transformation optimum yields 0.527, which lies above the linear error floor, against 0.256. Apart from XOR, whose reference is near one, only NARMA-10, whose memory is carried by the readout, crosses its reference, from $3.8\times10^{4}$ receptors onward. Two windows per symbol, on the other hand, restore most of what is lost with one window. In the mean-field model, the S1 optima yield 0.104 for forecasting, 0.336 for transformation, and 0.264 for MG-Cubed. Under noise, the S1 channel with two windows outperforms its reference on MG-Cubed from 2000 receptors, on transformation from 2600, on NARMA-10 from $2.9\times10^{4}$, on forecasting from $1.8\times10^{5}$, and on the threshold decision from $2.7\times10^{5}$, and it reaches an NRMSE of 0.25 on forecasting from 4600 receptors. When the channel is tunable as well, two windows suffice for transformation and MG-Cubed at every receptor number studied, and for forecasting from $4.3\times10^{4}$ receptors. However, within the sweep, two windows do not bring the bit tasks to an NRMSE of 0.25, and under S4, they lose the within-symbol resolution exploited by the sampled readout (0.567 against 0.161 for XOR at $5\times10^{5}$ receptors).

\FloatBarrier
\subsection{Case Study and Online Task Switching}
\label{sec:res_switch}

\begin{figure*}[t!]
\centering
\includegraphics[width=\textwidth]{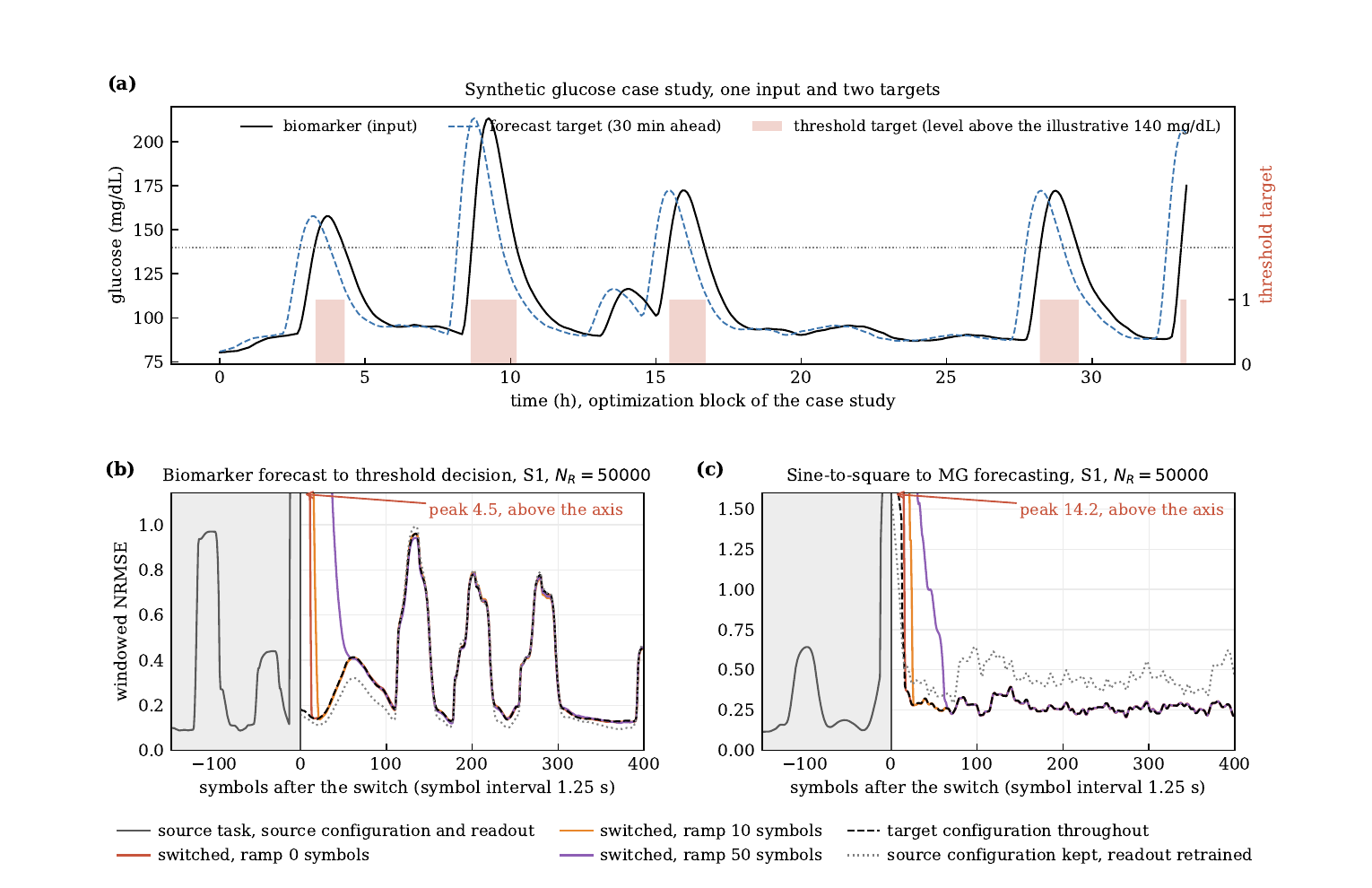}
\caption{Synthetic glucose case study and online task switching. a) The glucose signal over part of the optimization block on the source-series time axis, together with its forecasting target 30 minutes ahead, the illustrative threshold of 140~mg/dL (dotted), and the threshold-decision target (shaded). b) Error of the active task in a centered window of 25 symbols, computed afterwards from the recorded outputs, for a single S1 channel on the channel time scale with $T=1.25$~s. Values are averaged over 50 realizations of the stochastic receptor model with 50000 receptors. Before symbol 0, the channel forecasts the biomarker (gray). At symbol 0, its receptor rates are tuned to those of the threshold-decision optimum, either at once or along a ramp of 10 or 50 symbols, and its readout is replaced. The dashed line indicates a channel kept in the threshold-decision configuration throughout, and the dotted line indicates the forecasting channel kept with a readout retrained for the threshold decision. c) The same for the switch from sine-to-square transformation to MG forecasting, where the input changes with the task, and the retrained readout on the old channel (dotted) remains far above the retuned channel. Peaks above the displayed range are annotated. Recovery times refer to the windowed criterion, and they exclude the time required to train the replacement readouts or to actuate a physical receptor layer.}
\label{fig:switch}
\end{figure*}

The results presented above compare configurations that are tuned for benchmark tasks and evaluated in a static, fixed-task setting. However, the scenario that motivated this study in Section~\ref{sec:introduction}, i.e., a receiver that reads a physiological signal, requires the task to change during operation. Therefore, the last set of results concerns a case study on a synthetic blood glucose signal. The signal is generated from the minimal model of glucose and insulin kinetics of Bergman et al.\ \cite{bergman1979minimal}, driven by the meal-absorption model of Hovorka et al.\ \cite{hovorka2004nonlinear}, and it contains no patient data (Section~\ref{sec:met_case}). The same input serves two targets, i.e., the glucose level 30 minutes ahead (forecasting) and a binary indicator of whether the glucose level exceeds an illustrative threshold of 140~mg/dL (threshold decision), which correspond to the two requirements of the monitoring scenario introduced in Section~\ref{sec:introduction}. \textbf{Figure~\ref{fig:switch}}a shows the signal and the two targets. The signal is dominated by the rises and falls of the glucose level that follow each meal, and the receiver cannot anticipate the onset of a meal from the glucose signal alone, which is a plausible limit for any forecast of this signal. The best linear regression on the input achieves 0.332 for forecasting and 0.499 for the threshold decision.

The mean-field searches reproduce the benchmark results on this signal (Supporting Information Table~S26). For forecasting, the S1 and S4 optima improve on the nominal channel by only 2 and 3 percent, respectively, and both keep the receptor in the linear part of the binding curve ($c_{\mathrm{peak}}/K_D=0.13$ and 0.23). The best forecast found essentially matches the linear reference (0.329 against 0.332). The unobserved meal onsets are a plausible reason for this; however, the near-equal errors of the channel and the reference do not by themselves establish an irreducible forecasting floor. For the threshold decision, the S1 and S4 optima improve on the nominal channel by 11 and 18 percent, respectively. As in the transformation task, the S4 optimum drives the receptor into saturation ($c_{\mathrm{peak}}/K_D=36$) with a short diffusion memory ($\tau_D/T=0.06$), and both optima select $L=1$, since a decision on the current level requires no readout memory. In the mean-field model, the channel lies well below the linear reference on this task (0.344 against 0.499), since the decision corresponds to the nonlinearity that the saturating receptor provides. Under noise, the two targets separate as in the benchmark tasks. At 50000 receptors, the noise-aware S1 optimum achieves 0.383 for forecasting, against 0.424 for the nominal channel with a retrained readout, and 0.478 for the threshold decision, against 0.513. The S4 optimum achieves 0.419 on the decision, i.e., below the linear reference, whereas at 500 receptors, the S1 optimum, with 0.521, lies just above it. Hence, a decision task requires a few thousand receptors, in line with the benchmark transformation task. The forecast on this signal, whose mean-field optimum only reaches the linear reference, outperforms the reference under noise from about $4.1\times10^{8}$ receptors (Section~\ref{sec:res_budget}).

A receiver that has to switch from one target to the other during operation has two options, i.e., it can either keep its channel and retrain the readout, whose cost was quantified in Section~\ref{sec:res_reuse}, or retune the channel. Figure~\ref{fig:switch}b and c simulate the second option with the stochastic receptor model at 50000 receptors (Section~\ref{sec:met_switch}). A single S1 channel operates with the receptor rates of the source-task optimum. At the switch, its rates are tuned to those of the target-task optimum, either at once or along a ramp of 10 or 50 symbols, and its readout is replaced by the target-task readout, which was trained on a separate run of the target configuration. To display the transient, we compute the error of the task that is active at each symbol over a window of 25 symbols centered on that symbol. Note that this windowed error is a summary computed afterwards from the recorded outputs, and the receiver itself does not compute it during operation. On the biomarker signal (Figure~\ref{fig:switch}b), the channel forecasts at a steady error level of 0.289 before the switch. Immediately after an instantaneous switch, the error of the decision task rises to 4.5 times the standard deviation of its target, since the receptor occupancy still reflects the old rates, whereas the new readout expects the new ones. The error returns to within 10 percent of the steady level of the target configuration after 13 symbols. On the time scale of the simulated channel, i.e., with $T=1.25$~s, this corresponds to 16 seconds, and under the mapping of one symbol to one five-minute biomarker sample, it corresponds to 65 minutes. Note that this simulated time should not be interpreted as a measured physiological adaptation time. The two ramps require 21 and 55 symbols, i.e., 105 and 275 minutes on the biomarker time scale, respectively.

Across the 6 directed switches of the 3 task pairs studied, i.e., the biomarker pair, the two MG tasks, and sine-to-square with forecasting (Supporting Information Table~S28), the instantaneous switch recovers within 13 to 17 symbols, and the 50-symbol ramp within 52 to 65 symbols. The reported recovery counts depend on the prescribed ramp; on the receptor relaxation, the diffusion memory, and the $L$ symbols retained by the readout; and on the centered window of the criterion. The window centered on symbol $n$ spans symbols $n-12$ to $n+12$, and thus, the first window free of the switch symbol is centered 13 symbols after it. Therefore, the recovery counts of 13 to 17 symbols after an instantaneous switch reflect the window width as much as the transient itself, and they should not be interpreted as measurements of a receptor or actuator response time. For the benchmark pairs, a ramp also lowers the peak of the transient, e.g., from 14.2 to 3.3 times the target standard deviation for the switch from sine-to-square to forecasting. For the biomarker pair, on the other hand, the peak remains at 4.5 to 5.6, since the intermediate configurations serve neither target.

Whether the switch is worthwhile depends on how far apart the two configurations lie. The dotted curves in Figure~\ref{fig:switch} show the alternative of keeping the source channel and retraining the readout for the target task. For the biomarker pair, this alternative settles at 0.365, against 0.378 for the retuned channel, and in the reverse direction at 0.299, against 0.269, which corresponds to a penalty of 11 percent. For the two MG tasks, the retrained readout incurs no loss in one direction and a loss of 2 percent in the other. Hence, for these pairs, whose configurations lie close to each other, retuning offers little or no steady-state advantage over the refitted readout at this receptor number. Between sine-to-square and forecasting (Figure~\ref{fig:switch}c), on the other hand, the retrained readout settles at 0.453, against 0.276 for the retuned channel, which corresponds to a penalty of 64 percent, and the penalty is 15 percent in the reverse direction. In this case, retuning yields a substantially lower post-transient error than reusing the source configuration with a refitted readout, and the transient lasts tens of symbols. These results refine the rule stated in Section~\ref{sec:res_reuse}: the benefit of retuning tracks the reuse penalty measured at the operating receptor number, and it cannot be inferred from the task labels alone. In the benchmark pair, the two configurations lie on different sides of the boundary between memory-dominated and nonlinearity-dominated tasks, the penalty is large, and retuning is beneficial. The biomarker pair also combines a forecast with a threshold decision, but its noise-aware S1 configurations leave a penalty of at most 11 percent, and the refitted readout serves equally well. These comparisons suggest checking both the steady-state benefit and the transient before retuning, although they do not determine the task duration or the physical cost at which a switch becomes worthwhile.

\FloatBarrier
\section{Discussion}
\label{sec:discussion}

The mean-field results identify two quantities that a task-adaptive MC reservoir has to set, i.e., the receptor operating point, which is the ratio of the received concentration to the dissociation constant, and the receptor timescale relative to the symbol interval. The forecasting configurations select a receptor that follows the concentration within a symbol interval and remains in the responsive part of its binding curve, with a peak concentration of about three times $K_D$. The memory of such a reservoir is provided by the diffusion channel and by the $L$ past states that the readout retains. The transformation configurations, on the other hand, select a higher-affinity receptor that is driven into saturation, such that its occupancy approaches a threshold function of the recent concentration. A task that requires both lies in between. Notably, the receptor operating point is the quantity that living cells regulate when they adapt the affinity of their receptors to the ambient concentration \cite{barkai1997robustness,neumann2014gain}, and it is also the quantity that adaptive MC receivers adjust to the statistics of the received signal to maintain detection performance \cite{kuscu2026adaptive}. Therefore, task-adaptive reservoir computing and adaptive detection rest on the same physical mechanism. The interventions proposed for adaptive detection, i.e., thermal or electrical modulation of an aptamer layer and the selection of recognition elements with prescribed kinetics \cite{chen2022tunable,ma2017electrostatic,alkhamis2024slowoff}, are thus candidate mechanisms for task adaptation, although whether a single receptor layer can reach the selected kinetic pairs reversibly remains to be established. The dissociation constants tuned in this study, i.e., about 1.4~nM for forecasting and 0.27~nM for transformation, lie within the range of practical aptamers, and the difference between the tasks corresponds to a difference in operating point within the affinity range of a single class of recognition elements.

The absolute errors, however, should be interpreted on an appropriate scale. A digital echo state network of equal size outperforms the tuned channel by a factor of about 27 on forecasting and 51 on MG-Cubed, and by a factor of about 2.2 on transformation and 2.4 on NARMA-10. On the bit tasks, the network achieves errors below 0.001, against 0.003 and 0.006 for the S4 channel, and on the biomarker forecast, it matches the channel and the linear reference (Section~\ref{sec:res_budget}). The design rules derived in this paper compare configurations of a single channel, and thus, they do not depend on this scale. The MC reservoir offers a way to compute with a channel that already exists, such as inside tissue or between cells, and it does not replace a digital reservoir where one can be placed.

For IoBNT systems, the results obtained under partial tunability lead to three practical implications. First, an implanted biosensor can select and possibly modulate its receptor layer and its readout, whereas the source, the medium, and its placement are imposed, which corresponds to the S1 abstraction. In the mean-field model, such a biosensor lowers the error of the nominal channel by 12 to 24 percent, which amounts to 92 and 84 percent of the gain of tuning all parameters on the two forecasting tasks, and 61 percent on the transformation task. Second, an engineered source acting on native receptors gains little from tuning its release. Release tuning changes the concentration waveform, the operating point, and the concentration-dependent relaxation time; however, it leaves the association and dissociation constants fixed, and in the tested searches, this restricted access yielded smaller gains than receptor tuning. Third, an engineered transceiver pair should be co-designed at both ends, which improves every task, with the caveat that its larger search space converges more slowly at an equal budget. The convergence curves also have a practical implication for calibration. A receiver-only calibration achieved most of its gain within twenty evaluations, whereas the search over all parameters required the full budget, which suggests a smaller evaluation burden for receiver-only calibration. Note that the physical calibration time and reagent costs were not evaluated.

The results on reusing configurations across tasks reveal the limits of retraining the readout alone. Retraining the readout on a configuration tuned for another task left a penalty in every one of the 24 combinations, up to 58 percent, and in 6 of them, the reused configuration performed worse than the untuned channel. The penalty is small between related tasks and large across the boundary between memory-dominated and nonlinearity-dominated tasks, which is consistent with a saturated receptor compressing the amplitude differences that the refitted linear readout does not restore. Note, however, that the parameters were selected jointly, and thus, this comparison does not isolate saturation as the sole cause. Moreover, the penalty is asymmetric: a receptor tuned for forecasting serves the transformation task at a moderate cost, whereas a receptor tuned for transformation performs far worse at forecasting. Therefore, a minimax configuration, i.e., the configuration chosen for the worst case, lies near the forecasting optimum, and it remains within 4 to 8 percent of every task-specific optimum in the mean-field model. Hence, a system that must serve several tasks without retuning should be designed as such a minimax configuration from the outset, instead of being reused from a design tuned for a single task. This agrees with the observation of our previous study that this channel serves smoothly correlated inputs far better than independent ones \cite{uzun2025molecular}, as well as with the memory-nonlinearity trade-off in reservoir computing theory \cite{dambre2012capacity,inubushi2017beyond}.

The results obtained under receptor noise qualify all of the above findings. The stochastic receptor model reproduces the particle simulations, and it reveals that the failure of the mean-field optima in those simulations has two components, i.e., a readout tuned for a noiseless state and a channel tuned for a noiseless receiver. Retraining the readout removes the first component, and the ridge value, which the mean-field model leaves at its smallest value, becomes the first parameter to adjust for a receiver that counts bound receptors. Retuning the channel, on the other hand, removes part of the second component. When only the receptor is tunable, the noise-aware optima shift the forecasting receptor to a low occupancy, where the response is linear and can be averaged. When the channel is tunable as well, they shift it into saturation with the shortest diffusion memory. For the transformation task, they keep the receptor in saturation, where the threshold-like response is least affected by the count noise. These rules trade the sharpness of the nonlinearity and memory of the reservoir for averaging in the receptor, the channel, and the readout. The gain of the noise-aware design over the retrained mean-field design decreases as the receptor number increases. At $5\times10^{5}$ receptors, the forecasting error of the best noise-aware configuration is still 2.6 times its mean-field value, and the sweep to $10^{9}$ receptors shows that the mean-field errors are approached only between $10^{8}$ and $10^{9}$ receptors in this model.

The receptor number and the task together determine what a receiver counting bound receptors can compute, and the critical receptor number introduced in Section~\ref{sec:res_budget} quantifies this dependence. The tasks fall into three groups. First, smooth transformations and threshold decisions, which require a saturating nonlinearity and little memory, outperform the noiseless linear reference with about $10^{3}$ receptors, since the receptor keeps supplying the nonlinearity, and the count noise perturbs a threshold-like response the least. Second, nonlinear computations on a bit stream, which a linear map of the input computes poorly or not at all, reach an NRMSE of 0.25 only with about $2\times10^{5}$ receptors, since the noise obscures the within-symbol structure that encodes the bit combination. Third, memory tasks require fine amplitude differences held over many symbols, which the count noise removes before any readout can access them. With the readout of the noiseless design, forecasting outperforms its reference only at about $2.2\times10^{7}$ receptors.

The readout designed for noise changes this last group the most. Integrating nodes, a low-pass filter over one to two symbol intervals, and a readout memory of twenty symbols lower the critical receptor number of forecasting to $1.9\times10^{4}$ receptors, and the receptor number required for an NRMSE of 0.25 to 1100. They also bring MG-Cubed below its reference at every receptor number studied, and NARMA-10 from about $3.0\times10^{4}$ receptors. The biomarker forecast, on the other hand, whose mean-field optimum only matches its reference, gains little, and its readout grid does not cross the reference within the evaluated sweep. Note that this readout averages over symbols. Hence, it recovers the performance of the five-symbol design, whereas the performance of the twenty-symbol design, which the same readout reaches in the noiseless model, remains out of reach. Moreover, it does not help the bit tasks or the threshold decision, which are limited by the noise within the symbol.

This averaging also has a biological interpretation. A cell improves the precision of a concentration measurement by averaging its receptor occupancy over time \cite{berg1977physics,bialek2005physical}, and its downstream pathways, from second messengers to gene expression, respond with time constants of seconds to minutes, thereby low-pass filtering the occupancy before any decision is made. The integrating nodes and the low-pass filter used in Section~\ref{sec:res_budget} perform the same operation. Therefore, the critical receptor numbers obtained with the designed readout apply to a receiver with slow downstream processing, whereas the sampled readout describes an electronic receiver that resolves the transient within the symbol. The cell-like readout evaluated in Section~\ref{sec:res_budget} further reveals how much of the within-symbol structure a receiver has to retain. Apart from the uninformative XOR crossing, one integrated sample per symbol serves only the memory benchmark. Two samples per symbol, which a cell might form with two pathways of different time constants, recover transformation, MG-Cubed, and forecasting, although at higher receptor numbers than the sampled readout requires.

This distinction has direct consequences for the systems described in Section~\ref{sec:sets}. With a readout that resolves the transient within a symbol, a receiver with $10^{3}$ to $10^{4}$ receptors of one type serves threshold-like and classification tasks, and with averaging, it forecasts a smooth signal with a useful accuracy. A coarse-grained readout that integrates over the symbol crosses the reference on transformations and MG-Cubed with a few thousand receptors, and on forecasting with about $10^{5}$ receptors, only when it keeps two windows per symbol, within the stated model. In the coarse two-window comparison, the threshold-decision readout crosses its linear reference at about $2.7\times10^{5}$ receptors under S1 and 3700 under S4, and the two-window bit-task readouts do not reach an NRMSE of 0.25 within the evaluated sweep. Note that these results describe the tested acquisition and readout schemes, and they do not prescribe a particular electronic or cellular implementation. Even a cell at the upper end of its expression range, i.e., with about $10^{5}$ receptors \cite{lauffenburger1993receptors}, computes nonlinear functions of a bit stream only coarsely, as an NRMSE of 0.25 on XOR requires about $2.0\times10^{5}$ receptors. Electrochemical aptamer sensors carry probe densities on the order of $10^{11}$ to $10^{12}$ per cm$^2$ \cite{schoukroun2015heterogeneous,downs2022realtime}, which corresponds to $1.1\times10^{5}$ to $1.1\times10^{6}$ receptors on a receiver of the size simulated here, i.e., with a radius of 3~$\mu$m, and to $10^{7}$ to $10^{10}$ receptors on an electrode of 0.01 to 1~mm$^2$.

For the idealized point-concentration model, an illustrative similarity scaling can be established. With $D$ fixed, scaling $d$ by $\alpha$, $T$ by $\alpha^{2}$, both release endpoints $N_{\min}$ and $N_{\max}$ by $\alpha^{3}$, and both kinetic rates by $\alpha^{-2}$ preserves the mean occupancy trajectory as a function of the normalized time, provided that the acquisition and filter times are scaled consistently. Note that this is a mean-field similarity statement: it does not preserve the stochastic error when $N_R$ changes, it does not establish a uniform concentration over a larger electrode, and it does not account for receptor crowding or depletion. Therefore, the receiver-area and probe-density estimates that follow are illustrative mappings of the receptor count, and they do not validate a scaled system. For the biomarker case study, whose sampling interval of five minutes is 240 times the benchmark symbol interval, this scaling yields a receiver with a radius of about 46~$\mu$m at a distance of about 110~$\mu$m from the source. At the probe densities given above, such a receiver carries $2.7\times10^{7}$ to $2.7\times10^{8}$ receptors. Hence, the electrode band in Figures~\ref{fig:noise} and \ref{fig:taskmap} represents an illustrative receptor-count range for a biosensor that reads a slow physiological signal, whereas the cell band represents the range of fast signaling between cells. At the upper end of the electrode range, every task studied here approaches its mean-field performance in the model. Where a receiver cannot reach that receptor number, the noise-aware search described here provides the configuration that makes the best use of the available receptors, and the stochastic receptor model provides it within seconds per evaluation.

The case study and the switching simulation connect these results to the monitoring scenario introduced in Section~\ref{sec:introduction}. On a biomarker signal whose variability results from meals, the two targets select the same two operating regimes as the benchmarks, i.e., a receptor in its linear range for the forecast, and a saturated receptor for the threshold decision. A single channel switches between two configurations within tens of symbols when its receptor rates are retuned and its readout is replaced. Whether retuning is worthwhile depends on the distance between the configurations. For sine-to-square and forecasting, retuning yields a substantially lower post-transient error than reusing the source configuration with a refitted readout, which incurs a penalty of 15 to 64 percent. For the biomarker pair and the two MG tasks, on the other hand, the advantage is small, absent, or dependent on the switching direction, i.e., at most 11 percent. These comparisons suggest checking both the steady-state benefit and the transient before retuning, although they do not determine the task duration or the physical cost at which a switch becomes worthwhile. Therefore, a system whose task pair incurs a small reuse penalty at its receptor number should be designed as a minimax configuration and retrain its readout, whereas a system whose task pair incurs a large penalty, as in the case of transformation and forecasting, should retune its channel.

This study has several limitations, which concern the searches, the model, and the switching simulation. First, the searches are finite, and the reported optima are the best configurations found within 100 evaluations under each set. Several optima lie on a boundary of the search domain, i.e., at the fastest dissociation rate for forecasting and at the shortest distance for transformation, and therefore, the domain limits what can be said about the location of the true optimum in those directions. The searches minimized the error on the optimization block and selected the ridge value on the same block, and the held-out block confirms the directions of the comparisons without providing a prospective test on unseen data. The values quoted for the noise-aware optima are re-evaluations with independent realizations. Under the fixed-readout scheme, they estimate the error of the selected settings without the optimistic bias of the search objective. Under the retrained-readout scheme, on the other hand, $L$ and the ridge value are re-selected on the optimization block of these independent realizations, and thus, those values remain selection-biased estimates. Neither scheme constitutes an independent test on a new input sequence, and the choice of configuration carries the selection bias of the search.

Second, the retained concentration history of the channel model, which is kept from our previous study \cite{uzun2025molecular}, is a modeling assumption. The mean-field errors of the MG-Cubed optima change by up to 0.1373, and the noise-aware values by up to 0.058, when the complete history is retained (Supporting Information Section~S1), since the longer history raises the background concentration and shifts the operating point. Moreover, the stochastic receptor model treats the concentration as a deterministic field and the receiver as a point. Hence, it omits the molecule-count noise and the spatial nonuniformity of the concentration over the receiver, which the particle simulations capture. The particle comparison at 500 receptors found the two models in close agreement; however, it does not separate these contributions at other distances or receptor numbers. The model also omits flow, degradation, receptor heterogeneity, nonspecific binding, and sensor drift. Third, the switching simulation changes the receptor rates along a prescribed ramp and replaces the readout with a pre-trained one. It does not model the physical actuator, its energy or reagent consumption, or whether a particular receptor layer can realize that trajectory reversibly, and the case-study signal is synthetic. Whether a built system can switch between two configurations reversibly, and at what cost in time and reagents, is a property of its actuators that only experiments can establish.

The natural next step is to realize an experimental channel that incorporates the tuning mechanisms listed in Table~\ref{tab:tunable}. A microfluidic transmitter with programmable release \cite{bolhassan2024microfluidic} and an electrochemical aptamer-based receiver whose affinity is modulated by temperature or by an electric field \cite{chen2022tunable,ma2017electrostatic} would realize S3 with reversible control at both ends. In such a system, the impulse response and the binding kinetics would be measured directly, and the receptor number would be estimated from the probe density. The configuration for each task would then be selected with the stochastic receptor model at that receptor number on training and validation data, and frozen before testing. Finally, the switch between two configurations would be executed on a single system, and its transient would be recorded.

\section{Conclusion}
\label{sec:conclusion}

In this paper, we have investigated task-adaptive reservoir computing in a diffusion-based MC channel with ligand receptors. This channel is a physical reservoir whose two resources, i.e., the memory of diffusion and the nonlinearity of receptor saturation, are set by the physical parameters of its components, which living cells regulate, and which biosensor and synthetic-biology engineering can tune. Using a mean-field model, we have shown that the same channel can be tuned to different temporal tasks by shifting the receptor operating point along its binding curve and by changing the scaled dissociation rate, and that the best configurations found for the tasks lie in separate regions of the plane spanned by these two quantities. On the three primary benchmarks, tuning the receptor kinetics alone recovers most of the task-specific gain, whereas release tuning alone gains little. Retraining the readout on a configuration tuned for another task leaves a penalty in every tested mean-field pair, up to 58 percent across the boundary between memory-dominated and nonlinearity-dominated tasks, whereas a minimax configuration serves the three benchmarks within a few percent. Using a stochastic receptor model validated against particle-based simulations, we have further determined the critical receptor number of every task, i.e., the receptor number at which the channel outperforms a linear filter on its input. With a readout that resolves the transient within a symbol, transformations and threshold decisions outperform a linear filter on the input with the receptor numbers of a cell, whereas nonlinear bit computations require receptor numbers at or above the upper end of what a cell expresses. A readout that integrates the occupancy over the whole symbol interval, as a living cell naturally does, falls behind the linear filter on most tasks, and it outperforms the filter only when it resolves two half-symbol windows and is supported by more receptors. Memory tasks require the receptor numbers of an artificial engineered receiver, unless the readout is designed for noise, which lowers the critical receptor number of the memory benchmarks by orders of magnitude. At the receptor numbers of electrochemical aptamer electrodes, every task approaches the performance of the mean-field design in the model. When receptors are few, on the other hand, a noise-aware design shifts the operating point, the diffusion memory, and the readout regularization away from their mean-field values. Lastly, we have shown that a single channel can switch between two tasks within tens of symbols when its receptor rates are retuned and its readout is replaced, and that retuning outperforms readout retraining when the two task-specific configurations lie far apart, as in the case of transformation and forecasting. The resulting design procedure is physical and testable, and it can be summarized as follows: the receptor operating point and timescale should be matched to the task, with the receptor layer tuned first; the receptor number and the task determine which design applies; the readout should be designed for the noise; and the channel should be retuned when the reuse penalty is large, whereas the readout should be retrained when the penalty is small.
\section{Experimental Section}
\label{sec:methods}

\threesubsection{Evaluation of the Model}
\label{sec:met_numerics}

Equations~\eqref{eq:channel} and \eqref{eq:binding} are evaluated on a time grid whose step divides the node spacing $T/M$, i.e., $\Delta t=T/(Mn_{\mathrm{sub}})$, where $n_{\mathrm{sub}}$ is the smallest integer that gives $\Delta t\le10^{-3}$~s. Therefore, every virtual node lies on a grid point in every symbol interval, and no sampling instant is rounded. The concentration is computed by fast convolution as the discrete superposition of the impulse response over the release instants, and the contribution of each release is retained for $\operatorname{round}(100\,\tau_D/T)$ symbol intervals, with a minimum of one, as in our previous study \cite{uzun2025molecular}. The mean-field occupancy is advanced from $p_0=0$ with the exact solution of Equation~\eqref{eq:binding} for a concentration held constant over one grid step, i.e., $p_{k+1}=p_\infty(c_k)+[p_k-p_\infty(c_k)]\exp[-(k_{\mathrm{on}}c_k+k_{\mathrm{off}})\Delta t]$ with $p_\infty(c)=c/(c+K_D)$. In the stochastic receptor model, starting with all receptors free, i.e., $b(0)=0$, the numbers of receptors that bind and unbind in every grid step are drawn from binomial distributions. The success probabilities of these distributions are the exact transition probabilities of a two-state receptor over one step at the same constant concentration, and thus, the expectation of this process follows the mean-field recursion. For computational speed, binomial draws with a mean above 25 are replaced by their normal approximation; the exact sampler changes the errors of the optima by less than the variability across realizations (Supporting Information Section~S1). Refining the grid step fourfold changes the mean-field errors of the optima by at most 0.0001, which is the largest change observed under this refinement. Retaining the complete concentration history instead of the truncated one changes the mean-field errors by up to 0.1373 for the MG-Cubed optima, and the noise-aware values by up to 0.058, since the longer history raises the background concentration at the receiver (Supporting Information Section~S1). Therefore, the retained history, which is kept from our previous study \cite{uzun2025molecular}, is part of the channel model, and every comparison in this paper is made within this model.

\threesubsection{Benchmark Tasks, Data Blocks, and Readout}
\label{sec:met_tasks}

The Mackey-Glass series is generated from $dx/dt=\beta x(t-\tau)/[1+x(t-\tau)^{q}]-\gamma x(t)$, with $\beta=0.2$, $\gamma=0.1$, $q=10$, and $\tau=17$, using a fourth-order Runge-Kutta scheme with a unit step, and then normalized to $[0,1]$. MG forecasting uses the normalized series as the input and its value six samples ahead as the target. Sine-to-square transformation uses the input $u(n)=0.5+0.5\sin[2\pi(n-1)/200]$, and the target is 1 where $u>0.5$, and 0 where $u<0.5$. At the 40 samples where $u$ equals 0.5 exactly, the target series of our previous study \cite{uzun2025molecular} takes the values 0, 0.5, or 1 as generated, and it is used unchanged. MG-Cubed uses the MG input and the target $x(n+10)^3$, where the cube is taken after normalization. The three additional tasks share the blocks, the node count, and the readout with the MG tasks (Section~\ref{sec:res_budget}). An on-off keyed bit stream $b(n)\in\{0,1\}$ of independent and equiprobable bits serves as the input of temporal XOR, whose target is $b(n)\oplus b(n-1)$, and of pattern detection, whose target is 1 where $(b(n-2),b(n-1),b(n))=(1,0,1)$, and 0 otherwise. NARMA-10 \cite{atiya2000new} uses an independent input $u(n)$, uniformly distributed on $[0,0.5]$ and scaled to $[0,1]$ for the release encoding, and the recurrence $y(n+1)=0.3y(n)+0.05y(n)\sum_{i=0}^{9}y(n-i)+1.5u(n-9)u(n)+0.1$. The target at symbol $n$ is $y(n)$, which is paired with the input $u(n)$ of the same symbol. The bit streams and the NARMA series are generated from a single seed (2029), and their first 1700 symbols are used with the MG blocks.

The reservoir state uses $M=50$ virtual nodes per symbol for the two MG tasks and $M=100$ for sine-to-square, with $L\in\{1,\ldots,5\}$. The sequences consist of 1700 symbols for the MG tasks and 4000 symbols for sine-to-square. Symbols 1 to 500 form the washout block, and symbols 501 to 1000 (501 to 2500 for sine-to-square) form the training block. Symbols 1001 to 1200 (2501 to 3000) form the held-out block, which was not used in any search objective, and symbols 1201 to 1700 (3001 to 4000) form the optimization block. The error is quantified by the NRMSE on the evaluation block, i.e., $\operatorname{NRMSE}=\sqrt{|\mathcal T|^{-1}\sum_{n\in\mathcal T}[y(n)-\hat y(n)]^2}/s_{y,\mathcal T}$, where $\mathcal T$ denotes the set of symbols in the evaluation block, and $s_{y,\mathcal T}$ is the sample standard deviation of the target on that block; hence, a constant predictor equal to the target mean yields an NRMSE close to one. The readout is fitted on the training block only, with a penalized intercept, and refitted for every configuration, every task, and every realization \cite{lukosevicius2009reservoir,cucchi2022hands}. For every evaluation, the ridge parameter is chosen from the grid $\lambda\in\{10^{-6},10^{-5},\ldots,10^{3}\}$ as the value yielding the smallest error on the optimization block. Under noise, a single ridge value and a single $L$ are selected for the whole realization ensemble as the minimizers of the mean optimization-block error over the realizations, whereas the output weights are fitted separately on each realization. For the mean-field optima of the three benchmark tasks, this rule selects $\lambda=10^{-6}$, whereas the other tasks and schemes use the values listed in their parameter tables, e.g., $10^{-5}$ for the S4 threshold-decision optimum (Supporting Information Table~S26). Every search minimized the error on the optimization block, and every reported configuration was subsequently evaluated on the held-out block with the readout trained as before. The held-out block was excluded from every search objective; hence, it serves as a retrospective check of the comparisons, and it does not provide a prospective validation of the full design procedure \cite{cawley2010overfitting}. The inputs are normalized to $[0,1]$ over the whole sequence, and $L$ and $\lambda$ of every configuration are selected on the optimization block. As baselines, we use an input-tap regression without a channel and a current-symbol readout ablation of the tuned channel ($L=1$). The input-tap baseline regresses the target on the current input and the $k-1$ preceding inputs, for $k$ up to 50, and its best value over these tap lengths constitutes the linear reference of a task. Note that this is a digital baseline, and it does not bound the performance of all receiver algorithms. The ablation differs from S0, which retains the nominal physical parameters while tuning the readout memory and the regularization.

\threesubsection{Bayesian Optimization and Tunable-Parameter Sets}
\label{sec:met_bo}

The searches employ Bayesian optimization with a Gaussian-process surrogate and the expected-improvement acquisition function \cite{snoek2012practical,racca2021robust,griffith2019forecasting}, as implemented in scikit-optimize 0.10.2, with a Mat\'ern kernel and 100 evaluations per search, the first ten of which are Latin-hypercube points. The readout memory $L$ is treated as an integer variable, whereas the ridge parameter is chosen within every evaluation, as described above. The search domain spans $5\times10^{-20}$ to $2\times10^{-17}$~m$^3$~s$^{-1}$ for $k_{\mathrm{on}}$, 0.1 to 10~s$^{-1}$ for $k_{\mathrm{off}}$, and 0.5 to 2~s for $T$. It further spans 6 to 50~$\mu$m for $d$, 200 to 20000 molecules for $N_{\max}$, $0.5\times10^{-11}$ to $2\times10^{-10}$~m$^2$~s$^{-1}$ for $D$, and the integers 1 to 5 for $L$. Logarithmic transforms are applied to the parameters spanning several orders of magnitude (Supporting Information Table~S5). Under each of S1 to S4, ten independent mean-field searches with different random seeds were run for every task, i.e., 120 searches and 12000 evaluations in total for the three primary benchmark tasks, with the parameters outside the set held at their nominal values. Since S0 has no free physical parameter, it is enumerated over $L$. The result of each search is its best configuration. With equal evaluation budgets, the comparison concerns what can be found within a fixed number of objective evaluations. Note that this does not correspond to an equal physical calibration cost, since changing $T$ changes the duration of a transmission, and physical interventions differ in cost. The search domain is an idealized design domain. In particular, the rate domain, i.e., $3\times10^{7}$ to $1.2\times10^{10}$~M$^{-1}$s$^{-1}$ for $k_{\mathrm{on}}$, does not imply that every rate pair is realizable with a single ligand, surface, and medium, and several optima lie on a boundary of the domain (Section~\ref{sec:res_optima}). The wider domain used for the boundary check keeps the association rate at its diffusion-limited upper bound, and it extends the dissociation rate up to 100~s$^{-1}$, the symbol interval up to 5~s, the release amplitude up to $10^{5}$ molecules, and the distance down to 3.5~$\mu$m, i.e., just outside the reception shell of the particle model. In this wider domain, five mean-field searches and three noise-aware searches at 50000 receptors were run per task under S1 and S4.

\threesubsection{Noise-Aware Searches and Re-Evaluation}
\label{sec:met_noise}

The noise-aware searches use the same optimizer, domain, and budget, with the stochastic receptor model as the objective. Each evaluation simulates $R=10$ realizations of the receptor process with $N_R$ receptors, fits the readout on the training block of each realization, and returns the mean NRMSE over the realizations on the optimization block at the best ridge value. The random-number seed of the receptor process is fixed within a search, and the surrogate models the remaining fluctuation with an additive noise term. Under each of S1 to S4 and for each of the three primary tasks, five searches were run at each of $N_R=500$, 5000, and 50000, and three searches at $N_R=5\times10^{5}$, i.e., 216 searches in total, whereas S0 was enumerated over $L$ with the same objective. The result of each search is its best evaluation, and the best result across the five or three searches performed for a condition constitutes the noise-aware optimum of the set. Every optimum, whether mean-field or noise-aware, was then re-evaluated with 50 independent realizations at $N_R$ from 500 to $5\times10^{5}$ under two evaluation schemes. The fixed-readout scheme keeps $L$ and $\lambda$ of the optimum, whereas the retrained-readout scheme keeps the six physical parameters and re-selects $L$ from 1 to 5 and $\lambda$ on the optimization block of the same realizations, i.e., the readout that a user would fit on the noisy receiver. The retrained sweep of the mean-field optima of S0, S1, and S4 was continued to $N_R=10^{6}$, $10^{7}$, $10^{8}$, and $10^{9}$ with 20 independent realizations for every task (Supporting Information Section~S13). For the three additional tasks, five mean-field searches per task were run under S1 and S4, and three noise-aware searches per task, set, and receptor number were run under S0, S1, and S4 at 500 and 50000 receptors, with the optima re-evaluated in the same way (Supporting Information Section~S11). All values quoted for the noise-aware optima in Sections~\ref{sec:res_noise} and \ref{sec:res_noiseaware} are these re-evaluations with independent realizations. The search objectives themselves are biased downward by the selection of the best of 100 noisy evaluations, and hence, they are not quoted.

\threesubsection{Readout Designed for Noise, Joint Search, and Critical Receptor Numbers}
\label{sec:met_remedies}

The readout designed for noise changes three settings of the reservoir state (Section~\ref{sec:res_budget}). First, integrating nodes replace the occupancy at the node instant by its mean over the node window $T/M'$, with $M'\in\{50,25,10,5\}$ nodes per symbol for the tasks with $M=50$, and $M'\in\{100,50,20,10\}$ for sine-to-square. Second, a causal first-order low-pass filter with time constant $\tau_f$ is applied to the node sequence, with $\tau_f/T\in\{0,0.1,0.25,0.5,1,2,5\}$, where 0 denotes no filter. Third, the readout memory takes the values $L\in\{5,10,20\}$, with the number of features $M'L$ capped at 1000. Every combination was evaluated for the S1 and S4 mean-field optima of the eight tasks, both in the mean-field model and at $N_R=500$, 5000, 50000, and $5\times10^{5}$ with 20 independent realizations, with the ridge value chosen as before. The best combination at each receptor number gives the value of the readout designed for noise, and the sampled readout with $M$ nodes, the $L$ of the optimum, and a re-selected ridge value on the same realizations constitutes the default of the same run. Note that this grid default differs from the retrained-readout scheme of the receptor-number sweep, which also re-selects $L$, and uses its own realizations and a denser receptor grid. Therefore, differences between the crossing estimates of the two schemes cannot be attributed to the readout alone. Supporting Information Table~S34 compares the grid-run sampled default with the designed readout on the same grid and realizations.

The joint search adds $M'$ as a categorical variable, $\tau_f/T$ between 0.05 and 5 on a logarithmic scale, and $L$ between 1 and 20 to the channel parameters of S1 and S3. Under S1, the receptor rates keep the default domain, whereas under S3, the symbol interval was allowed to vary from 0.5 to 5~s and the release amplitude from 200 to 20000 molecules (Supporting Information Table~S5). The joint search has no cap on the feature count $M'L$, whereas the readout grid above caps it at 1000. These searches are distinct from the default physical-parameter searches and from the readout grid on fixed configurations, and the improvements of the joint S3 design reflect the combined search over physical and readout settings in this wider domain. The joint search minimizes the stochastic objective described in Section~\ref{sec:met_noise} at 5000 and 50000 receptors, with three searches per set, task, and receptor number for MG forecasting, sine-to-square, MG-Cubed, and NARMA-10, i.e., 48 searches in total. Each joint search starts from twelve Latin-hypercube points instead of ten. The result of each search is its best evaluation, and the optima were re-evaluated with 50 independent realizations.

The critical receptor number $N_R^{*}$ is an estimated criterion-crossing receptor count for the selected S4 mean-field configuration, readout scheme, and evaluation block. Specifically, it is the receptor number at which the retrained error falls below the linear reference of the task, obtained by linear interpolation of the error against $\log N_R$ between the evaluated receptor numbers. It is reported as $\le500$ when the criterion is met at the smallest evaluated receptor count, and as not reached when no crossing is observed within the stated sweep. The second criterion replaces the reference with an illustrative NRMSE of 0.25. The designed readout is the best of the integrating-node grid and the sampled default, evaluated on the same realizations up to $5\times10^{5}$ receptors, and beyond that receptor number, the sampled default, which is one of its candidates, continues the curve. The cell-like readout restricts the grid to one or two integrating windows per symbol ($M'=1$ or 2), i.e., the occupancy averaged over the whole symbol interval or over each of its halves, with the same low-pass constants and readout memories. It was evaluated for the S1 and S4 optima of the eight tasks, both in the mean-field model and at the same four receptor numbers with 20 independent realizations. Its critical receptor numbers follow the same two criteria within the sweep up to $5\times10^{5}$ receptors (Supporting Information Section~S12).

\threesubsection{Reference Echo State Network}
\label{sec:met_esn}

For each task, a feature-count-matched echo state network serves as a digital reference \cite{jaeger2001echo,lukosevicius2009reservoir}. Here, equal size means that the number of units equals the $5M$ features of the default channel readout; it implies neither equal hardware resources nor equal independent dynamical degrees of freedom. The network has $N=5M$ leaky units, i.e., 250 for the tasks with $M=50$ virtual nodes and 500 for sine-to-square, which equals the number of features of the default channel readout with $L=5$. The recurrent weights are drawn uniformly from $[-1,1]$ with a connection density of 0.1 and scaled to the spectral radius $\rho$, whereas the input weights and the biases are drawn uniformly from $[-1,1]$ and scaled by $s_{\mathrm{in}}$ and $s_b$, respectively. The state evolves as $\mathbf x_{n}=(1-a)\,\mathbf x_{n-1}+a\tanh(\mathbf W\mathbf x_{n-1}+\mathbf w_{\mathrm{in}}u_n+\mathbf b)$, where $a$ is the leak rate, and the input $u_n$ is scaled to $[0,1]$ as in the release encoding. The readout regresses the target on the state, the input, and a constant, using the same blocks, ridge grid, and ridge rule as the channel readout. The four hyperparameters, i.e., $\rho$ from 0.1 to 1.5, $s_{\mathrm{in}}$ from 0.01 to 10 on a logarithmic scale, $a$ from 0.05 to 1, and $s_b$ from 0 to 1, are tuned by the optimizer described in Section~\ref{sec:met_bo} with the same budget, in three searches per task with the weight seed fixed within a search. The best configuration of each search is re-evaluated with ten independent weight seeds, and the best of the three searches provides the reference value of the task (Supporting Information Section~S14).

\threesubsection{Error Surfaces and Metrics}
\label{sec:met_landscapes}

The receiver error surface is evaluated on a 25 by 25 grid over $k_{\mathrm{on}}$ and $k_{\mathrm{off}}$ at the nominal transmitter and channel settings. Similarly, the transmitter error surface is evaluated on a 25 by 25 grid over $T$ and $N_{\max}$ at the nominal receiver and channel settings, and the channel error surface shown in the Supporting Information on a 21 by 21 grid over $d$ and $D$. At every grid point, the occupancy is simulated once, the readout is fitted for $L=1$ to 5 and for every ridge value, and the best pair is reported. The noise-aware receiver error surfaces shown in Section~\ref{sec:res_noiseaware} use a 17 by 17 grid with $R=10$ realizations per point. The metrics of every configuration are computed from its own mean-field simulation under the task input, where $c_{\mathrm{peak}}$ is the maximum concentration at the receiver, the occupancy statistics are taken over the optimization block, and $K_D$ is converted to molar units using Avogadro's number.

\threesubsection{Cross-Task Reuse}
\label{sec:met_reuse}

A reuse evaluation retains the six physical parameters of a configuration tuned for the source task, and takes the input series, the target, the number of virtual nodes $M$, and the data blocks from the target task. The readout is retrained on the training block of the target task, with $L$ and $\lambda$ re-selected on its optimization block. In the mean-field model, the best configuration of each of the ten source-task searches under a set is evaluated on the two other tasks. The penalty in percent is given by $100\,(E_{\mathrm{reuse}}/E_{\mathrm{ref}}-1)$, where $E_{\mathrm{reuse}}$ is the median error over the ten reused configurations, and $E_{\mathrm{ref}}$ is the median error of the ten searches tuned for the target task within the same set. Under noise, the noise-aware optimum of each set and design receptor number $N_R$ is evaluated on the other tasks at the same $N_R$ with 20 independent realizations, and the reference is the noise-aware optimum of the target task under the same scheme.

\threesubsection{Particle-Based Simulations}
\label{sec:met_particle}

The particle-based simulations that validate the stochastic receptor model were run with Smoldyn 2.75 \cite{andrews2017smoldyn} for twelve configurations, i.e., the best configuration of one search under each set for each task in a preliminary set of mean-field searches. The domain of those preliminary searches admitted transmitter-receiver distances below the receiver radius, which the domain described in Section~\ref{sec:met_bo} excludes, and their configurations serve only this validation. The simulation region is a cube with a half-width of 25~$\mu$m, whose faces carry Smoldyn's source-dependent unbounded-emitter absorption, such that the field of molecules emitted at the transmitter position emulates an unbounded medium. The receiver is a reflecting sphere of radius 3~$\mu$m with $N_R=500$ immobile receptors placed randomly on its surface, located at the center-to-center distance of the configuration from a point transmitter. Binding converts a free molecule and an unbound receptor into a bound receptor with rate $k_{\mathrm{on}}$, and unbinding releases both with rate $k_{\mathrm{off}}$. The task-encoded integer number of molecules is released at the transmitter at every $(n-1)T$, the time step is 0.01~s, and the number of bound receptors is recorded at every step. Five realizations with different random seeds were run for each configuration, i.e., 60 runs in total. The recorded occupancy was smoothed by a trailing moving average over $K+1$ samples, with $K=2000$ (20~s) for the MG tasks and $K=6000$ (60~s) for sine-to-square. The smoothed occupancy was sampled at the virtual-node instants on the 0.01~s grid, and read out with $\lambda=10^{-6}$ by a readout fitted on the training block of each realization. The unfiltered NRMSE values are obtained with the same pipeline without the moving average. For the comparison in Section~\ref{sec:res_noise}, the stochastic receptor model was run for the same twelve configurations with 50 realizations, and processed with the same pipeline, including the filter and the ridge value.

\threesubsection{Minimax Configuration}
\label{sec:met_compromise}

The minimax search uses the same domain and optimizer as the task-specific searches. Its objective, which is minimized over the configurations, is the largest ratio, over the three benchmark tasks, between the NRMSE of a single configuration on a task and the error of the task-specific optimum of the same set on that task. The physical parameters and $L$ are shared across the tasks, whereas the readout, including the ridge value, is fitted for every task. Five mean-field searches and three noise-aware searches at 50000 receptors were run under S1 and S4, where the noise-aware searches were run against the noise-aware task-specific optima at that receptor number.

\threesubsection{Synthetic Glucose Case Study}
\label{sec:met_case}

The case-study input is a synthetic blood glucose signal generated from the minimal model of glucose and insulin kinetics \cite{bergman1979minimal}, driven by meals. The model equations are
\begin{equation}
\begin{aligned}
\frac{dG}{dt}&=-p_1(G-G_b)-XG+\frac{R_a(t)}{V_G},\\
\frac{dX}{dt}&=-p_2X+p_3s(t)(I-I_b),\\
\frac{dI}{dt}&=-n(I-I_b)+\sigma\max(G-G_b,0),
\end{aligned}
\label{eq:glucose}
\end{equation}
where $G$, $X$, and $I$ denote the glucose level, the insulin action, and the insulin level, respectively. The glucose appearance rate $R_a(t)$ is the sum of the contributions of all meals, where, following Hovorka et al.\ \cite{hovorka2004nonlinear}, a meal of size $D_G$ that starts at $t_m$ contributes $D_GA_G(t-t_m)\exp[-(t-t_m)/t_{\max}]/t_{\max}^2$ for $t\geq t_m$, and $A_G$ and $t_{\max}$ denote the absorbed fraction of the meal and the time to peak absorption, respectively. Three meals per day, i.e., in the morning, at midday, and in the evening, have random timing and size. An optional afternoon snack, a 24-hour modulation of insulin sensitivity, and a slow fluctuation of the basal glucose represent the variability of a real glucose signal. The model is integrated with a fourth-order Runge-Kutta scheme at a one-minute step over 14 days, and sampled every five minutes, which yields one sample per symbol of the channel. The parameters of the generator are listed in Supporting Information Table~S25, and the whole signal is generated from a single random seed (2028).

Under S0 and S1, as well as in the switching runs, the channel run is a time-compressed and encoded transmission of this signal at the nominal symbol interval $T=1.25$~s. Under S4, the symbol interval is tunable, and the selected values are listed in Supporting Information Table~S26. Note that the five-minute sampling interval of the source series is distinct from the symbol interval of the channel, and times in symbols map to minutes at the source sampling interval. For sampled nodes, the last feature of symbol $n$ is acquired at $nT-T/M$, and the lead time to a target $h$ symbols ahead is $(h-1)T+T/M$, whereas for integrating nodes, the last window ends at $nT$, and the lead time is $(h-1)T$, before any implementation latency. The forecasting target is the glucose level six samples (30 minutes) ahead, and the threshold-decision target is one where the current level exceeds 140~mg/dL, and zero otherwise. The input and the forecasting target are normalized to $[0,1]$ over the signal. The blocks, the node count, and the readout follow those of the MG tasks (Section~\ref{sec:met_tasks}). The channel symbol interval represents the sampling interval of the biomarker, and the tuned rates translate to a deployed system through the dimensionless groups defined in Section~\ref{sec:system}. Under S0, S1, and S4, five mean-field searches and three noise-aware searches at 50000 receptors were run for each target, together with three noise-aware searches at 500 receptors under S0 and S1.

\threesubsection{Online Task Switching}
\label{sec:met_switch}

A switching run is one simulation of a single S1 channel over 3400 symbols. The transmitter and channel parameters of this channel are nominal, and its receptor rates are those of the noise-aware S1 optimum of the source task at 50000 receptors. This optimum is the S1 configuration with the lowest noise-aware error at 50000 receptors among the searches in the default domain and in the wider domain (Supporting Information Section~S7). For MG forecasting and MG-Cubed, it is the wider-domain optimum, whose dissociation rate lies above the default range, whereas for sine-to-square and the two biomarker targets, it is the default-domain optimum. After the first 1700 symbols, i.e., at the boundary between symbols 1700 and 1701, the rates are changed to those of the target-task optimum, either at once or along a log-linear ramp of 10 or 50 symbols that starts at that boundary. The readout is replaced by the target-task readout at the same boundary, and symbol 1701 is the first target-task symbol. Both readouts were trained beforehand on standalone runs of their own configurations, on the training block of the standard task, with the ridge value of the corresponding optimum. Two reference runs use the same input and the same random numbers, i.e., one with the target configuration throughout, and one with the source configuration throughout and a readout retrained for the target task on the source channel. The error is the root-mean-square error in a centered window of 25 symbols, normalized by the standard deviation of the active target over the sequence, and averaged over 50 realizations of the receptor process. The adaptation time is defined as the first window after the switch at which the error of the switched run stays within 10 percent of the reference level given by the reference run for the next 100 symbols. The task pairs are the two case-study targets, MG forecasting and MG-Cubed, which share their input series, and sine-to-square and MG forecasting, for which the input changes with the task, each in both directions, i.e., six directed switches in total. The windows after the switch use the target of the new task and its standard deviation over the sequence, and the windows that overlap the switching instant are considered part of the transient.

\threesubsection{Memory and Nonlinear Capacity}
\label{sec:met_capacity}

The linear memory capacity and the second-order capacity follow the construction of Dambre et al.\ \cite{dambre2012capacity}, restricted to linear delays of 1 to 30 symbols, and to second-order Legendre terms of delays 0 to 15 with their pairwise products up to delay 5. The input is an independent, uniformly distributed sequence of 2200 symbols, encoded as in the tasks, and the readout is fitted on 1000 symbols, with the ridge value selected on the following 700 symbols. Each capacity is the sum of the squared correlations between the prediction and the target on these 700 symbols, averaged over three input sequences. The second-order terms are the Legendre polynomial $P_2$ of the input at delays 0 to 15 and the products of $P_1$ at two different delays up to 5. Note that these are finite-basis estimates with validation-selected regularization, and thus, they are neither independent-test values nor exhaustive total capacities. The readout uses the $L$ of the evaluated configuration, i.e., 5 for the nominal configuration. The mean-field S1 optima of the three tasks and the nominal configuration are evaluated both in the mean-field model and with the stochastic receptor model at 500 to 500000 receptors, with five realizations per input sequence.

\threesubsection{Statistics and Reproducibility}
\label{sec:met_stats}

The mean-field search results are reported as means and sample standard deviations over ten searches, and the reuse results as medians and interquartile ranges over ten configurations. The noise-aware results are reported as means and sample standard deviations over independent realizations, and the particle results as means and sample standard deviations over five realizations. Every number in the text, tables, and figures is generated from the result files by analysis scripts. The simulator, the optimization scripts, the analysis scripts, and the result files are available as described in the Data Availability Statement.

\medskip
\textbf{Supporting Information}\par
Supporting Information is available from the Wiley Online Library or from the author. It contains the simulator validation and robustness checks, the task definitions and data blocks, the search domain, the complete results of the mean-field and noise-aware searches, the error-surface summaries, the reuse tables on both evaluation blocks, the particle-simulation results, the receptor-number sweeps to $10^{9}$ receptors, the wider-domain and minimax searches, the tables of the case study and of the online task switching, the capacity analysis, the additional tasks, the readout grid and joint searches, and the reproducibility notes.

\medskip
\begin{minipage}{\textwidth}
\textbf{Acknowledgements}\par
This work was supported by the Scientific and Technological Research Council of Turkey (TUBITAK) under Grants \#123E516 and \#123C592.

During the preparation of this work, generative AI tools were used to revise parts of the text and to write and check some of the scripts, and the authors take full responsibility for the content of this article.

\medskip
\textbf{Author Contributions}\par
S.Y., K.B.I., and M.K. designed the study and carried out the simulations and analyses. M.K. supervised the work. All authors wrote and revised the manuscript.

\medskip
\textbf{Conflict of Interest}\par
The authors declare no conflict of interest.

\medskip
\textbf{Data Availability Statement}\par
The data and code that support the findings of this study are openly available in GitHub at \url{https://github.com/calico-lab/task-adaptive-mc-prc}. These include the simulator, the optimization scripts, the scripts for the re-evaluation, reuse, error-surface, validation, robustness, readout, and critical-receptor-number analyses, the figure and table generation scripts, and the result files of all optimization runs. An archived copy with a DOI will be deposited upon acceptance.
\end{minipage}

\begin{figure}[p]
\textbf{Table of Contents}\par
\medskip
\centering
\includegraphics[width=110mm]{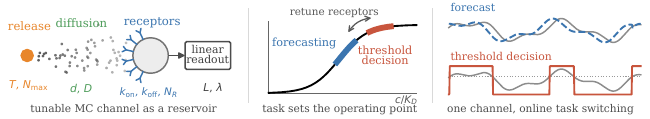}
\medskip
\caption*{A molecular communication channel with ligand receptors serves as a physical reservoir computer, with memory from diffusion and nonlinearity from receptor saturation. Both are set by parameters that living cells regulate. Different computing tasks require different configurations, tuning the receptor kinetics alone provides most of the benefit, and a single channel can switch between tasks during operation.}
\end{figure}


\begin{thebibliography}{99}

\bibitem{akyildiz2015iobnt}
I. F. Akyildiz, M. Pierobon, S. Balasubramaniam, and Y. Koucheryavy, ``The Internet of Bio-Nano Things,'' \textit{IEEE Communications Magazine}, vol.~53, no.~3, pp.~32--40, 2015. \url{https://doi.org/10.1109/MCOM.2015.7060516}

\bibitem{farsad2016survey}
N. Farsad, H. B. Yilmaz, A. Eckford, C.-B. Chae, and W. Guo, ``A comprehensive survey of recent advancements in molecular communication,'' \textit{IEEE Communications Surveys \& Tutorials}, vol.~18, no.~3, pp.~1887--1919, 2016. \url{https://doi.org/10.1109/COMST.2016.2527741}

\bibitem{akyildiz2019moving}
I. F. Akyildiz, M. Pierobon, and S. Balasubramaniam, ``Moving forward with molecular communication: From theory to human health applications,'' \textit{Proceedings of the IEEE}, vol.~107, no.~5, pp.~858--865, 2019.

\bibitem{kuscu2021iobnt}
M. Kuscu and B. D. Unluturk, ``Internet of Bio-Nano Things: A review of applications, enabling technologies and key challenges,'' \textit{ITU Journal on Future and Evolving Technologies}, vol.~2, no.~3, 2021.

\bibitem{pierobon2010physical}
M. Pierobon and I. F. Akyildiz, ``A physical end-to-end model for molecular communication in nanonetworks,'' \textit{IEEE Journal on Selected Areas in Communications}, vol.~28, no.~4, pp.~602--611, 2010. \url{https://doi.org/10.1109/JSAC.2010.100509}

\bibitem{nakano2013book}
T. Nakano, A. W. Eckford, and T. Haraguchi, \textit{Molecular Communication}. Cambridge University Press, 2013.

\bibitem{jamali2019channel}
V. Jamali, A. Ahmadzadeh, W. Wicke, A. Noel, and R. Schober, ``Channel modeling for diffusive molecular communication: A tutorial review,'' \textit{Proceedings of the IEEE}, vol.~107, no.~7, pp.~1256--1301, 2019. \url{https://doi.org/10.1109/JPROC.2019.2919455}

\bibitem{kilinc2013receiver}
D. Kilinc and O. B. Akan, ``Receiver design for molecular communication,'' \textit{IEEE Journal on Selected Areas in Communications}, vol.~31, no.~12, pp.~705--714, 2013.

\bibitem{noel2014enzymes}
A. Noel, K. C. Cheung, and R. Schober, ``Improving receiver performance of diffusive molecular communication with enzymes,'' \textit{IEEE Transactions on NanoBioscience}, vol.~13, no.~1, pp.~31--43, 2014.

\bibitem{pierobon2011noise}
M. Pierobon and I. F. Akyildiz, ``Noise analysis in ligand-binding reception for molecular communication in nanonetworks,'' \textit{IEEE Transactions on Signal Processing}, vol.~59, no.~9, pp.~4168--4182, 2011. \url{https://doi.org/10.1109/TSP.2011.2159497}

\bibitem{chou2015receiver}
C. T. Chou, ``Impact of receiver reaction mechanisms on the performance of molecular communication networks,'' \textit{IEEE Transactions on Nanotechnology}, vol.~14, no.~2, pp.~304--317, 2015.

\bibitem{kuscu2016sinw}
M. Kuscu and O. B. Akan, ``Modeling and analysis of SiNW FET-based molecular communication receiver,'' \textit{IEEE Transactions on Communications}, vol.~64, no.~9, pp.~3708--3721, 2016.

\bibitem{kuscu2018ml}
M. Kuscu and O. B. Akan, ``Maximum likelihood detection with ligand receptors for diffusion-based molecular communications in Internet of Bio-Nano Things,'' \textit{IEEE Transactions on NanoBioscience}, vol.~17, no.~1, pp.~44--54, 2018.

\bibitem{kuscu2022interference}
M. Kuscu and O. B. Akan, ``Detection in molecular communications with ligand receptors under molecular interference,'' \textit{Digital Signal Processing}, vol.~124, art.~103186, 2022.

\bibitem{kuscu2019transmitter}
M. Kuscu, E. Dinc, B. A. Bilgin, H. Ramezani, and O. B. Akan, ``Transmitter and receiver architectures for molecular communications: A survey on physical design with modulation, coding, and detection techniques,'' \textit{Proceedings of the IEEE}, vol.~107, no.~7, pp.~1302--1341, 2019.

\bibitem{jaeger2001echo}
H. Jaeger, ``The `echo state' approach to analysing and training recurrent neural networks,'' GMD Report 148, German National Research Center for Information Technology, 2001.

\bibitem{maass2002realtime}
W. Maass, T. Natschl\"ager, and H. Markram, ``Real-time computing without stable states: A new framework for neural computation based on perturbations,'' \textit{Neural Computation}, vol.~14, no.~11, pp.~2531--2560, 2002.

\bibitem{lukosevicius2009reservoir}
M. Luko\v{s}evi\v{c}ius and H. Jaeger, ``Reservoir computing approaches to recurrent neural network training,'' \textit{Computer Science Review}, vol.~3, no.~3, pp.~127--149, 2009.

\bibitem{dambre2012capacity}
J. Dambre, D. Verstraeten, B. Schrauwen, and S. Massar, ``Information processing capacity of dynamical systems,'' \textit{Scientific Reports}, vol.~2, art.~514, 2012. \url{https://doi.org/10.1038/srep00514}

\bibitem{inubushi2017beyond}
M. Inubushi and K. Yoshimura, ``Reservoir computing beyond memory-nonlinearity trade-off,'' \textit{Scientific Reports}, vol.~7, art.~10199, 2017. \url{https://doi.org/10.1038/s41598-017-10257-6}

\bibitem{tanaka2019recent}
G. Tanaka, T. Yamane, J.-B. H\'eroux, R. Nakane, N. Kanazawa, S. Takeda, H. Numata, D. Nakano, and A. Hirose, ``Recent advances in physical reservoir computing: A review,'' \textit{Neural Networks}, vol.~115, pp.~100--123, 2019. \url{https://doi.org/10.1016/j.neunet.2019.03.005}

\bibitem{nakajima2020physical}
K. Nakajima, ``Physical reservoir computing: An introductory perspective,'' \textit{Japanese Journal of Applied Physics}, vol.~59, art.~060501, 2020.

\bibitem{liang2024emerging}
X. Liang, J. Tang, Y. Zhong, B. Gao, H. Qian, and H. Wu, ``Physical reservoir computing with emerging electronics,'' \textit{Nature Electronics}, vol.~7, pp.~193--206, 2024.

\bibitem{baltussen2024chemical}
M.~G. Baltussen, T.~J. de~Jong, Q.~Duez, W.~E. Robinson, and W.~T.~S. Huck, ``Chemical reservoir computation in a self-organizing reaction network,'' \textit{Nature}, vol.~631, pp.~549--555, 2024.

\bibitem{appeltant2011information}
L. Appeltant, M. C. Soriano, G. Van der Sande, J. Danckaert, S. Massar, J. Dambre, B. Schrauwen, C. R. Mirasso, and I. Fischer, ``Information processing using a single dynamical node as complex system,'' \textit{Nature Communications}, vol.~2, art.~468, 2011.

\bibitem{uzun2025molecular}
M. Uzun, K. B. Ikiz, and M. Kuscu, ``Molecular communication channel as a physical reservoir computer,'' \textit{IEEE Transactions on Emerging Topics in Computing}, early access, 2026. \url{https://doi.org/10.1109/TETC.2026.3708167}

\bibitem{lee2023task}
O. Lee, T. Wei, K. D. Stenning, J. C. Gartside, D. Prestwood, S. Seki, A. Aqeel, K. Karube, N. Kanazawa, Y. Taguchi, C. Back, Y. Tokura, W. R. Branford, and H. Kurebayashi, ``Task-adaptive physical reservoir computing,'' \textit{Nature Materials}, vol.~23, no.~1, pp.~79--87, 2024. \url{https://doi.org/10.1038/s41563-023-01698-8}

\bibitem{maraj2023sensory}
J.~J. Maraj, K.~P.~T. Haughn, D.~J. Inman, and S.~A. Sarles, ``Sensory adaptation in biomolecular memristors improves reservoir computing performance,'' \textit{Advanced Intelligent Systems}, vol.~5, no.~8, art.~2300049, 2023.

\bibitem{mohamed2025memimpedance}
A. S. Mohamed, N. X. Armendarez, M. S. Hasan, and J. S. Najem, ``Memimpedance-based biomolecular device for adaptive physical reservoir computing,'' \textit{Advanced Intelligent Systems}, vol.~7, art.~2500281, 2025. \url{https://doi.org/10.1002/aisy.202500281}

\bibitem{ghosh2026recursive}
S. Ghosh, M. G. Baltussen, A. C. Knox, R. Haije, Q. Duez, A. T. Tsitsimeli, M. H. Chak, J. E. Beves, and W. T. S. Huck, ``A recursive enzymatic competition network capable of multitask molecular information processing,'' \textit{Nature Chemistry}, vol.~18, pp.~302--308, 2026. \url{https://doi.org/10.1038/s41557-025-01981-y}

\bibitem{barkai1997robustness}
N. Barkai and S. Leibler, ``Robustness in simple biochemical networks,'' \textit{Nature}, vol.~387, pp.~913--917, 1997.

\bibitem{sartori2011noise}
P. Sartori and Y. Tu, ``Noise filtering strategies in adaptive biochemical signaling networks: Application to E. coli chemotaxis,'' \textit{Journal of Statistical Physics}, vol.~142, no.~6, pp.~1206--1217, 2011. \url{https://doi.org/10.1007/s10955-011-0169-z}

\bibitem{neumann2014gain}
S. Neumann, L. L\o vdok, K. Bentele, J. Meisig, E. Ullner, F. S. Paldy, V. Sourjik, and M. Kollmann, ``Exponential signaling gain at the receptor level enhances signal-to-noise ratio in bacterial chemotaxis,'' \textit{PLOS ONE}, vol.~9, no.~4, art.~e87815, 2014. \url{https://doi.org/10.1371/journal.pone.0087815}

\bibitem{kuscu2026adaptive}
M. Kuscu, ``Adaptive molecular communication receivers with tunable ligand-receptor interactions,'' \textit{IEEE Transactions on Molecular, Biological, and Multi-Scale Communications}, vol.~12, pp.~482--492, 2026. \url{https://doi.org/10.1109/TMBMC.2026.3681866}

\bibitem{ding2024kineticitc}
Y. Ding and J. Liu, ``Kinetic ITC of DNA aptamers binding for small molecules and implications for binding assays and biosensors,'' \textit{ChemBioChem}, vol.~25, no.~15, art.~e202400225, 2024. \url{https://doi.org/10.1002/cbic.202400225}

\bibitem{ding2023pushing}
Y. Ding and J. Liu, ``Pushing adenosine and ATP SELEX for DNA aptamers with nanomolar affinity,'' \textit{Journal of the American Chemical Society}, vol.~145, no.~13, pp.~7540--7547, 2023. \url{https://doi.org/10.1021/jacs.3c00848}

\bibitem{alkhamis2024slowoff}
O. Alkhamis, J. Canoura, L. Wang, and Y. Xiao, ``Nuclease-assisted selection of slow-off rate aptamers,'' \textit{Science Advances}, vol.~10, no.~24, art.~eadl3426, 2024. \url{https://doi.org/10.1126/sciadv.adl3426}

\bibitem{chen2022tunable}
Z.-M. Chen, Q. Mou, S.-H. Wu, Y. Xie, K. Salminen, and J.-J. Sun, ``Real-time tunable dynamic range for calibration-free biomolecular measurements with a temperature-modulated electrochemical aptamer-based sensor in an unprocessed actual sample,'' \textit{Analytical Chemistry}, vol.~94, no.~2, pp.~1397--1405, 2022. \url{https://doi.org/10.1021/acs.analchem.1c04697}

\bibitem{ma2017electrostatic}
X. Ma, A. Gosai, G. Balasubramanian, and P. Shrotriya, ``Aptamer based electrostatic-stimuli responsive surfaces for on-demand binding/unbinding of a specific ligand,'' \textit{Journal of Materials Chemistry B}, vol.~5, no.~20, pp.~3675--3685, 2017. \url{https://doi.org/10.1039/C6TB02386J}

\bibitem{bolhassan2024microfluidic}
E. Akyol, A. B. Ozturk, I. M. Bolhassan, and M. Kuscu, ``Experimental characterization of hydrodynamic gating-based molecular communication transmitter,'' in \textit{Proc. 11th ACM Int. Conf. Nanoscale Computing and Communication (NanoCom)}, pp.~60--65, 2024. \url{https://doi.org/10.1145/3686015.3689353}

\bibitem{smith2023engineering}
A.-M. Makri Pistikou, G. A. O. Cremers, B. L. Nathalia, T. J. Meuleman, B. W. A. B\"ogels, B. V. Eijkens, A. de Dreu, M. T. H. Bezembinder, O. M. J. A. Stassen, C. C. V. Bouten, M. Merkx, R. Jerala, and T. F. A. de Greef, ``Engineering a scalable and orthogonal platform for synthetic communication in mammalian cells,'' \textit{Nature Communications}, vol.~14, art.~7001, 2023. \url{https://doi.org/10.1038/s41467-023-42810-5}

\bibitem{monck2024synthetic}
C. Monck, Y. Elani, and F. Ceroni, ``Genetically programmed synthetic cells for thermo-responsive protein synthesis and cargo release,'' \textit{Nature Chemical Biology}, vol.~20, pp.~1380--1386, 2024. \url{https://doi.org/10.1038/s41589-024-01673-7}

\bibitem{downs2022realtime}
A.~M. Downs and K.~W. Plaxco, ``Real-time, in vivo molecular monitoring using electrochemical aptamer-based sensors: opportunities and challenges,'' \textit{ACS Sensors}, vol.~7, no.~10, pp.~2823--2832, 2022.

\bibitem{kuscu2021graphene}
M. Kuscu, H. Ramezani, E. Dinc, S. Akhavan, and O. B. Akan, ``Fabrication and microfluidic analysis of graphene-based molecular communication receiver for Internet of Nano Things (IoNT),'' \textit{Scientific Reports}, vol.~11, art.~19600, 2021. \url{https://doi.org/10.1038/s41598-021-98609-1}

\bibitem{arjmandi2016modulator}
H. Arjmandi, A. Ahmadzadeh, R. Schober, and M. Nasiri-Kenari, ``Ion channel based bio-synthetic modulator for diffusive molecular communication,'' \textit{IEEE Transactions on NanoBioscience}, vol.~15, no.~5, pp.~418--432, 2016.

\bibitem{zhu2022modular}
I. Zhu, R. Liu, J. M. Garcia, A. Hyrenius-Wittsten, D. I. Piraner, J. Alavi, D. V. Israni, B. Liu, A. S. Khalil, and K. T. Roybal, ``Modular design of synthetic receptors for programmed gene regulation in cell therapies,'' \textit{Cell}, vol.~185, no.~8, pp.~1431--1443.e16, 2022. \url{https://doi.org/10.1016/j.cell.2022.03.023}

\bibitem{heuberger2024protocell}
L. Heuberger, M. Korpidou, A. Guinart, D. Doellerer, D. Monserrat L\'opez, C.-A. Schoenenberger, D. Milinkovic, E. L\"ortscher, B. L. Feringa, and C. G. Palivan, ``Photoreceptor-like signal transduction between polymer-based protocells,'' \textit{Advanced Materials}, vol.~37, no.~3, art.~2413981, 2025. \url{https://doi.org/10.1002/adma.202413981}

\bibitem{mashima2022shuttling}
T. Mashima, M. H. M. E. van Stevendaal, F. R. A. Cornelissens, A. F. Mason, B. J. H. M. Rosier, W. J. Altenburg, K. Oohora, S. Hirayama, T. Hayashi, J. C. M. van Hest, and L. Brunsveld, ``DNA-mediated protein shuttling between coacervate-based artificial cells,'' \textit{Angewandte Chemie International Edition}, vol.~61, art.~e202115041, 2022. \url{https://doi.org/10.1002/anie.202115041}

\bibitem{schoukroun2015heterogeneous}
L. R. Schoukroun-Barnes, E. P. Glaser, and R. J. White, ``Heterogeneous electrochemical aptamer-based sensor surfaces for controlled sensor response,'' \textit{Langmuir}, vol.~31, no.~23, pp.~6563--6569, 2015. \url{https://doi.org/10.1021/acs.langmuir.5b01418}

\bibitem{abeykoon2025interrogation}
S. W. Abeykoon, W. N. Dikella, M. Santos-Cancel, R. A. Lazenby, and R. J. White, ``Interrogation of small molecule to surface-bound aptamer binding kinetics with electrochemical aptamer-based sensors using intermittent pulse amperometry,'' \textit{Analytical Chemistry}, vol.~97, no.~46, pp.~25391--25397, 2025. \url{https://doi.org/10.1021/acs.analchem.5c01604}

\bibitem{filius2025decoding}
M. Filius, L. Fasching, R. van Wee, A. Y. Rwei, and C. Joo, ``Decoding aptamer-protein binding kinetics for continuous biosensing using single-molecule techniques,'' \textit{Science Advances}, vol.~11, no.~7, art.~eads9687, 2025. \url{https://doi.org/10.1126/sciadv.ads9687}

\bibitem{thompson2020ph}
I. A. P. Thompson, L. Zheng, M. Eisenstein, and H. T. Soh, ``Rational design of aptamer switches with programmable pH response,'' \textit{Nature Communications}, vol.~11, art.~2946, 2020. \url{https://doi.org/10.1038/s41467-020-16808-2}

\bibitem{sun2024allosteric}
H. Sun, D. Zhao, Y. He, H.-M. Meng, and Z. Li, ``Aptamer-based DNA allosteric switch for regulation of protein activity,'' \textit{Advanced Science}, vol.~11, art.~2402531, 2024. \url{https://doi.org/10.1002/advs.202402531}

\bibitem{cucchi2022hands}
M.~Cucchi, S.~Abreu, G.~Ciccone, D.~Brunner, and H.~Kleemann, ``Hands-on reservoir computing: a tutorial for practical implementation,'' \textit{Neuromorphic Computing and Engineering}, vol.~2, no.~3, art.~032002, 2022.

\bibitem{snoek2012practical}
J. Snoek, H. Larochelle, and R. P. Adams, ``Practical Bayesian optimization of machine learning algorithms,'' in \textit{Proc. Neural Information Processing Systems (NIPS)}, 2012.

\bibitem{racca2021robust}
A. Racca and L. Magri, ``Robust optimization and validation of echo state networks for learning chaotic dynamics,'' \textit{Neural Networks}, vol.~142, pp.~252--268, 2021. \url{https://doi.org/10.1016/j.neunet.2021.05.004}

\bibitem{griffith2019forecasting}
A. Griffith, A. Pomerance, and D. J. Gauthier, ``Forecasting chaotic systems with very low connectivity reservoir computers,'' \textit{Chaos}, vol.~29, art.~123108, 2019.

\bibitem{cawley2010overfitting}
G. C. Cawley and N. L. C. Talbot, ``On over-fitting in model selection and subsequent selection bias in performance evaluation,'' \textit{Journal of Machine Learning Research}, vol.~11, pp.~2079--2107, 2010.

\bibitem{andrews2017smoldyn}
S. S. Andrews, ``Smoldyn: Particle-based simulation with rule-based modeling, improved molecular interaction and a library interface,'' \textit{Bioinformatics}, vol.~33, no.~5, pp.~710--717, 2017.

\bibitem{bergman1979minimal}
R. N. Bergman, Y. Z. Ider, C. R. Bowden, and C. Cobelli, ``Quantitative estimation of insulin sensitivity,'' \textit{American Journal of Physiology-Endocrinology and Metabolism}, vol.~236, no.~6, pp.~E667--E677, 1979.

\bibitem{hovorka2004nonlinear}
R. Hovorka, V. Canonico, L. J. Chassin, U. Haueter, M. Massi-Benedetti, M. Orsini Federici, T. R. Pieber, H. C. Schaller, L. Schaupp, T. Vering, and M. E. Wilinska, ``Nonlinear model predictive control of glucose concentration in subjects with type 1 diabetes,'' \textit{Physiological Measurement}, vol.~25, no.~4, pp.~905--920, 2004.

\bibitem{atiya2000new}
A. F. Atiya and A. G. Parlos, ``New results on recurrent network training: unifying the algorithms and accelerating convergence,'' \textit{IEEE Transactions on Neural Networks}, vol.~11, no.~3, pp.~697--713, 2000. \url{https://doi.org/10.1109/72.846741}

\bibitem{berg1977physics}
H. C. Berg and E. M. Purcell, ``Physics of chemoreception,'' \textit{Biophysical Journal}, vol.~20, no.~2, pp.~193--219, 1977. \url{https://doi.org/10.1016/S0006-3495(77)85544-6}

\bibitem{bialek2005physical}
W. Bialek and S. Setayeshgar, ``Physical limits to biochemical signaling,'' \textit{Proceedings of the National Academy of Sciences}, vol.~102, no.~29, pp.~10040--10045, 2005. \url{https://doi.org/10.1073/pnas.0504321102}

\bibitem{lauffenburger1993receptors}
D. A. Lauffenburger and J. J. Linderman, \textit{Receptors: Models for Binding, Trafficking, and Signaling}, Oxford University Press, New York, 1993.

\bibitem{li2004cellular}
M. Li and G. L. Hazelbauer, ``Cellular stoichiometry of the components of the chemotaxis signaling complex,'' \textit{Journal of Bacteriology}, vol.~186, no.~12, pp.~3687--3694, 2004. \url{https://doi.org/10.1128/JB.186.12.3687-3694.2004}

\end{thebibliography}
\end{document}


\pagestyle{fancy}
\fancyhf{}
\renewcommand{\headrulewidth}{0pt}
\cfoot{\thepage}

\title{Supporting Information\\Task-Adaptive Physical Reservoir Computing via Tunable Molecular Communication Dynamics}
\maketitle

\author{Saad Yousuf, Kaan Burak Ikiz, and Murat Kuscu*}

\begin{affiliations}
S. Yousuf, K. B. Ikiz, M. Kuscu\\
Nano/Bio/Physical Information and Communications Laboratory (CALICO Lab),\\
Department of Electrical and Electronics Engineering, Ko\c{c} University,\\
Istanbul 34450, Turkey\\
Email: mkuscu@ku.edu.tr
\end{affiliations}

\section*{S1. Simulator, Validation, and Numerical Robustness}

\textbf{Table~\ref{tab:si_settings}} lists the fixed settings of the simulator used for all results reported in the main article. The time grid is aligned with the virtual nodes, the concentration is computed by fast convolution as a discrete superposition, the mean-field occupancy is obtained from the exact one-step solution for a piecewise-constant concentration, and the stochastic receptor model draws the exact one-step transition counts of a two-state receptor.

\begin{table}[H]
\centering
\caption{Fixed settings of the simulator.}
\label{tab:si_settings}
\small
\begin{tabularx}{\textwidth}{@{}p{4.2cm}p{4.2cm}X@{}}
\toprule
\textbf{Setting} & \textbf{Value} & \textbf{Role} \\
\midrule
Minimum release $N_{\min}$ & 100 & lower endpoint of the release encoding \\
Grid step $\Delta t$ & $T/(Mn_{\mathrm{sub}})\le10^{-3}$~s & smallest $n_{\mathrm{sub}}$ with $\Delta t\le1$~ms; every node on a grid point \\
Concentration history & $\max[\operatorname{round}(100\,\tau_D/T),1]$ symbols & retained contribution of each release \\
Initial occupancy & $p_0=0$ and $b(0)=0$ & all receptors free at the start, mean-field and stochastic model \\
Mean-field update & exact exponential step & piecewise-constant concentration over one step \\
Receptor transitions & binomial, exact one-step probabilities & bind from free, unbind from bound, same step \\
Normal approximation & binomial mean $\ge25$ & speed; exact sampler checked in Table~\ref{tab:si_robustness} \\
Virtual nodes & $M=50$ (MG tasks), $M=100$ (sine-to-square) & occupancy at $t_n+iT/M$ \\
Readout memory & $L\in\{1,\ldots,5\}$ & concatenated past states \\
Ridge grid & $10^{-6}$ to $10^{3}$ in decades & chosen by the optimization-block error \\
Intercept & penalized & included in the ridge penalty \\
Error & NRMSE with the sample standard deviation & evaluation block \\
\bottomrule
\end{tabularx}
\end{table}

\textbf{Internal checks:} The concentration superposition computed by fast convolution agrees with the direct per-release sum to within a relative error of $10^{-15}$. The mean of the stochastic receptor model over 200 realizations agrees with the mean-field occupancy at every node within the sampling uncertainty, and the standard deviation across realizations equals the binomial value $\sqrt{p(t)[1-p(t)]/N_R}$ to within one percent.

\textbf{Robustness:} \textbf{Table~\ref{tab:si_robustness}} re-evaluates the mean-field and noise-aware optima of the three primary benchmark tasks under a four times finer grid step, under the complete concentration history instead of the truncated one, and, for the stochastic receptor model, with the exact binomial sampler at every step. Grid refinement changes the mean-field errors by at most 0.0001, and the exact sampler changes the stochastic errors by at most 0.005. The complete history changes the mean-field errors by up to 0.1373 and the stochastic errors by up to 0.058, where the largest changes belong to the MG-Cubed optima, whose target is the most sensitive to the background concentration added by the longer history. Note that the retained history is a property of the channel model adopted from our previous study, and every comparison in the main article is made within this model.

\scriptsize
\setlength{\tabcolsep}{3pt}
\begin{longtable}{@{}llllccccccc@{}}
\caption{Numerical robustness of the optima. Mean-field NRMSE under the default grid, a fourfold finer grid, and the complete concentration history, and stochastic NRMSE (mean $\pm$ standard deviation over 20 realizations, same seed) under the same three variants and with the exact binomial sampler. Mean-field optima are evaluated at $N_R=500$ and noise-aware optima at their design $N_R$, optimization block, fixed readout. MF denotes the mean-field model.}
\label{tab:si_robustness}\\
\toprule
\textbf{Design} & $N_R$ & \textbf{Set} & \textbf{Task} & \textbf{MF default} & \textbf{MF fine} & \textbf{MF full history} & \textbf{Stochastic default} & \textbf{fine} & \textbf{full history} & \textbf{exact sampler} \\
\midrule
\endfirsthead
\toprule
\textbf{Design} & $N_R$ & \textbf{Set} & \textbf{Task} & \textbf{MF default} & \textbf{MF fine} & \textbf{MF full history} & \textbf{Stochastic default} & \textbf{fine} & \textbf{full history} & \textbf{exact sampler} \\
\midrule
\endhead
\bottomrule
\endfoot
mean-field &  & S0 & MG & 0.1215 & 0.1215 & 0.1341 & 1.011 $\pm$ 0.053 & 0.985 & 0.971 & 1.011 \\
mean-field &  & S0 & Sine-to-square & 0.2929 & 0.2929 & 0.2985 & 0.633 $\pm$ 0.018 & 0.616 & 0.627 & 0.633 \\
mean-field &  & S0 & MG-Cubed & 0.3465 & 0.3464 & 0.4263 & 1.199 $\pm$ 0.088 & 1.195 & 1.191 & 1.199 \\
mean-field &  & S1 & MG & 0.0883 & 0.0883 & 0.1109 & 0.855 $\pm$ 0.061 & 0.835 & 0.856 & 0.855 \\
mean-field &  & S1 & Sine-to-square & 0.2559 & 0.2559 & 0.2670 & 0.539 $\pm$ 0.016 & 0.533 & 0.579 & 0.539 \\
mean-field &  & S1 & MG-Cubed & 0.2986 & 0.2986 & 0.4186 & 1.117 $\pm$ 0.066 & 1.103 & 1.129 & 1.117 \\
mean-field &  & S2 & MG & 0.1109 & 0.1109 & 0.1260 & 0.878 $\pm$ 0.074 & 0.914 & 0.906 & 0.878 \\
mean-field &  & S2 & Sine-to-square & 0.2726 & 0.2726 & 0.2870 & 0.537 $\pm$ 0.018 & 0.538 & 0.547 & 0.537 \\
mean-field &  & S2 & MG-Cubed & 0.3388 & 0.3388 & 0.4550 & 1.179 $\pm$ 0.082 & 1.189 & 1.183 & 1.179 \\
mean-field &  & S3 & MG & 0.0880 & 0.0880 & 0.1077 & 0.785 $\pm$ 0.050 & 0.787 & 0.805 & 0.785 \\
mean-field &  & S3 & Sine-to-square & 0.2482 & 0.2482 & 0.2602 & 0.492 $\pm$ 0.019 & 0.488 & 0.535 & 0.492 \\
mean-field &  & S3 & MG-Cubed & 0.2872 & 0.2872 & 0.4245 & 1.026 $\pm$ 0.082 & 1.017 & 1.042 & 1.026 \\
mean-field &  & S4 & MG & 0.0872 & 0.0872 & 0.1086 & 0.771 $\pm$ 0.032 & 0.765 & 0.781 & 0.771 \\
mean-field &  & S4 & Sine-to-square & 0.2312 & 0.2313 & 0.2419 & 0.436 $\pm$ 0.014 & 0.430 & 0.449 & 0.436 \\
mean-field &  & S4 & MG-Cubed & 0.2847 & 0.2848 & 0.4023 & 1.020 $\pm$ 0.059 & 1.013 & 1.019 & 1.020 \\
noise-aware & 500 & S0 & MG & 0.6023 & 0.6023 & 0.5972 & 0.687 $\pm$ 0.008 & 0.686 & 0.682 & 0.687 \\
noise-aware & 500 & S0 & Sine-to-square & 0.4299 & 0.4299 & 0.4367 & 0.529 $\pm$ 0.015 & 0.532 & 0.533 & 0.529 \\
noise-aware & 500 & S0 & MG-Cubed & 0.7905 & 0.7905 & 0.7980 & 0.841 $\pm$ 0.012 & 0.843 & 0.844 & 0.841 \\
noise-aware & 500 & S1 & MG & 0.4937 & 0.4937 & 0.4966 & 0.535 $\pm$ 0.005 & 0.537 & 0.543 & 0.535 \\
noise-aware & 500 & S1 & Sine-to-square & 0.4098 & 0.4097 & 0.4314 & 0.423 $\pm$ 0.006 & 0.423 & 0.451 & 0.423 \\
noise-aware & 500 & S1 & MG-Cubed & 0.6894 & 0.6894 & 0.6919 & 0.738 $\pm$ 0.018 & 0.740 & 0.763 & 0.738 \\
noise-aware & 500 & S2 & MG & 0.5110 & 0.5110 & 0.5134 & 0.594 $\pm$ 0.011 & 0.600 & 0.604 & 0.594 \\
noise-aware & 500 & S2 & Sine-to-square & 0.3770 & 0.3770 & 0.3803 & 0.454 $\pm$ 0.016 & 0.452 & 0.478 & 0.454 \\
noise-aware & 500 & S2 & MG-Cubed & 0.7502 & 0.7502 & 0.7508 & 0.795 $\pm$ 0.008 & 0.795 & 0.804 & 0.795 \\
noise-aware & 500 & S3 & MG & 0.4829 & 0.4829 & 0.4863 & 0.515 $\pm$ 0.008 & 0.515 & 0.521 & 0.515 \\
noise-aware & 500 & S3 & Sine-to-square & 0.3968 & 0.3968 & 0.4096 & 0.402 $\pm$ 0.004 & 0.402 & 0.420 & 0.402 \\
noise-aware & 500 & S3 & MG-Cubed & 0.6656 & 0.6656 & 0.6663 & 0.728 $\pm$ 0.025 & 0.715 & 0.741 & 0.728 \\
noise-aware & 500 & S4 & MG & 0.4056 & 0.4056 & 0.4181 & 0.457 $\pm$ 0.007 & 0.457 & 0.471 & 0.457 \\
noise-aware & 500 & S4 & Sine-to-square & 0.3462 & 0.3462 & 0.3530 & 0.377 $\pm$ 0.005 & 0.377 & 0.382 & 0.377 \\
noise-aware & 500 & S4 & MG-Cubed & 0.6692 & 0.6692 & 0.7001 & 0.692 $\pm$ 0.008 & 0.690 & 0.721 & 0.692 \\
noise-aware & 5000 & S0 & MG & 0.3194 & 0.3194 & 0.3293 & 0.563 $\pm$ 0.023 & 0.574 & 0.580 & 0.563 \\
noise-aware & 5000 & S0 & Sine-to-square & 0.3870 & 0.3870 & 0.3930 & 0.465 $\pm$ 0.011 & 0.462 & 0.462 & 0.465 \\
noise-aware & 5000 & S0 & MG-Cubed & 0.6118 & 0.6118 & 0.6456 & 0.743 $\pm$ 0.021 & 0.741 & 0.757 & 0.743 \\
noise-aware & 5000 & S1 & MG & 0.3422 & 0.3422 & 0.3581 & 0.433 $\pm$ 0.009 & 0.434 & 0.451 & 0.435 \\
noise-aware & 5000 & S1 & Sine-to-square & 0.3645 & 0.3645 & 0.3739 & 0.396 $\pm$ 0.006 & 0.396 & 0.404 & 0.396 \\
noise-aware & 5000 & S1 & MG-Cubed & 0.5600 & 0.5600 & 0.5787 & 0.617 $\pm$ 0.009 & 0.615 & 0.630 & 0.615 \\
noise-aware & 5000 & S2 & MG & 0.3178 & 0.3178 & 0.3115 & 0.491 $\pm$ 0.014 & 0.501 & 0.501 & 0.491 \\
noise-aware & 5000 & S2 & Sine-to-square & 0.3626 & 0.3626 & 0.3668 & 0.395 $\pm$ 0.011 & 0.393 & 0.401 & 0.395 \\
noise-aware & 5000 & S2 & MG-Cubed & 0.5672 & 0.5672 & 0.5779 & 0.659 $\pm$ 0.016 & 0.663 & 0.681 & 0.659 \\
noise-aware & 5000 & S3 & MG & 0.3145 & 0.3145 & 0.3210 & 0.369 $\pm$ 0.007 & 0.368 & 0.382 & 0.369 \\
noise-aware & 5000 & S3 & Sine-to-square & 0.3461 & 0.3461 & 0.3550 & 0.374 $\pm$ 0.006 & 0.373 & 0.387 & 0.377 \\
noise-aware & 5000 & S3 & MG-Cubed & 0.5406 & 0.5406 & 0.5544 & 0.584 $\pm$ 0.010 & 0.587 & 0.599 & 0.584 \\
noise-aware & 5000 & S4 & MG & 0.3333 & 0.3333 & 0.3092 & 0.359 $\pm$ 0.004 & 0.361 & 0.350 & 0.359 \\
noise-aware & 5000 & S4 & Sine-to-square & 0.2858 & 0.2859 & 0.2944 & 0.337 $\pm$ 0.008 & 0.338 & 0.349 & 0.337 \\
noise-aware & 5000 & S4 & MG-Cubed & 0.5455 & 0.5455 & 0.5600 & 0.587 $\pm$ 0.009 & 0.586 & 0.604 & 0.589 \\
noise-aware & 50000 & S0 & MG & 0.2534 & 0.2534 & 0.2781 & 0.392 $\pm$ 0.015 & 0.398 & 0.407 & 0.389 \\
noise-aware & 50000 & S0 & Sine-to-square & 0.3685 & 0.3685 & 0.3789 & 0.409 $\pm$ 0.007 & 0.409 & 0.415 & 0.409 \\
noise-aware & 50000 & S0 & MG-Cubed & 0.4696 & 0.4696 & 0.5206 & 0.568 $\pm$ 0.019 & 0.573 & 0.623 & 0.573 \\
noise-aware & 50000 & S1 & MG & 0.2698 & 0.2698 & 0.2680 & 0.308 $\pm$ 0.006 & 0.306 & 0.314 & 0.306 \\
noise-aware & 50000 & S1 & Sine-to-square & 0.3344 & 0.3344 & 0.3494 & 0.361 $\pm$ 0.006 & 0.360 & 0.373 & 0.360 \\
noise-aware & 50000 & S1 & MG-Cubed & 0.4977 & 0.4977 & 0.5338 & 0.532 $\pm$ 0.009 & 0.533 & 0.556 & 0.529 \\
noise-aware & 50000 & S2 & MG & 0.2385 & 0.2385 & 0.2543 & 0.357 $\pm$ 0.010 & 0.352 & 0.354 & 0.356 \\
noise-aware & 50000 & S2 & Sine-to-square & 0.3610 & 0.3610 & 0.3663 & 0.370 $\pm$ 0.004 & 0.367 & 0.374 & 0.368 \\
noise-aware & 50000 & S2 & MG-Cubed & 0.4926 & 0.4926 & 0.5329 & 0.545 $\pm$ 0.012 & 0.544 & 0.571 & 0.545 \\
noise-aware & 50000 & S3 & MG & 0.2596 & 0.2596 & 0.2595 & 0.283 $\pm$ 0.004 & 0.282 & 0.286 & 0.281 \\
noise-aware & 50000 & S3 & Sine-to-square & 0.3150 & 0.3150 & 0.3352 & 0.340 $\pm$ 0.005 & 0.342 & 0.354 & 0.341 \\
noise-aware & 50000 & S3 & MG-Cubed & 0.4795 & 0.4795 & 0.5262 & 0.521 $\pm$ 0.009 & 0.519 & 0.556 & 0.516 \\
noise-aware & 50000 & S4 & MG & 0.2584 & 0.2584 & 0.2566 & 0.282 $\pm$ 0.006 & 0.281 & 0.282 & 0.283 \\
noise-aware & 50000 & S4 & Sine-to-square & 0.2861 & 0.2862 & 0.2952 & 0.301 $\pm$ 0.004 & 0.301 & 0.311 & 0.301 \\
noise-aware & 50000 & S4 & MG-Cubed & 0.5036 & 0.5036 & 0.5266 & 0.522 $\pm$ 0.007 & 0.519 & 0.541 & 0.522 \\
noise-aware & 500000 & S0 & MG & 0.1930 & 0.1930 & 0.2085 & 0.300 $\pm$ 0.009 & 0.298 & 0.317 & 0.300 \\
noise-aware & 500000 & S0 & Sine-to-square & 0.3466 & 0.3466 & 0.3605 & 0.379 $\pm$ 0.005 & 0.378 & 0.387 & 0.376 \\
noise-aware & 500000 & S0 & MG-Cubed & 0.4696 & 0.4696 & 0.5206 & 0.495 $\pm$ 0.008 & 0.494 & 0.545 & 0.491 \\
noise-aware & 500000 & S1 & MG & 0.2404 & 0.2404 & 0.2378 & 0.257 $\pm$ 0.003 & 0.257 & 0.256 & 0.256 \\
noise-aware & 500000 & S1 & Sine-to-square & 0.3104 & 0.3104 & 0.3306 & 0.338 $\pm$ 0.005 & 0.335 & 0.348 & 0.336 \\
noise-aware & 500000 & S1 & MG-Cubed & 0.4498 & 0.4498 & 0.4929 & 0.465 $\pm$ 0.009 & 0.467 & 0.509 & 0.468 \\
noise-aware & 500000 & S2 & MG & 0.1751 & 0.1751 & 0.2131 & 0.273 $\pm$ 0.005 & 0.273 & 0.287 & 0.273 \\
noise-aware & 500000 & S2 & Sine-to-square & 0.3144 & 0.3144 & 0.3252 & 0.354 $\pm$ 0.006 & 0.353 & 0.358 & 0.354 \\
noise-aware & 500000 & S2 & MG-Cubed & 0.4566 & 0.4566 & 0.5121 & 0.483 $\pm$ 0.008 & 0.483 & 0.541 & 0.483 \\
noise-aware & 500000 & S3 & MG & 0.2011 & 0.2011 & 0.2195 & 0.224 $\pm$ 0.004 & 0.224 & 0.240 & 0.224 \\
noise-aware & 500000 & S3 & Sine-to-square & 0.2911 & 0.2911 & 0.3086 & 0.311 $\pm$ 0.004 & 0.310 & 0.333 & 0.311 \\
noise-aware & 500000 & S3 & MG-Cubed & 0.4355 & 0.4355 & 0.4889 & 0.455 $\pm$ 0.006 & 0.456 & 0.511 & 0.454 \\
noise-aware & 500000 & S4 & MG & 0.2062 & 0.2062 & 0.2172 & 0.226 $\pm$ 0.004 & 0.226 & 0.234 & 0.225 \\
noise-aware & 500000 & S4 & Sine-to-square & 0.2724 & 0.2724 & 0.2762 & 0.281 $\pm$ 0.003 & 0.281 & 0.286 & 0.281 \\
noise-aware & 500000 & S4 & MG-Cubed & 0.4398 & 0.4398 & 0.4895 & 0.456 $\pm$ 0.006 & 0.457 & 0.508 & 0.456 \\
\end{longtable}
\normalsize
\setlength{\tabcolsep}{6pt}

\FloatBarrier
\section*{S2. Benchmark Tasks and Data Blocks}

\begin{table}[H]
\centering
\caption{Benchmark tasks and data blocks. Indices are one-based, inclusive, and refer to symbols. Inputs are scaled to $[0,1]$.}
\label{tab:si_tasks}
\small
\begin{tabularx}{\textwidth}{@{}p{2.4cm}XXX@{}}
\toprule
 & \textbf{MG forecasting} & \textbf{Sine-to-square} & \textbf{MG-Cubed} \\
\midrule
Input & normalized Mackey-Glass series ($\beta=0.2$, $\gamma=0.1$, $q=10$, $\tau=17$, unit step) & $0.5+0.5\sin[2\pi(n-1)/200]$ & normalized Mackey-Glass series \\
Target & input six samples ahead & 1 where the input exceeds 0.5, 0 otherwise & cube of the normalized input ten samples ahead \\
Virtual nodes $M$ & 50 & 100 & 50 \\
Symbols & 1700 & 4000 & 1700 \\
Washout & 1 to 500 & 1 to 500 & 1 to 500 \\
Training block & 501 to 1000 & 501 to 2500 & 501 to 1000 \\
Held-out block & 1001 to 1200 & 2501 to 3000 & 1001 to 1200 \\
Optimization block & 1201 to 1700 & 3001 to 4000 & 1201 to 1700 \\
Linear error floor & none & 0.435 for any linear map of the input & none \\
\bottomrule
\end{tabularx}
\end{table}

The input-tap baseline used in the main article regresses the target on the current input and the preceding inputs with the same blocks and the same ridge rule (\textbf{Table~\ref{tab:si_taps}}). For sine-to-square, a single sinusoid carries a fraction $8/\pi^2$ of the power of the square wave; therefore, any linear map of the input has an NRMSE of at least $\sqrt{1-8/\pi^2}=0.435$.

\begin{table}[H]
\centering
\caption{Input-tap baseline, NRMSE on the optimization block for 1, 2, 3, 5, 10, 20, and 50 taps.}
\label{tab:si_taps}
\small
\begin{tabular}{@{}lccccccc@{}}
\toprule
\textbf{Task} & 1 & 2 & 3 & 5 & 10 & 20 & 50 \\
\midrule
MG & 0.750 & 0.370 & 0.317 & 0.187 & 0.163 & 0.154 & 0.147 \\
Sine-to-square & 0.435 & 0.435 & 0.435 & 0.435 & 0.435 & 0.435 & 0.435 \\
MG-Cubed & 0.964 & 0.716 & 0.711 & 0.598 & 0.555 & 0.552 & 0.526 \\
\bottomrule
\end{tabular}
\end{table}

\FloatBarrier
\section*{S3. Mean-Field Searches and Error Surfaces}

\textbf{Table~\ref{tab:si_domain}} gives the search domain of the Bayesian optimization.

\begin{table}[H]
\centering
\caption{Search domain of the Bayesian optimization (Experimental Section of the main article). Logarithmic transforms are applied to the parameters spanning several orders of magnitude.}
\label{tab:si_domain}
\small
\begin{tabularx}{\textwidth}{@{}p{2.0cm}p{4.6cm}p{1.6cm}X@{}}
\toprule
\textbf{Parameter} & \textbf{Range} & \textbf{Scale} & \textbf{Role} \\
\midrule
$k_{\mathrm{on}}$ & $5\times10^{-20}$ to $2\times10^{-17}$ m$^3$ s$^{-1}$ & log & receptor association rate \\
$k_{\mathrm{off}}$ & 0.1 to 10 s$^{-1}$ & log & receptor dissociation rate \\
$T$ & 0.5 to 2 s & linear & symbol interval \\
$d$ & 6 to 50 $\mu$m & log & transmitter-receiver distance \\
$L$ & integers 1 to 5 & integer & readout memory \\
$N_{\max}$ & 200 to 20000 & linear & release amplitude \\
$D$ & $0.5\times10^{-11}$ to $2\times10^{-10}$ m$^2$ s$^{-1}$ & log & diffusion coefficient \\
$\lambda$ & $10^{-6}$ to $10^{3}$, decades & grid & ridge parameter, chosen inside every evaluation \\
$M'$ & 50, 25, 10, 5 (100, 50, 20, 10 for sine-to-square) & categorical & nodes per symbol, joint search only \\
$\tau_f/T$ & 0.05 to 5 & log & low-pass time constant, joint search only \\
$L$, joint search & integers 1 to 20 & integer & readout memory in the joint search \\
$T$, joint S3 search & 0.5 to 5 s & linear & symbol interval in the joint S3 search (default searches 0.5 to 2 s) \\
$M'L$, joint search & uncapped & & the readout grid caps the feature count at 1000, the joint search does not \\
\bottomrule
\end{tabularx}
\end{table}

\begin{figure}[H]
\centering
\includegraphics[width=\textwidth]{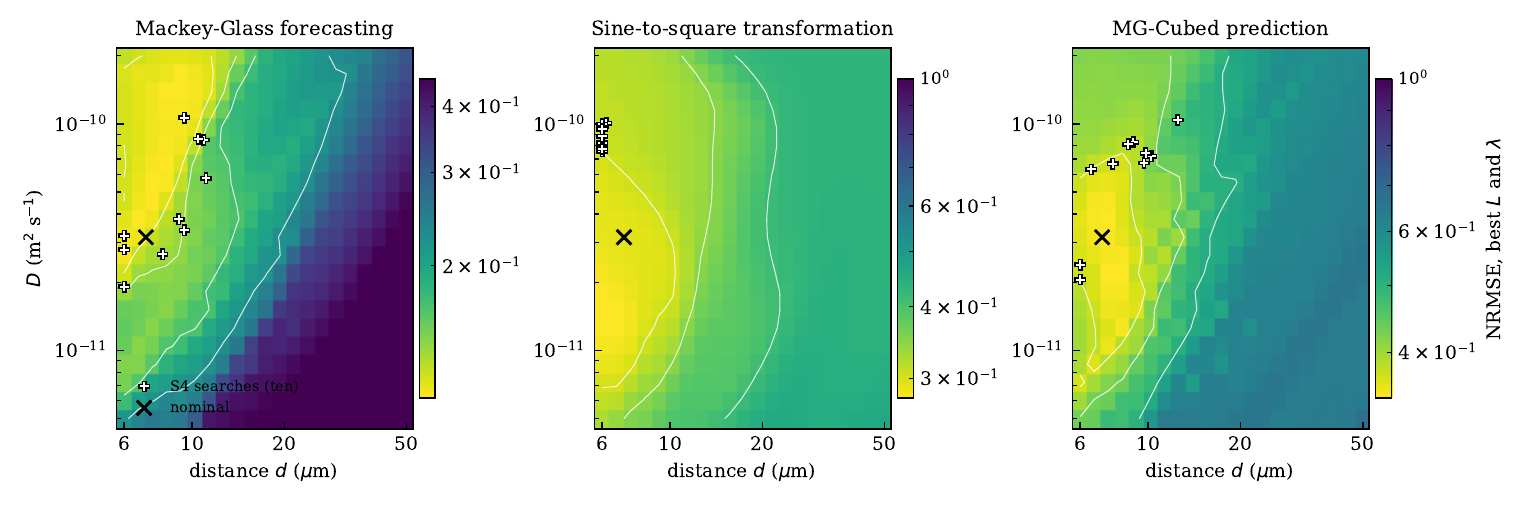}
\caption{Channel error surface. NRMSE over the transmitter-receiver distance and the diffusion coefficient at the nominal transmitter and receiver settings, with the best $L$ and ridge value at every grid point, mean field. Crosses mark the projections of the ten S4 configurations onto this plane, with the other parameters at their nominal values; note that these projections are not optima of this two-dimensional slice. The black cross marks the nominal channel.}
\label{fig:si_channel}
\end{figure}

\textbf{Table~\ref{tab:si_sets}} gives the errors under the tunable-parameter sets on both evaluation blocks (Figure 4 of the main article), and \textbf{Table~\ref{tab:si_s0}} enumerates the nominal configuration over $L$. \textbf{Table~\ref{tab:si_det_all}} lists the best configuration of every mean-field search, i.e., 120 searches in total. \textbf{Tables~\ref{tab:si_land_rec}} to \ref{tab:si_land_ch} summarize the three error surfaces, and \textbf{Figure~\ref{fig:si_channel}} shows the channel error surface.

\begin{table}[H]
\centering
\caption{NRMSE under the tunable-parameter sets on the optimization block and on the held-out block, which was not used by any search (mean-field model). For S1 to S4, the mean and sample standard deviation over ten independent searches of 100 evaluations are given, and standard deviations below 0.0005 are omitted. S0 is the nominal channel with the best readout.}
\label{tab:si_sets}
\small
\setlength{\tabcolsep}{5pt}
\begin{tabular}{@{}lccccc@{}}
\toprule
\textbf{Task, block} & \textbf{S0} & \textbf{S1} & \textbf{S2} & \textbf{S3} & \textbf{S4} \\
\midrule
MG, optimization & 0.122 & 0.092 $\pm$ 0.004 & 0.111 & 0.089 $\pm$ 0.001 & 0.090 $\pm$ 0.003 \\
MG, held-out & 0.129 & 0.100 $\pm$ 0.004 & 0.118 & 0.096 $\pm$ 0.001 & 0.097 $\pm$ 0.003 \\
Sine-to-square, optimization & 0.293 & 0.257 & 0.273 & 0.249 & 0.233 $\pm$ 0.002 \\
Sine-to-square, held-out & 0.297 & 0.261 $\pm$ 0.001 & 0.276 & 0.253 & 0.238 $\pm$ 0.002 \\
MG-Cubed, optimization & 0.346 & 0.300 $\pm$ 0.001 & 0.339 & 0.289 $\pm$ 0.001 & 0.290 $\pm$ 0.007 \\
MG-Cubed, held-out & 0.396 & 0.349 $\pm$ 0.001 & 0.387 $\pm$ 0.001 & 0.339 $\pm$ 0.002 & 0.339 $\pm$ 0.003 \\
\bottomrule
\end{tabular}
\end{table}

\begin{table}[H]
\centering
\caption{S0, nominal configuration, NRMSE on the optimization block / held-out block for $L=1$ to 5, mean field.}
\label{tab:si_s0}
\small
\begin{tabular}{@{}lccccc@{}}
\toprule
\textbf{Task} & $L=1$ & $L=2$ & $L=3$ & $L=4$ & $L=5$ \\
\midrule
MG & 0.3619 / 0.3758 & 0.2153 / 0.2216 & 0.1894 / 0.1999 & 0.1311 / 0.1398 & 0.1215 / 0.1295 \\
Sine-to-square & 0.3202 / 0.3306 & 0.2957 / 0.3004 & 0.2941 / 0.2985 & 0.2931 / 0.2974 & 0.2929 / 0.2974 \\
MG-Cubed & 0.6362 / 0.6952 & 0.5123 / 0.5613 & 0.4712 / 0.5286 & 0.3689 / 0.4183 & 0.3465 / 0.3957 \\
\bottomrule
\end{tabular}
\end{table}

\scriptsize
\setlength{\tabcolsep}{2.5pt}
\begin{longtable}{@{}lllcccccccccccc@{}}
\caption{Best configuration of every mean-field search. $d$ in $\mu$m, NRMSE on the optimization block (opt) and on the held-out block (held). Units are m$^3$~s$^{-1}$ for $k_{\mathrm{on}}$, s$^{-1}$ for $k_{\mathrm{off}}$, s for $T$, $\mu$m for $d$, and m$^2$~s$^{-1}$ for $D$.}
\label{tab:si_det_all}\\
\toprule
\textbf{Set} & \textbf{Task} & \textbf{Seed} & $k_{\mathrm{on}}$ & $k_{\mathrm{off}}$ & $T$ & $d$ & $D$ & $N_{\max}$ & $L$ & \textbf{opt} & \textbf{held} & $k_{\mathrm{off}}T$ & $c_{\mathrm{peak}}/K_D$ & $\tau_D/T$ \\
\midrule
\endfirsthead
\toprule
\textbf{Set} & \textbf{Task} & \textbf{Seed} & $k_{\mathrm{on}}$ & $k_{\mathrm{off}}$ & $T$ & $d$ & $D$ & $N_{\max}$ & $L$ & \textbf{opt} & \textbf{held} & $k_{\mathrm{off}}T$ & $c_{\mathrm{peak}}/K_D$ & $\tau_D/T$ \\
\midrule
\endhead
\bottomrule
\endfoot
S1 & MG & 1 & $2.00\times10^{-17}$ & 10.00 & 1.25 & 7.1 & $3.16\times10^{-11}$ & 10100 & 5 & 0.0994 & 0.1062 & 12.5 & 6.9 & 0.21 \\
S1 & MG & 2 & $8.23\times10^{-18}$ & 8.45 & 1.25 & 7.1 & $3.16\times10^{-11}$ & 10100 & 5 & 0.0883 & 0.0958 & 10.6 & 3.4 & 0.21 \\
S1 & MG & 3 & $2.00\times10^{-17}$ & 10.00 & 1.25 & 7.1 & $3.16\times10^{-11}$ & 10100 & 5 & 0.0994 & 0.1062 & 12.5 & 6.9 & 0.21 \\
S1 & MG & 4 & $1.09\times10^{-17}$ & 9.66 & 1.25 & 7.1 & $3.16\times10^{-11}$ & 10100 & 5 & 0.0890 & 0.0968 & 12.1 & 3.9 & 0.21 \\
S1 & MG & 5 & $7.35\times10^{-18}$ & 6.88 & 1.25 & 7.1 & $3.16\times10^{-11}$ & 10100 & 5 & 0.0898 & 0.0976 & 8.6 & 3.7 & 0.21 \\
S1 & MG & 6 & $9.07\times10^{-18}$ & 8.75 & 1.25 & 7.1 & $3.16\times10^{-11}$ & 10100 & 5 & 0.0885 & 0.0961 & 10.9 & 3.6 & 0.21 \\
S1 & MG & 7 & $1.19\times10^{-17}$ & 9.38 & 1.25 & 7.1 & $3.16\times10^{-11}$ & 10100 & 5 & 0.0903 & 0.0983 & 11.7 & 4.4 & 0.21 \\
S1 & MG & 8 & $5.78\times10^{-18}$ & 4.79 & 1.25 & 7.1 & $3.16\times10^{-11}$ & 10100 & 5 & 0.0930 & 0.1012 & 6.0 & 4.2 & 0.21 \\
\end{longtable}
\normalsize
\setlength{\tabcolsep}{6pt}

\begin{table}[H]
\centering
\caption{Receiver error surface (Figure 2a of the main article). Location of the grid minimum, its readout memory, the minimum and maximum NRMSE on the grid, and the percentage of grid points within 10 percent of the minimum.}
\label{tab:si_land_rec}
\small
\begin{tabular}{@{}lcccccc@{}}
\toprule
\textbf{Task} & $k_{\mathrm{on}}$ (m$^3$ s$^{-1}$) & $k_{\mathrm{off}}$ (s$^{-1}$) & $L$ & \textbf{min} & \textbf{max} & \textbf{within 10\%} \\
\midrule
MG & $9.46\times10^{-18}$ & 10.00 & 5 & 0.088 & 0.354 & 4 \\
Sine-to-square & $1.21\times10^{-17}$ & 2.15 & 5 & 0.256 & 0.435 & 29 \\
MG-Cubed & $5.74\times10^{-18}$ & 3.16 & 5 & 0.299 & 0.582 & 15 \\
\bottomrule
\end{tabular}
\end{table}

\begin{table}[H]
\centering
\caption{Transmitter error surface (Figure 2b of the main article), same columns.}
\label{tab:si_land_tra}
\small
\begin{tabular}{@{}lcccccc@{}}
\toprule
\textbf{Task} & $T$ (s) & $N_{\max}$ & $L$ & \textbf{min} & \textbf{max} & \textbf{within 10\%} \\
\midrule
MG & 1.81 & 7625 & 5 & 0.111 & 0.272 & 41 \\
Sine-to-square & 2.00 & 20000 & 5 & 0.273 & 0.435 & 62 \\
MG-Cubed & 1.75 & 8450 & 5 & 0.339 & 0.599 & 51 \\
\bottomrule
\end{tabular}
\end{table}

\begin{table}[H]
\centering
\caption{Channel error surface (Figure~\ref{fig:si_channel}), same columns.}
\label{tab:si_land_ch}
\small
\begin{tabular}{@{}lcccccc@{}}
\toprule
\textbf{Task} & $d$ ($\mu$m) & $D$ (m$^2$ s$^{-1}$) & $L$ & \textbf{min} & \textbf{max} & \textbf{within 10\%} \\
\midrule
MG & 7.4 & $4.57\times10^{-11}$ & 5 & 0.113 & 0.659 & 12 \\
Sine-to-square & 6.0 & $1.51\times10^{-11}$ & 5 & 0.277 & 0.464 & 12 \\
MG-Cubed & 7.4 & $2.19\times10^{-11}$ & 5 & 0.343 & 0.681 & 10 \\
\bottomrule
\end{tabular}
\end{table}

\FloatBarrier
\section*{S4. Cross-Task Reuse, Mean Field}

\begin{figure}[H]
\centering
\includegraphics[width=\textwidth]{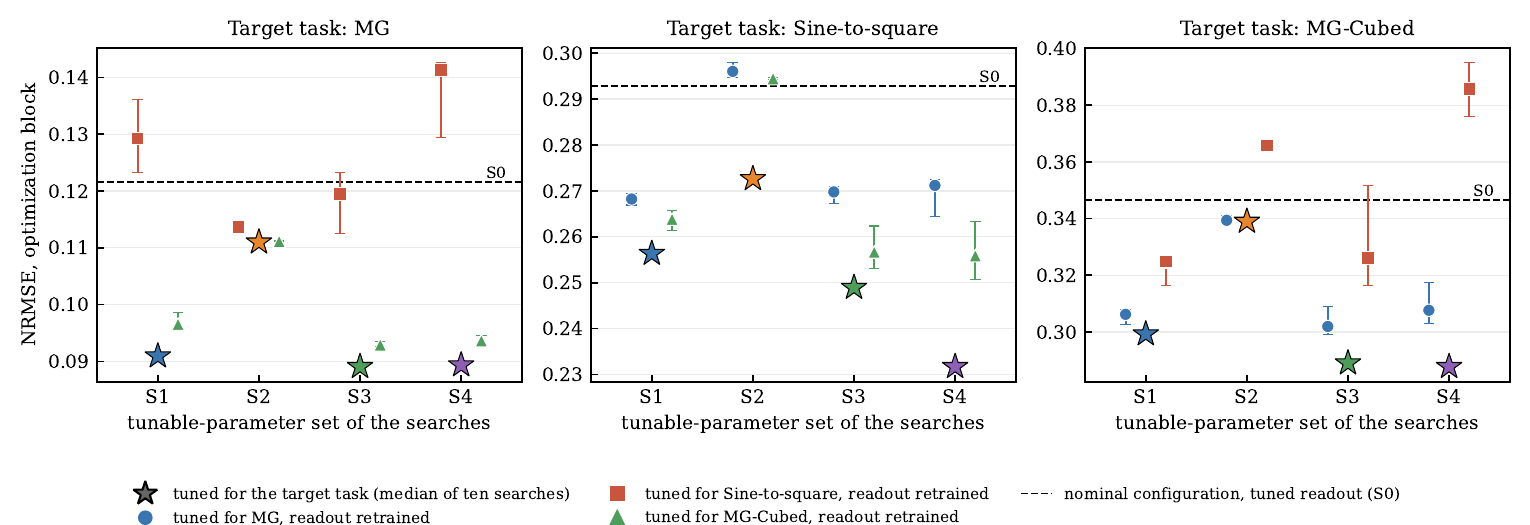}
\caption{Cost of reusing a configuration tuned for another task, mean-field model, optimization block. Panels correspond to the target tasks, and the horizontal axis indicates the tunable-parameter set. Stars indicate the median NRMSE of the ten searches tuned for the target. Colored markers indicate the median NRMSE when the best configuration of each of the ten searches tuned for another task is retained and only the readout is retrained on the target, with $L$ and the ridge value re-selected. Bars span the interquartile range. The dashed line is the nominal channel with a tuned readout (S0).}
\label{fig:si_reuse}
\end{figure}

\begin{figure}[H]
\centering
\includegraphics[width=\textwidth]{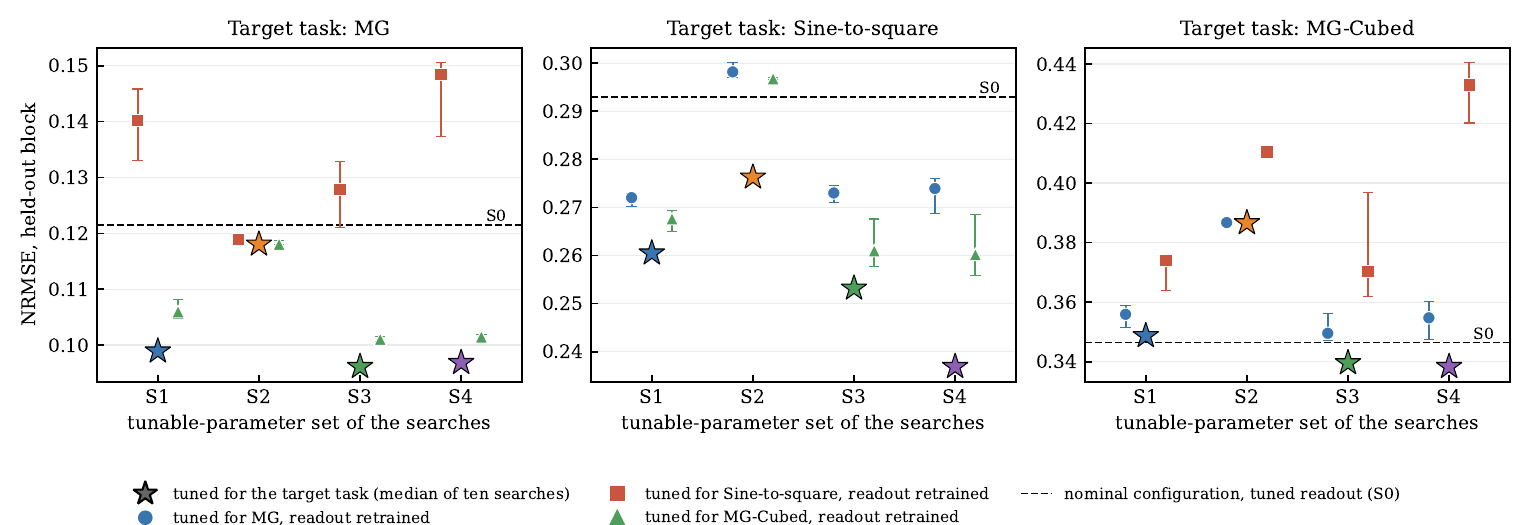}
\caption{Cost of reusing a configuration tuned for another task, held-out block. Same construction as Figure~\ref{fig:si_reuse}.}
\label{fig:si_reuse_held}
\end{figure}

\textbf{Tables~\ref{tab:si_reuse_opt}} and \ref{tab:si_reuse_held} give the reuse results reported in Section 3.3 of the main article on the optimization block and on the held-out block, respectively. \textbf{Figures~\ref{fig:si_reuse}} and \textbf{\ref{fig:si_reuse_held}} show the cost of reuse on the same two blocks.

\begin{table}[H]
\centering
\caption{Reuse of a configuration tuned for another task, under each tunable-parameter set, optimization block. Median NRMSE over the ten reused configurations with the interquartile range in brackets, median of the ten searches tuned for the target (reference), relative penalty in percent, and whether the reused median exceeds the nominal channel with a tuned readout (S0).}
\label{tab:si_reuse_opt}
\scriptsize
\begin{tabular}{@{}lllcccc@{}}
\toprule
\textbf{Set} & \textbf{Tuned for} & \textbf{Target} & \textbf{Reused median [IQR]} & \textbf{Reference} & \textbf{Penalty (\%)} & \textbf{Worse than S0} \\
\midrule
S1 & Sine-to-square & MG & $0.129\,[0.123,\,0.136]$ & 0.091 & 42.2 & yes \\
S1 & MG-Cubed & MG & $0.097\,[0.096,\,0.099]$ & 0.091 & 6.2 &  \\
S1 & MG & Sine-to-square & $0.268\,[0.267,\,0.269]$ & 0.256 & 4.6 &  \\
S1 & MG-Cubed & Sine-to-square & $0.264\,[0.261,\,0.266]$ & 0.256 & 2.9 &  \\
S1 & MG & MG-Cubed & $0.306\,[0.303,\,0.308]$ & 0.299 & 2.3 &  \\
S1 & Sine-to-square & MG-Cubed & $0.325\,[0.316,\,0.327]$ & 0.299 & 8.5 &  \\
S2 & Sine-to-square & MG & $0.114\,[0.114,\,0.114]$ & 0.111 & 2.4 &  \\
S2 & MG-Cubed & MG & $0.111\,[0.111,\,0.111]$ & 0.111 & 0.1 &  \\
S2 & MG & Sine-to-square & $0.296\,[0.295,\,0.298]$ & 0.273 & 8.6 & yes \\
S2 & MG-Cubed & Sine-to-square & $0.294\,[0.294,\,0.295]$ & 0.273 & 8.0 & yes \\
S2 & MG & MG-Cubed & $0.339\,[0.339,\,0.341]$ & 0.339 & 0.1 &  \\
S2 & Sine-to-square & MG-Cubed & $0.366\,[0.366,\,0.366]$ & 0.339 & 7.9 & yes \\
S3 & Sine-to-square & MG & $0.120\,[0.113,\,0.123]$ & 0.089 & 34.2 &  \\
S3 & MG-Cubed & MG & $0.093\,[0.093,\,0.094]$ & 0.089 & 4.3 &  \\
S3 & MG & Sine-to-square & $0.270\,[0.267,\,0.271]$ & 0.249 & 8.4 &  \\
S3 & MG-Cubed & Sine-to-square & $0.257\,[0.253,\,0.262]$ & 0.249 & 3.1 &  \\
S3 & MG & MG-Cubed & $0.302\,[0.299,\,0.309]$ & 0.289 & 4.5 &  \\
S3 & Sine-to-square & MG-Cubed & $0.326\,[0.316,\,0.352]$ & 0.289 & 12.9 &  \\
S4 & Sine-to-square & MG & $0.141\,[0.129,\,0.143]$ & 0.089 & 58.2 & yes \\
S4 & MG-Cubed & MG & $0.094\,[0.093,\,0.095]$ & 0.089 & 4.8 &  \\
S4 & MG & Sine-to-square & $0.271\,[0.264,\,0.272]$ & 0.232 & 17.0 &  \\
S4 & MG-Cubed & Sine-to-square & $0.256\,[0.251,\,0.263]$ & 0.232 & 10.4 &  \\
S4 & MG & MG-Cubed & $0.308\,[0.303,\,0.318]$ & 0.288 & 6.9 &  \\
S4 & Sine-to-square & MG-Cubed & $0.386\,[0.376,\,0.395]$ & 0.288 & 34.0 & yes \\
\bottomrule
\end{tabular}
\end{table}

\begin{table}[H]
\centering
\caption{Same as Table~\ref{tab:si_reuse_opt} on the held-out block.}
\label{tab:si_reuse_held}
\scriptsize
\begin{tabular}{@{}lllcccc@{}}
\toprule
\textbf{Set} & \textbf{Tuned for} & \textbf{Target} & \textbf{Reused median [IQR]} & \textbf{Reference} & \textbf{Penalty (\%)} & \textbf{Worse than S0} \\
\midrule
S1 & Sine-to-square & MG & $0.140\,[0.133,\,0.146]$ & 0.099 & 41.7 & yes \\
S1 & MG-Cubed & MG & $0.106\,[0.105,\,0.108]$ & 0.099 & 7.2 &  \\
S1 & MG & Sine-to-square & $0.272\,[0.270,\,0.273]$ & 0.260 & 4.4 &  \\
S1 & MG-Cubed & Sine-to-square & $0.268\,[0.265,\,0.269]$ & 0.260 & 2.8 &  \\
S1 & MG & MG-Cubed & $0.356\,[0.352,\,0.359]$ & 0.349 & 2.1 &  \\
S1 & Sine-to-square & MG-Cubed & $0.374\,[0.364,\,0.375]$ & 0.349 & 7.2 &  \\
S2 & Sine-to-square & MG & $0.119\,[0.119,\,0.119]$ & 0.118 & 0.7 &  \\
S2 & MG-Cubed & MG & $0.118\,[0.118,\,0.119]$ & 0.118 & 0.0 &  \\
S2 & MG & Sine-to-square & $0.298\,[0.297,\,0.300]$ & 0.276 & 7.9 & yes \\
S2 & MG-Cubed & Sine-to-square & $0.297\,[0.297,\,0.297]$ & 0.276 & 7.4 &  \\
S2 & MG & MG-Cubed & $0.387\,[0.386,\,0.387]$ & 0.387 & 0.0 &  \\
S2 & Sine-to-square & MG-Cubed & $0.410\,[0.410,\,0.410]$ & 0.387 & 6.2 & yes \\
S3 & Sine-to-square & MG & $0.128\,[0.121,\,0.133]$ & 0.096 & 33.0 &  \\
S3 & MG-Cubed & MG & $0.101\,[0.101,\,0.102]$ & 0.096 & 5.1 &  \\
S3 & MG & Sine-to-square & $0.273\,[0.271,\,0.275]$ & 0.253 & 7.8 &  \\
S3 & MG-Cubed & Sine-to-square & $0.261\,[0.258,\,0.268]$ & 0.253 & 3.1 &  \\
S3 & MG & MG-Cubed & $0.350\,[0.347,\,0.356]$ & 0.340 & 2.9 &  \\
S3 & Sine-to-square & MG-Cubed & $0.370\,[0.362,\,0.397]$ & 0.340 & 9.0 &  \\
S4 & Sine-to-square & MG & $0.148\,[0.137,\,0.151]$ & 0.097 & 53.3 & yes \\
S4 & MG-Cubed & MG & $0.101\,[0.100,\,0.102]$ & 0.097 & 4.8 &  \\
S4 & MG & Sine-to-square & $0.274\,[0.269,\,0.276]$ & 0.237 & 15.6 &  \\
S4 & MG-Cubed & Sine-to-square & $0.260\,[0.256,\,0.269]$ & 0.237 & 9.9 &  \\
S4 & MG & MG-Cubed & $0.355\,[0.348,\,0.360]$ & 0.338 & 4.8 &  \\
S4 & Sine-to-square & MG-Cubed & $0.433\,[0.420,\,0.440]$ & 0.338 & 28.0 & yes \\
\bottomrule
\end{tabular}
\end{table}

\FloatBarrier
\section*{S5. Particle-Based Simulations}

\begin{figure}[H]
\centering
\includegraphics[width=0.9\textwidth]{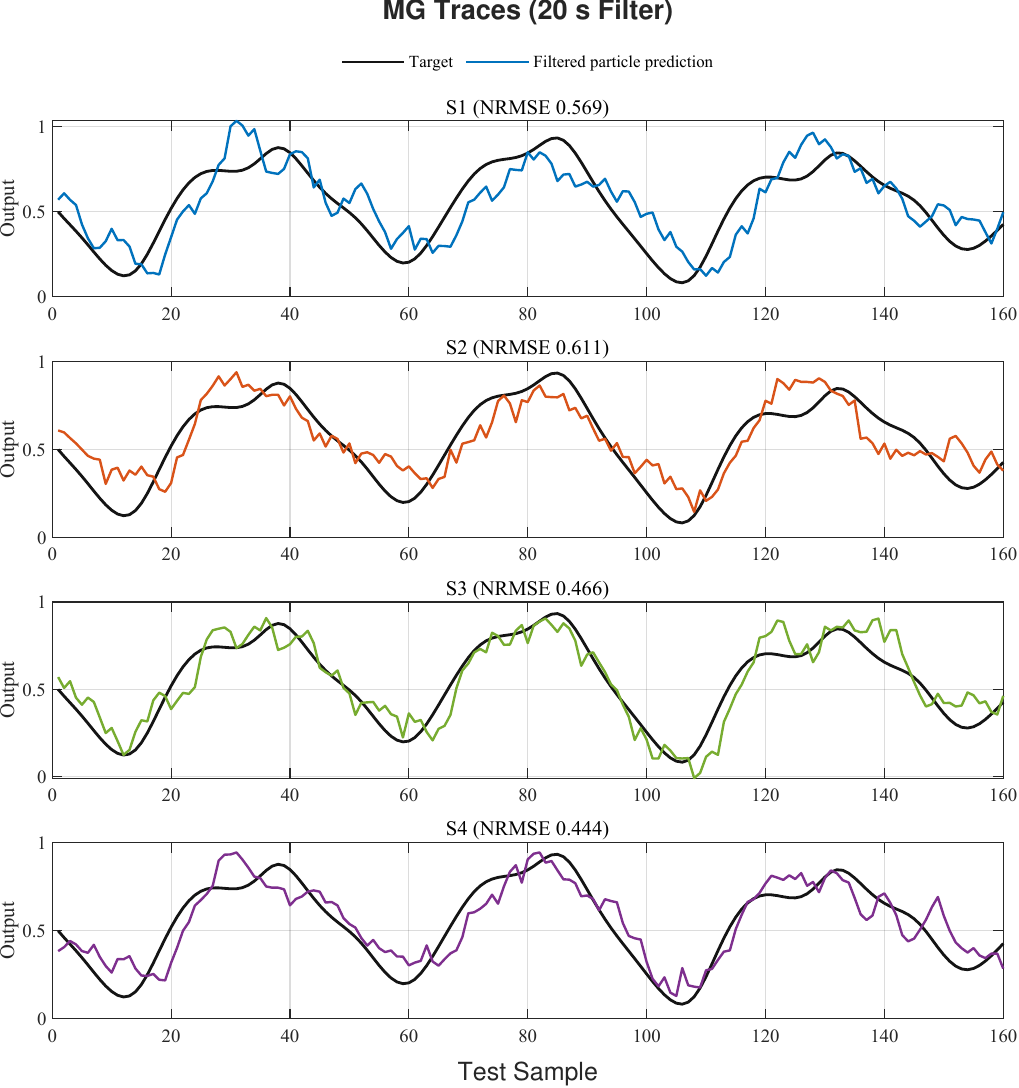}
\caption{Representative MG particle predictions for the four MG validation configurations of S1 to S4 (20~s trailing average, $\lambda=10^{-6}$). Each panel shows the same first 160 symbols of the optimization block, and the displayed realization is the one with the median filtered NRMSE among the five. The horizontal axis counts samples within the optimization block.}
\label{fig:si_mg_traces}
\end{figure}

The particle-based simulations that validate the stochastic receptor model were run for twelve configurations selected in a preliminary set of mean-field searches, whose domain admitted transmitter-receiver distances below the receiver radius. Note that these configurations are not the optima reported in the main article, and they serve only for the validation presented in Section 3.4 of the main article. \textbf{Table~\ref{tab:si_particle_runs}} lists the sixty per-realization values, and \textbf{Figures~\ref{fig:si_mg_traces}} to \ref{fig:si_mgc_traces} show representative particle predictions after the trailing moving-average filter described in the Experimental Section of the main article. \textbf{Table~\ref{tab:si_particle}} summarizes the comparison for each configuration.

\begin{table}[H]
\centering
\caption{Validation of the stochastic receptor model against the particle-based simulations (Figure 5a of the main article), optimization block, $N_R=500$. Particle entries give the mean and sample standard deviation over five realizations, and stochastic entries over 50 realizations, before (raw) and after the trailing moving-average filter, both with $\lambda=10^{-6}$ as in the processing of the particle-simulation outputs. The last column reads out the same realizations with the standard readout of the main article ($L$ and $\lambda$ re-selected, no filter). The mean-field column gives the mean-field error of the same configuration.}
\label{tab:si_particle}
\scriptsize
\setlength{\tabcolsep}{3pt}
\begin{tabular}{@{}llcccccc@{}}
\toprule
\textbf{Task} & \textbf{Set} & \textbf{Mean field} & \textbf{Particle, raw} & \textbf{Stochastic, raw} & \textbf{Particle, filtered} & \textbf{Stochastic, filtered} & \textbf{Stochastic, retrained} \\
\midrule
MG & S1 & 0.087 & 0.859 $\pm$ 0.064 & 0.829 $\pm$ 0.041 & 0.580 $\pm$ 0.026 & 0.551 $\pm$ 0.020 & 0.603 \\
MG-Cubed & S1 & 0.298 & 1.169 $\pm$ 0.093 & 1.110 $\pm$ 0.069 & 0.761 $\pm$ 0.021 & 0.776 $\pm$ 0.027 & 0.800 \\
Sine-to-square & S1 & 0.256 & 0.611 $\pm$ 0.030 & 0.543 $\pm$ 0.017 & 0.506 $\pm$ 0.010 & 0.483 $\pm$ 0.010 & 0.469 \\
MG & S2 & 0.111 & 0.917 $\pm$ 0.046 & 0.910 $\pm$ 0.052 & 0.614 $\pm$ 0.020 & 0.616 $\pm$ 0.016 & 0.638 \\
MG-Cubed & S2 & 0.355 & 1.151 $\pm$ 0.078 & 1.174 $\pm$ 0.059 & 0.749 $\pm$ 0.030 & 0.743 $\pm$ 0.026 & 0.843 \\
Sine-to-square & S2 & 0.273 & 0.577 $\pm$ 0.029 & 0.538 $\pm$ 0.021 & 0.483 $\pm$ 0.031 & 0.461 $\pm$ 0.013 & 0.454 \\
MG & S3 & 0.090 & 0.779 $\pm$ 0.028 & 0.749 $\pm$ 0.044 & 0.480 $\pm$ 0.027 & 0.414 $\pm$ 0.019 & 0.548 \\
MG-Cubed & S3 & 0.293 & 1.074 $\pm$ 0.092 & 1.042 $\pm$ 0.055 & 0.656 $\pm$ 0.047 & 0.595 $\pm$ 0.021 & 0.768 \\
Sine-to-square & S3 & 0.250 & 0.483 $\pm$ 0.007 & 0.464 $\pm$ 0.011 & 0.435 $\pm$ 0.008 & 0.401 $\pm$ 0.009 & 0.414 \\
MG & S4 & 0.092 & 0.801 $\pm$ 0.045 & 0.703 $\pm$ 0.039 & 0.446 $\pm$ 0.017 & 0.359 $\pm$ 0.018 & 0.530 \\
MG-Cubed & S4 & 0.299 & 1.202 $\pm$ 0.054 & 1.112 $\pm$ 0.063 & 0.771 $\pm$ 0.013 & 0.675 $\pm$ 0.037 & 0.811 \\
Sine-to-square & S4 & 0.235 & 0.482 $\pm$ 0.013 & 0.442 $\pm$ 0.016 & 0.399 $\pm$ 0.005 & 0.359 $\pm$ 0.007 & 0.378 \\
\bottomrule
\end{tabular}
\end{table}

\begin{table}[H]
\centering
\caption{The sixty particle NRMSE values, optimization block, five realizations per configuration, unfiltered (raw) and after the trailing moving-average filter (filtered). The mean-field value is the mean-field error of the configuration.}
\label{tab:si_particle_runs}
\scriptsize
\begin{tabular}{@{}llccc@{}}
\toprule
\textbf{Set} & \textbf{Task} & \textbf{Mean field} & \textbf{Raw, five realizations} & \textbf{Filtered, five realizations} \\
\midrule
S1 & MG & 0.087 & 0.843, 0.855, 0.762, 0.914, 0.921 & 0.569, 0.614, 0.599, 0.551, 0.565 \\
S1 & Sine-to-square & 0.255 & 0.599, 0.628, 0.591, 0.655, 0.584 & 0.490, 0.516, 0.505, 0.508, 0.511 \\
S1 & MG-Cubed & 0.298 & 1.309, 1.070, 1.158, 1.106, 1.202 & 0.786, 0.754, 0.772, 0.730, 0.763 \\
S2 & MG & 0.111 & 0.978, 0.939, 0.906, 0.906, 0.854 & 0.611, 0.619, 0.601, 0.646, 0.594 \\
S2 & Sine-to-square & 0.276 & 0.537, 0.571, 0.569, 0.617, 0.589 & 0.454, 0.469, 0.461, 0.525, 0.505 \\
S2 & MG-Cubed & 0.349 & 1.260, 1.154, 1.087, 1.186, 1.068 & 0.794, 0.726, 0.765, 0.734, 0.725 \\
S3 & MG & 0.096 & 0.772, 0.772, 0.790, 0.819, 0.742 & 0.506, 0.460, 0.514, 0.457, 0.466 \\
S3 & Sine-to-square & 0.239 & 0.487, 0.476, 0.491, 0.475, 0.485 & 0.426, 0.443, 0.440, 0.438, 0.428 \\
S3 & MG-Cubed & 0.306 & 1.202, 1.056, 1.022, 0.965, 1.126 & 0.599, 0.693, 0.689, 0.611, 0.688 \\
S4 & MG & 0.098 & 0.800, 0.829, 0.818, 0.725, 0.835 & 0.424, 0.467, 0.457, 0.436, 0.444 \\
S4 & Sine-to-square & 0.238 & 0.499, 0.491, 0.476, 0.478, 0.468 & 0.395, 0.399, 0.408, 0.397, 0.397 \\
S4 & MG-Cubed & 0.312 & 1.120, 1.248, 1.172, 1.235, 1.234 & 0.769, 0.773, 0.749, 0.782, 0.780 \\
\bottomrule
\end{tabular}
\end{table}

\begin{figure}[H]
\centering
\includegraphics[width=0.9\textwidth]{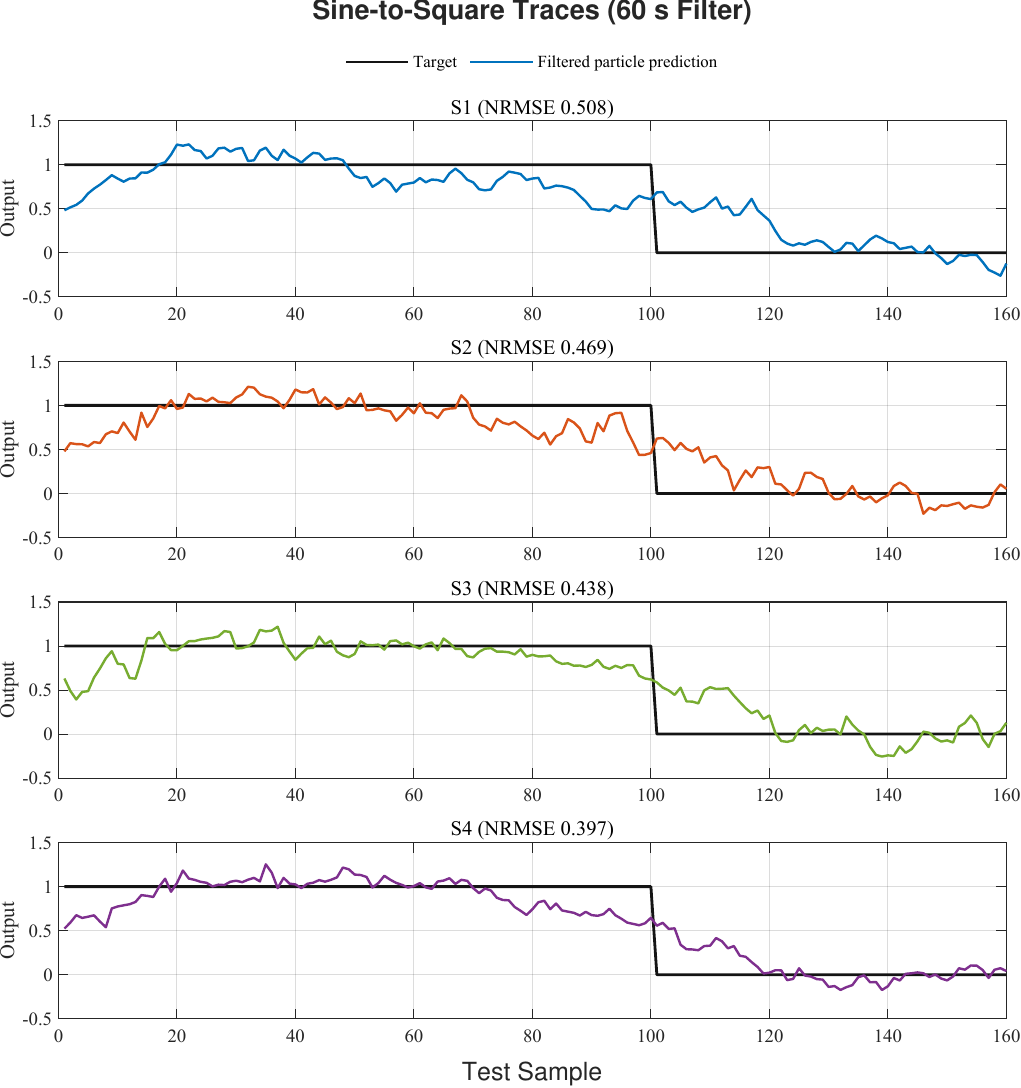}
\caption{Sine-to-square particle traces under the 60~s trailing moving-average filter. Each panel shows the same interval of the optimization block and a representative realization selected by median filtered NRMSE. The horizontal axis counts samples within the optimization block.}
\label{fig:si_sine_traces}
\end{figure}

\begin{figure}[H]
\centering
\includegraphics[width=0.9\textwidth]{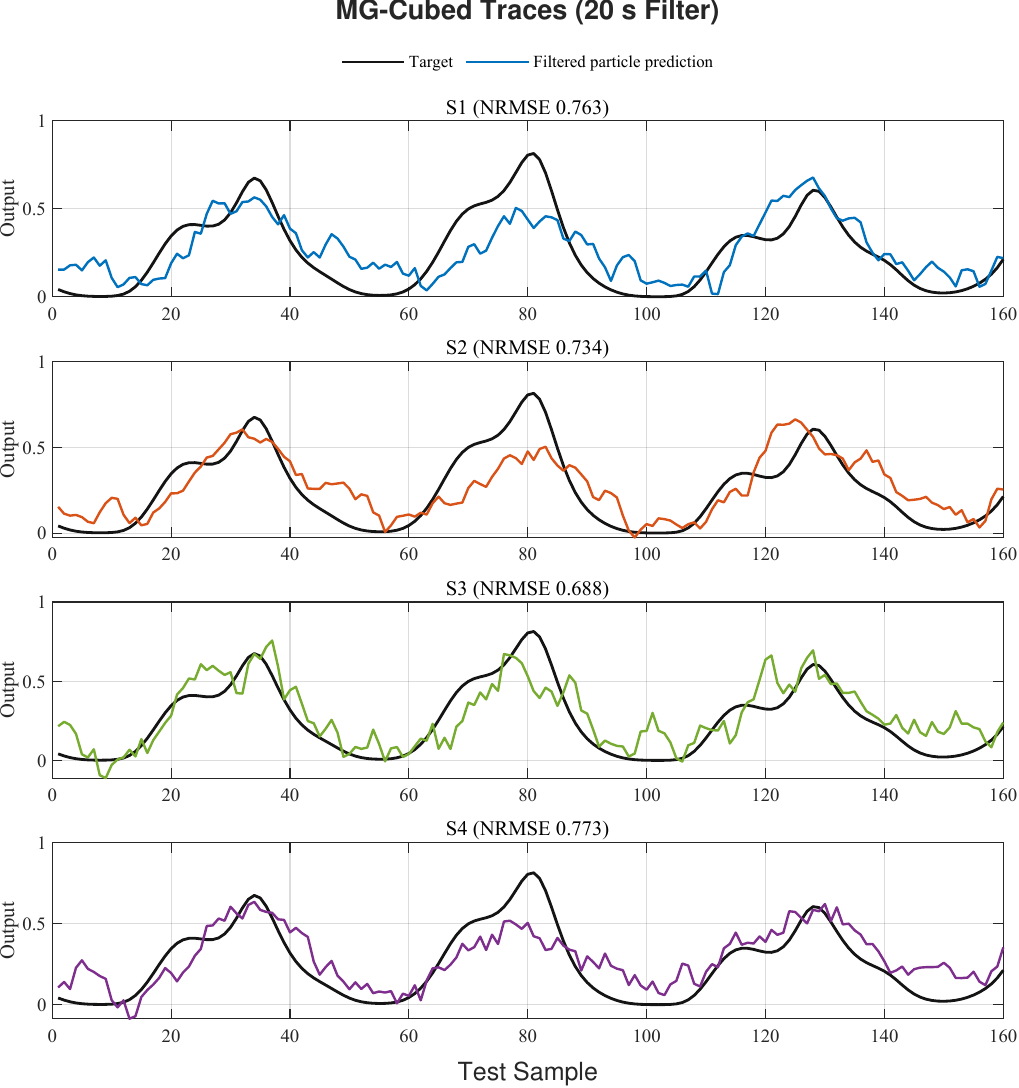}
\caption{MG-Cubed particle traces under the 20~s trailing moving-average filter. Each panel shows a representative realization selected by median filtered NRMSE over the same optimization-block interval. The horizontal axis counts samples within the optimization block.}
\label{fig:si_mgc_traces}
\end{figure}

\FloatBarrier
\section*{S6. Noise-Aware Searches, Receptor-Number Sweeps, and Reuse under Noise}

\begin{figure}[H]
\centering
\includegraphics[width=\textwidth]{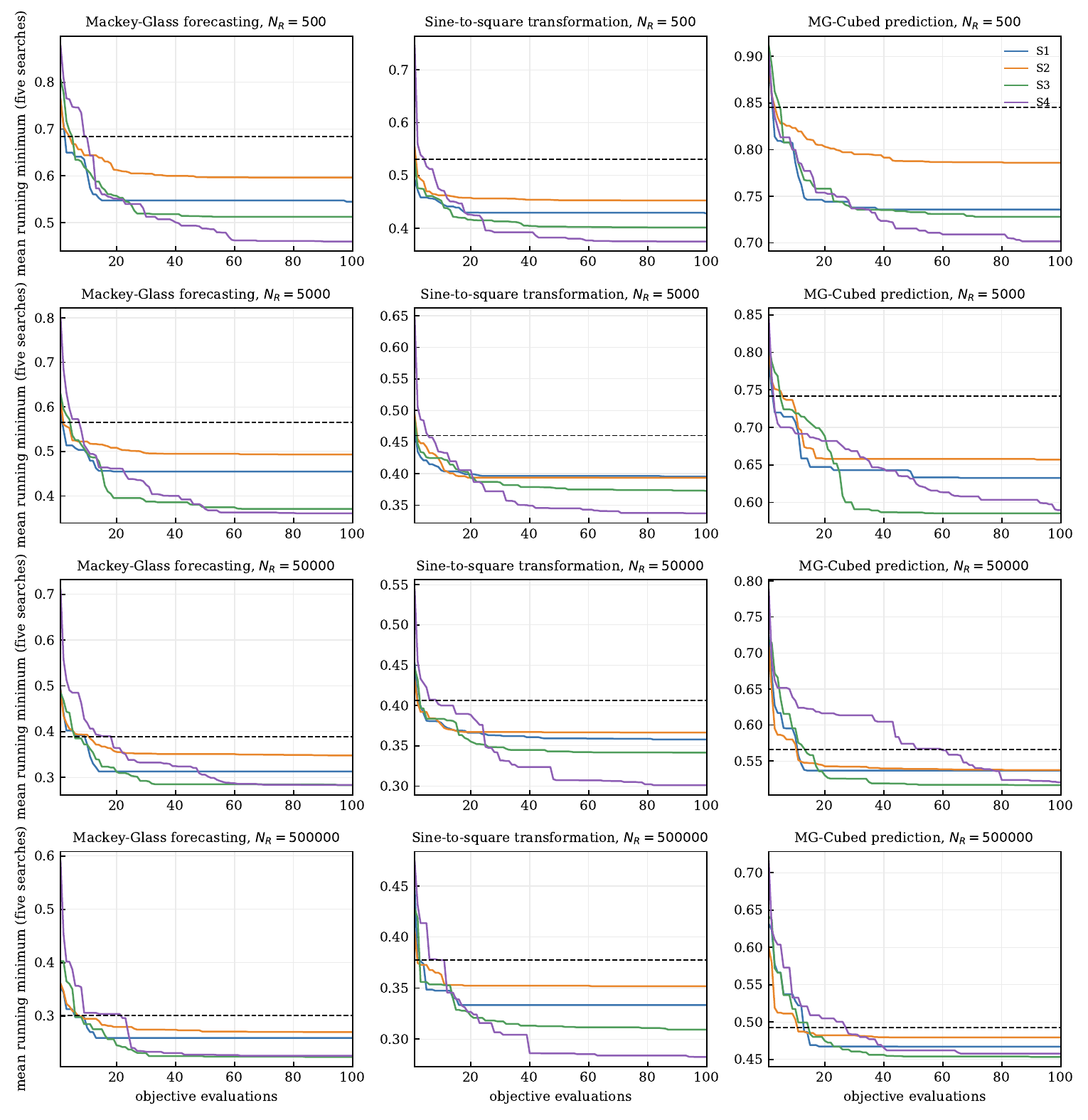}
\caption{Convergence of the noise-aware searches. Mean over the searches of the lowest objective found within the first $n$ evaluations, for the four tunable-parameter sets, at the four design receptor numbers (rows, five searches at the first three counts and three at $5\times10^{5}$) and for the three tasks (columns). The dashed line is S0 with the best $L$ and ridge value under the same noise.}
\label{fig:si_conv_hyb}
\end{figure}

\begin{figure}[H]
\centering
\includegraphics[width=\textwidth]{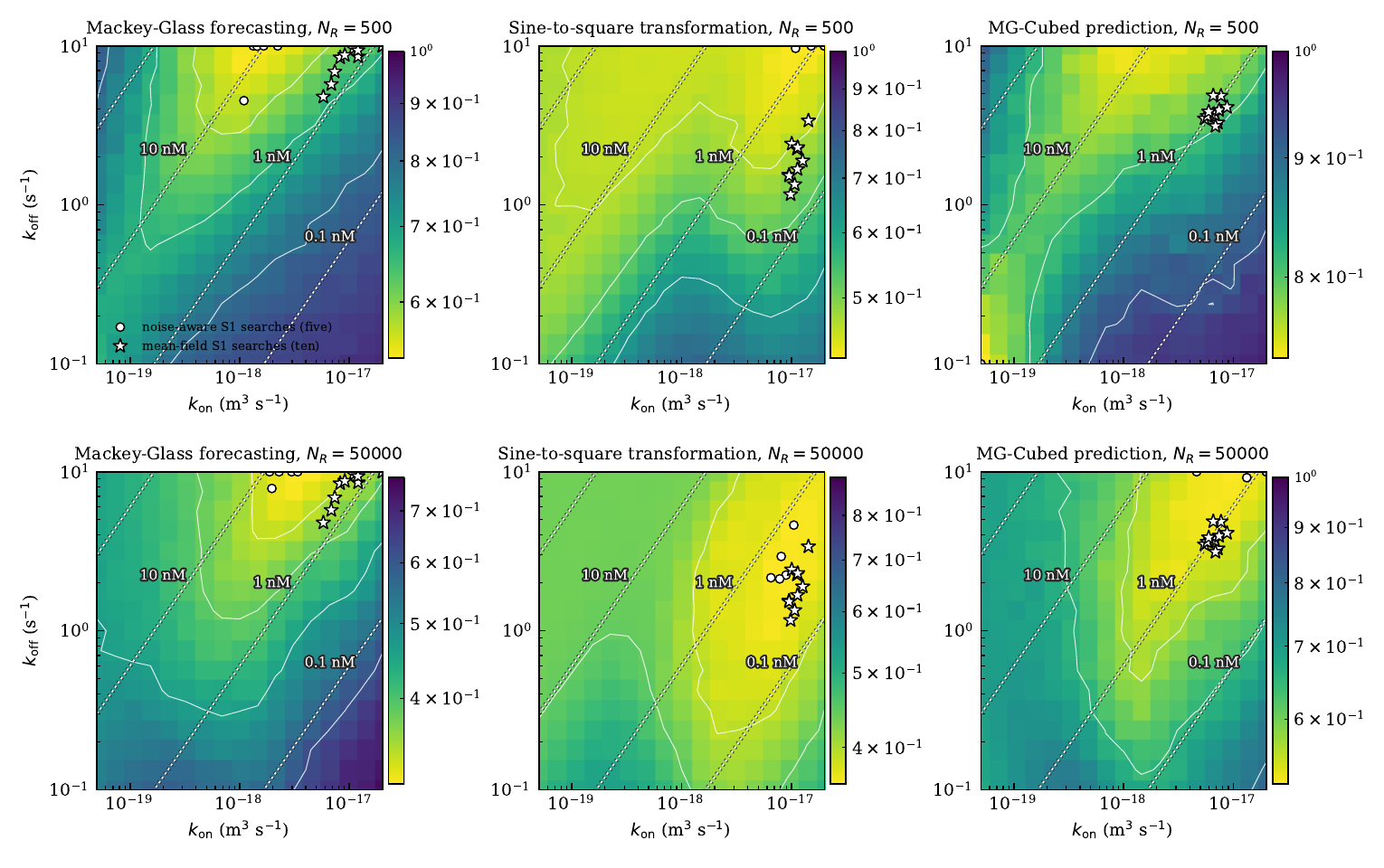}
\caption{Receiver error surfaces under receptor-count noise at $N_R=500$ (top) and $N_R=50000$ (bottom), at the nominal transmitter and channel settings, with the best $L$ and ridge value at every grid point and ten realizations per point. Circles mark the optima of the five noise-aware S1 searches, and stars mark those of the ten mean-field S1 searches.}
\label{fig:si_land_hyb}
\end{figure}

\textbf{Table~\ref{tab:si_hyb_all}} lists the best evaluation of every noise-aware search, i.e., 216 searches in total, together with the search objective (mean over the ten realizations of the search) and the held-out value. \textbf{Table~\ref{tab:si_na_optima}} gives the parameters of the noise-aware optima of S1 and S4 with their re-evaluation with independent realizations. \textbf{Figure~\ref{fig:si_conv_hyb}} shows the convergence of the noise-aware searches, and \textbf{Figure~\ref{fig:si_land_hyb}} shows the receiver error surfaces under noise for the three tasks at $N_R=500$ and 50000.

\scriptsize
\setlength{\tabcolsep}{2.2pt}
\begin{longtable}{@{}rlllccccccccccccc@{}}
\caption{Best evaluation of every noise-aware search. $d$ in $\mu$m. The objective is the mean NRMSE over the ten realizations of the search on the optimization block, and held is the corresponding held-out value. Units are m$^3$~s$^{-1}$ for $k_{\mathrm{on}}$, s$^{-1}$ for $k_{\mathrm{off}}$, s for $T$, $\mu$m for $d$, and m$^2$~s$^{-1}$ for $D$.}
\label{tab:si_hyb_all}\\
\toprule
$N_R$ & \textbf{Set} & \textbf{Task} & \textbf{Seed} & $k_{\mathrm{on}}$ & $k_{\mathrm{off}}$ & $T$ & $d$ & $D$ & $N_{\max}$ & $L$ & $\lambda$ & \textbf{objective} & \textbf{held} & $k_{\mathrm{off}}T$ & $c_{\mathrm{peak}}/K_D$ \\
\midrule
\endfirsthead
\toprule
$N_R$ & \textbf{Set} & \textbf{Task} & \textbf{Seed} & $k_{\mathrm{on}}$ & $k_{\mathrm{off}}$ & $T$ & $d$ & $D$ & $N_{\max}$ & $L$ & $\lambda$ & \textbf{objective} & \textbf{held} & $k_{\mathrm{off}}T$ & $c_{\mathrm{peak}}/K_D$ \\
\midrule
\endhead
\bottomrule
\endfoot
500 & S1 & MG & 1 & $1.33\times10^{-18}$ & 10.00 & 1.25 & 7.1 & $3.16\times10^{-11}$ & 10100 & 5 & $1$ & 0.535 & 0.542 & 12.5 & 0.46 \\
500 & S1 & MG & 2 & $2.22\times10^{-18}$ & 10.00 & 1.25 & 7.1 & $3.16\times10^{-11}$ & 10100 & 5 & $1$ & 0.541 & 0.547 & 12.5 & 0.76 \\
500 & S1 & MG & 3 & $1.44\times10^{-18}$ & 9.96 & 1.25 & 7.1 & $3.16\times10^{-11}$ & 10100 & 5 & $1$ & 0.536 & 0.537 & 12.5 & 0.5 \\
500 & S1 & MG & 4 & $1.09\times10^{-18}$ & 4.52 & 1.25 & 7.1 & $3.16\times10^{-11}$ & 10100 & 4 & $1$ & 0.576 & 0.577 & 5.7 & 0.83 \\
500 & S1 & MG & 5 & $1.65\times10^{-18}$ & 10.00 & 1.25 & 7.1 & $3.16\times10^{-11}$ & 10100 & 5 & $1$ & 0.535 & 0.535 & 12.5 & 0.57 \\
500 & S1 & Sine-to-square & 1 & $2.00\times10^{-17}$ & 10.00 & 1.25 & 7.1 & $3.16\times10^{-11}$ & 10100 & 5 & $1$ & 0.428 & 0.447 & 12.5 & 7.1 \\
500 & S1 & Sine-to-square & 2 & $2.00\times10^{-17}$ & 10.00 & 1.25 & 7.1 & $3.16\times10^{-11}$ & 10100 & 4 & $1$ & 0.428 & 0.449 & 12.5 & 7.1 \\
500 & S1 & Sine-to-square & 3 & $2.00\times10^{-17}$ & 10.00 & 1.25 & 7.1 & $3.16\times10^{-11}$ & 10100 & 5 & $10^{1}$ & 0.432 & 0.453 & 12.5 & 7.1 \\
\end{longtable}
\normalsize
\setlength{\tabcolsep}{6pt}

\begin{table}[H]
\centering
\caption{Noise-aware optima of S1 and S4 at each design receptor number. $d$ in $\mu$m. NRMSE is the re-evaluation with 50 independent realizations at the design $N_R$, with the fixed readout. Units are m$^3$~s$^{-1}$ for $k_{\mathrm{on}}$, s$^{-1}$ for $k_{\mathrm{off}}$, s for $T$, $\mu$m for $d$, and m$^2$~s$^{-1}$ for $D$.}
\label{tab:si_na_optima}
\scriptsize
\setlength{\tabcolsep}{2.5pt}
\begin{tabular}{@{}llrccccccccccc@{}}
\toprule
\textbf{Task} & \textbf{Set} & $N_R$ & $k_{\mathrm{on}}$ & $k_{\mathrm{off}}$ & $T$ & $d$ & $D$ & $N_{\max}$ & $L$ & $\lambda$ & $k_{\mathrm{off}}T$ & $c_{\mathrm{peak}}/K_D$ & \textbf{NRMSE} \\
\midrule
MG & S1 & 500 & $1.65\times10^{-18}$ & 10.00 & 1.25 & 7.1 & $3.16\times10^{-11}$ & 10100 & 5 & $1$ & 12.5 & 0.57 & 0.536 \\
MG & S1 & 5000 & $4.69\times10^{-18}$ & 10.00 & 1.25 & 7.1 & $3.16\times10^{-11}$ & 10100 & 5 & $10^{-1}$ & 12.5 & 1.6 & 0.433 \\
MG & S1 & 50000 & $3.38\times10^{-18}$ & 10.00 & 1.25 & 7.1 & $3.16\times10^{-11}$ & 10100 & 5 & $10^{-2}$ & 12.5 & 1.2 & 0.307 \\
MG & S1 & 500000 & $2.34\times10^{-18}$ & 10.00 & 1.25 & 7.1 & $3.16\times10^{-11}$ & 10100 & 5 & $10^{-3}$ & 12.5 & 0.81 & 0.257 \\
MG & S4 & 500 & $2.00\times10^{-17}$ & 10.00 & 2.00 & 6.0 & $1.35\times10^{-10}$ & 17048 & 5 & $1$ & 20.0 & 12 & 0.456 \\
MG & S4 & 5000 & $2.00\times10^{-17}$ & 10.00 & 2.00 & 6.0 & $1.33\times10^{-10}$ & 17861 & 5 & $10^{-1}$ & 20.0 & 12 & 0.360 \\
MG & S4 & 50000 & $1.21\times10^{-17}$ & 10.00 & 2.00 & 7.7 & $4.72\times10^{-11}$ & 4496 & 5 & $10^{-2}$ & 20.0 & 1.1 & 0.282 \\
\bottomrule
\end{tabular}
\end{table}

\textbf{Table~\ref{tab:si_noiseaware}} compares the noise-aware optima with the mean-field optima whose readout is retrained under the same noise (Figure 6a of the main article).

\begin{table}[H]
\centering
\caption{NRMSE under receptor-count noise, optimization block, mean over 50 independent realizations at the stated receptor number. S0 is the nominal channel with $L$ and $\lambda$ retrained under noise. The S1 and S4 mean-field columns give the mean-field optima with the readout retrained under noise. The noise-aware columns give the optima of the searches with the stochastic objective at that receptor number, with the fixed readout and the standard deviation over realizations.}
\label{tab:si_noiseaware}
\scriptsize
\setlength{\tabcolsep}{3.5pt}
\begin{tabular}{@{}lrccccccc@{}}
\toprule
\textbf{Task} & $N_R$ & \textbf{S0} & \textbf{S1, mean field} & \textbf{S4, mean field} & \textbf{S1, noise-aware} & \textbf{S2, noise-aware} & \textbf{S3, noise-aware} & \textbf{S4, noise-aware} \\
\midrule
MG & 500 & 0.687 & 0.615 & 0.575 & 0.536 $\pm$ 0.009 & 0.602 $\pm$ 0.013 & 0.515 $\pm$ 0.008 & 0.456 $\pm$ 0.007 \\
MG & 5000 & 0.566 & 0.449 & 0.405 & 0.433 $\pm$ 0.008 & 0.499 $\pm$ 0.018 & 0.370 $\pm$ 0.007 & 0.360 $\pm$ 0.005 \\
MG & 50000 & 0.387 & 0.338 & 0.315 & 0.307 $\pm$ 0.005 & 0.357 $\pm$ 0.011 & 0.281 $\pm$ 0.005 & 0.282 $\pm$ 0.004 \\
MG & 500000 & 0.298 & 0.249 & 0.242 & 0.257 $\pm$ 0.003 & 0.271 $\pm$ 0.008 & 0.223 $\pm$ 0.005 & 0.224 $\pm$ 0.003 \\
Sine-to-square & 500 & 0.531 & 0.468 & 0.381 & 0.422 $\pm$ 0.006 & 0.453 $\pm$ 0.012 & 0.402 $\pm$ 0.004 & 0.377 $\pm$ 0.006 \\
Sine-to-square & 5000 & 0.460 & 0.404 & 0.348 & 0.395 $\pm$ 0.007 & 0.395 $\pm$ 0.007 & 0.374 $\pm$ 0.007 & 0.338 $\pm$ 0.008 \\
Sine-to-square & 50000 & 0.409 & 0.364 & 0.311 & 0.361 $\pm$ 0.007 & 0.368 $\pm$ 0.005 & 0.341 $\pm$ 0.006 & 0.300 $\pm$ 0.005 \\
Sine-to-square & 500000 & 0.378 & 0.333 & 0.283 & 0.333 $\pm$ 0.005 & 0.354 $\pm$ 0.005 & 0.311 $\pm$ 0.004 & 0.281 $\pm$ 0.003 \\
MG-Cubed & 500 & 0.843 & 0.805 & 0.741 & 0.739 $\pm$ 0.016 & 0.798 $\pm$ 0.011 & 0.722 $\pm$ 0.024 & 0.692 $\pm$ 0.013 \\
MG-Cubed & 5000 & 0.740 & 0.648 & 0.630 & 0.614 $\pm$ 0.011 & 0.665 $\pm$ 0.015 & 0.586 $\pm$ 0.009 & 0.589 $\pm$ 0.010 \\
MG-Cubed & 50000 & 0.572 & 0.538 & 0.538 & 0.531 $\pm$ 0.010 & 0.547 $\pm$ 0.011 & 0.518 $\pm$ 0.010 & 0.522 $\pm$ 0.008 \\
MG-Cubed & 500000 & 0.491 & 0.471 & 0.480 & 0.466 $\pm$ 0.006 & 0.482 $\pm$ 0.008 & 0.456 $\pm$ 0.007 & 0.456 $\pm$ 0.007 \\
\bottomrule
\end{tabular}
\end{table}

\begin{figure}[H]
\centering
\includegraphics[width=\textwidth]{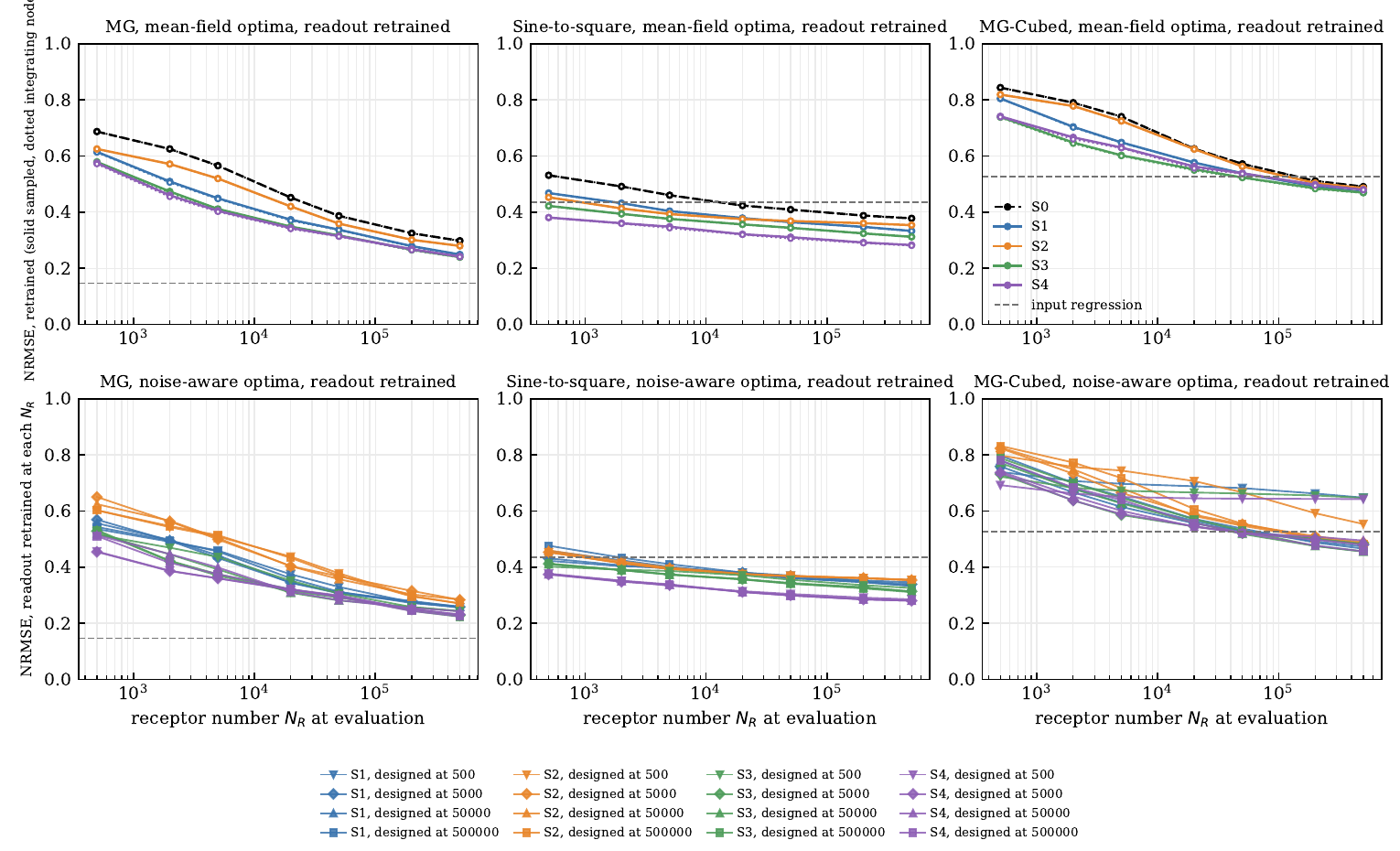}
\caption{Receptor-number sweeps. Top: the mean-field optimum of every set re-evaluated under noise with the readout retrained (solid lines for sampled nodes, dotted lines for integrating nodes). Bottom: the noise-aware optima designed at $N_R=500$, 5000, and 50000 (marker shape), evaluated at every $N_R$ with the readout retrained. The dashed gray line indicates the best input-tap baseline.}
\label{fig:si_nr_sweep}
\end{figure}

\textbf{Receptor-number sweeps:} \textbf{Table~\ref{tab:si_nr_mf}} re-evaluates the mean-field optimum of every set with 50 independent realizations at $N_R=500, 2000, 5000, 20000, 50000, 200000, 500000$, both with the fixed readout and with the readout retrained under noise. \textbf{Table~\ref{tab:si_nr_na}} does the same for the noise-aware optima evaluated away from their design receptor number, with the readout retrained. \textbf{Figure~\ref{fig:si_nr_sweep}} plots both, together with the integrating-node variant, in which every virtual node takes the mean occupancy over its node window instead of the occupancy at the node instant.

\begin{table}[H]
\centering
\caption{Mean-field optima under receptor-count noise. NRMSE on the optimization block, mean over 50 independent realizations, at each $N_R$, with the fixed readout and with $L$ and $\lambda$ retrained under noise. The mean-field column is the noise-free value.}
\label{tab:si_nr_mf}
\scriptsize
\setlength{\tabcolsep}{3pt}
\begin{tabular}{@{}lllcrrrrrrr@{}}
\toprule
\textbf{Set} & \textbf{Task} & \textbf{Readout} & \textbf{Mean field} & 500 & 2000 & 5000 & 20000 & 50000 & 200000 & 500000 \\
\midrule
S0 & MG & designed & 0.122 & 0.970 & 0.890 & 0.783 & 0.633 & 0.534 & 0.436 & 0.378 \\
S0 & MG & retrained & 0.122 & 0.687 & 0.625 & 0.566 & 0.452 & 0.387 & 0.325 & 0.298 \\
S0 & Sine-to-square & designed & 0.293 & 0.621 & 0.577 & 0.546 & 0.500 & 0.481 & 0.454 & 0.436 \\
S0 & Sine-to-square & retrained & 0.293 & 0.531 & 0.491 & 0.460 & 0.423 & 0.409 & 0.388 & 0.378 \\
S0 & MG-Cubed & designed & 0.346 & 1.197 & 1.121 & 1.034 & 0.868 & 0.803 & 0.695 & 0.633 \\
S0 & MG-Cubed & retrained & 0.346 & 0.843 & 0.790 & 0.740 & 0.626 & 0.572 & 0.511 & 0.491 \\
S1 & MG & designed & 0.088 & 0.849 & 0.695 & 0.616 & 0.514 & 0.457 & 0.384 & 0.336 \\
S1 & MG & retrained & 0.088 & 0.615 & 0.509 & 0.449 & 0.373 & 0.338 & 0.279 & 0.249 \\
S1 & Sine-to-square & designed & 0.256 & 0.541 & 0.496 & 0.466 & 0.433 & 0.418 & 0.400 & 0.382 \\
S1 & Sine-to-square & retrained & 0.256 & 0.468 & 0.432 & 0.404 & 0.379 & 0.364 & 0.348 & 0.333 \\
S1 & MG-Cubed & designed & 0.299 & 1.108 & 0.986 & 0.903 & 0.810 & 0.752 & 0.692 & 0.653 \\
S1 & MG-Cubed & retrained & 0.299 & 0.805 & 0.704 & 0.648 & 0.577 & 0.538 & 0.492 & 0.471 \\
S2 & MG & designed & 0.111 & 0.882 & 0.809 & 0.722 & 0.590 & 0.495 & 0.407 & 0.359 \\
S2 & MG & retrained & 0.111 & 0.625 & 0.571 & 0.520 & 0.420 & 0.359 & 0.301 & 0.280 \\
S2 & Sine-to-square & designed & 0.273 & 0.535 & 0.488 & 0.466 & 0.443 & 0.433 & 0.423 & 0.412 \\
S2 & Sine-to-square & retrained & 0.273 & 0.453 & 0.413 & 0.393 & 0.375 & 0.368 & 0.361 & 0.354 \\
S2 & MG-Cubed & designed & 0.339 & 1.169 & 1.113 & 1.017 & 0.871 & 0.784 & 0.689 & 0.625 \\
S2 & MG-Cubed & retrained & 0.339 & 0.819 & 0.778 & 0.725 & 0.624 & 0.563 & 0.504 & 0.485 \\
S3 & MG & designed & 0.088 & 0.776 & 0.652 & 0.561 & 0.475 & 0.430 & 0.365 & 0.320 \\
S3 & MG & retrained & 0.088 & 0.579 & 0.474 & 0.411 & 0.348 & 0.317 & 0.266 & 0.240 \\
S3 & Sine-to-square & designed & 0.248 & 0.486 & 0.451 & 0.435 & 0.409 & 0.396 & 0.372 & 0.356 \\
S3 & Sine-to-square & retrained & 0.248 & 0.422 & 0.394 & 0.376 & 0.356 & 0.344 & 0.324 & 0.312 \\
S3 & MG-Cubed & designed & 0.287 & 1.031 & 0.907 & 0.847 & 0.770 & 0.726 & 0.671 & 0.666 \\
S3 & MG-Cubed & retrained & 0.287 & 0.739 & 0.648 & 0.602 & 0.552 & 0.523 & 0.485 & 0.469 \\
S4 & MG & designed & 0.087 & 0.784 & 0.629 & 0.561 & 0.479 & 0.429 & 0.366 & 0.318 \\
S4 & MG & retrained & 0.087 & 0.575 & 0.460 & 0.405 & 0.344 & 0.315 & 0.268 & 0.242 \\
S4 & Sine-to-square & designed & 0.231 & 0.434 & 0.411 & 0.391 & 0.361 & 0.348 & 0.332 & 0.321 \\
S4 & Sine-to-square & retrained & 0.231 & 0.381 & 0.361 & 0.348 & 0.321 & 0.311 & 0.292 & 0.283 \\
S4 & MG-Cubed & designed & 0.285 & 1.010 & 0.907 & 0.871 & 0.787 & 0.739 & 0.685 & 0.669 \\
S4 & MG-Cubed & retrained & 0.285 & 0.741 & 0.666 & 0.630 & 0.563 & 0.538 & 0.498 & 0.480 \\
\bottomrule
\end{tabular}
\end{table}

\begin{table}[H]
\centering
\caption{Noise-aware optima evaluated at every $N_R$ with the readout retrained, mean over 50 independent realizations. The mean-field column is the noise-free value of the same configuration.}
\label{tab:si_nr_na}
\scriptsize
\setlength{\tabcolsep}{3pt}
\begin{tabular}{@{}rllcrrrrrrr@{}}
\toprule
\textbf{Design} $N_R$ & \textbf{Set} & \textbf{Task} & \textbf{Mean field} & 500 & 2000 & 5000 & 20000 & 50000 & 200000 & 500000 \\
\midrule
500 & S1 & MG & 0.494 & 0.536 & 0.491 & 0.459 & 0.374 & 0.329 & 0.275 & 0.255 \\
500 & S1 & Sine-to-square & 0.410 & 0.422 & 0.403 & 0.398 & 0.380 & 0.370 & 0.353 & 0.346 \\
500 & S1 & MG-Cubed & 0.689 & 0.739 & 0.707 & 0.697 & 0.688 & 0.681 & 0.662 & 0.647 \\
500 & S2 & MG & 0.511 & 0.602 & 0.542 & 0.509 & 0.437 & 0.378 & 0.298 & 0.270 \\
500 & S2 & Sine-to-square & 0.377 & 0.453 & 0.415 & 0.394 & 0.375 & 0.369 & 0.361 & 0.354 \\
500 & S2 & MG-Cubed & 0.750 & 0.798 & 0.757 & 0.744 & 0.707 & 0.666 & 0.592 & 0.553 \\
500 & S3 & MG & 0.483 & 0.515 & 0.469 & 0.436 & 0.352 & 0.306 & 0.258 & 0.243 \\
500 & S3 & Sine-to-square & 0.397 & 0.402 & 0.391 & 0.386 & 0.371 & 0.355 & 0.336 & 0.326 \\
500 & S3 & MG-Cubed & 0.666 & 0.722 & 0.683 & 0.672 & 0.666 & 0.662 & 0.655 & 0.647 \\
500 & S4 & MG & 0.406 & 0.456 & 0.387 & 0.360 & 0.320 & 0.299 & 0.250 & 0.231 \\
500 & S4 & Sine-to-square & 0.346 & 0.377 & 0.351 & 0.334 & 0.314 & 0.304 & 0.291 & 0.285 \\
500 & S4 & MG-Cubed & 0.669 & 0.692 & 0.662 & 0.650 & 0.645 & 0.644 & 0.643 & 0.642 \\
5000 & S1 & MG & 0.342 & 0.569 & 0.494 & 0.433 & 0.345 & 0.310 & 0.279 & 0.258 \\
5000 & S1 & Sine-to-square & 0.364 & 0.431 & 0.406 & 0.395 & 0.381 & 0.367 & 0.350 & 0.339 \\
5000 & S1 & MG-Cubed & 0.560 & 0.759 & 0.666 & 0.614 & 0.556 & 0.528 & 0.501 & 0.480 \\
5000 & S2 & MG & 0.318 & 0.650 & 0.563 & 0.499 & 0.404 & 0.367 & 0.315 & 0.283 \\
5000 & S2 & Sine-to-square & 0.363 & 0.453 & 0.414 & 0.395 & 0.374 & 0.368 & 0.361 & 0.353 \\
5000 & S2 & MG-Cubed & 0.567 & 0.823 & 0.733 & 0.665 & 0.587 & 0.552 & 0.510 & 0.487 \\
5000 & S3 & MG & 0.315 & 0.528 & 0.421 & 0.370 & 0.314 & 0.292 & 0.251 & 0.228 \\
5000 & S3 & Sine-to-square & 0.346 & 0.412 & 0.389 & 0.374 & 0.357 & 0.344 & 0.328 & 0.313 \\
5000 & S3 & MG-Cubed & 0.541 & 0.732 & 0.638 & 0.586 & 0.545 & 0.522 & 0.501 & 0.483 \\
5000 & S4 & MG & 0.333 & 0.454 & 0.386 & 0.360 & 0.319 & 0.296 & 0.249 & 0.231 \\
5000 & S4 & Sine-to-square & 0.286 & 0.374 & 0.348 & 0.336 & 0.312 & 0.300 & 0.285 & 0.280 \\
5000 & S4 & MG-Cubed & 0.545 & 0.736 & 0.637 & 0.589 & 0.546 & 0.521 & 0.498 & 0.480 \\
50000 & S1 & MG & 0.270 & 0.556 & 0.497 & 0.442 & 0.344 & 0.307 & 0.276 & 0.260 \\
50000 & S1 & Sine-to-square & 0.334 & 0.458 & 0.424 & 0.401 & 0.376 & 0.361 & 0.346 & 0.333 \\
50000 & S1 & MG-Cubed & 0.498 & 0.780 & 0.680 & 0.630 & 0.565 & 0.531 & 0.492 & 0.473 \\
50000 & S2 & MG & 0.239 & 0.624 & 0.565 & 0.504 & 0.403 & 0.357 & 0.303 & 0.284 \\
50000 & S2 & Sine-to-square & 0.361 & 0.460 & 0.421 & 0.398 & 0.376 & 0.368 & 0.360 & 0.355 \\
50000 & S2 & MG-Cubed & 0.493 & 0.826 & 0.748 & 0.681 & 0.582 & 0.547 & 0.503 & 0.484 \\
50000 & S3 & MG & 0.260 & 0.520 & 0.445 & 0.392 & 0.309 & 0.281 & 0.257 & 0.243 \\
50000 & S3 & Sine-to-square & 0.315 & 0.410 & 0.390 & 0.373 & 0.356 & 0.341 & 0.327 & 0.313 \\
50000 & S3 & MG-Cubed & 0.479 & 0.771 & 0.679 & 0.627 & 0.559 & 0.518 & 0.475 & 0.455 \\
50000 & S4 & MG & 0.258 & 0.514 & 0.447 & 0.399 & 0.312 & 0.282 & 0.255 & 0.243 \\
50000 & S4 & Sine-to-square & 0.286 & 0.374 & 0.350 & 0.337 & 0.311 & 0.300 & 0.285 & 0.280 \\
50000 & S4 & MG-Cubed & 0.504 & 0.740 & 0.652 & 0.601 & 0.544 & 0.522 & 0.508 & 0.495 \\
500000 & S1 & MG & 0.240 & 0.543 & 0.494 & 0.456 & 0.360 & 0.312 & 0.272 & 0.257 \\
500000 & S1 & Sine-to-square & 0.310 & 0.476 & 0.434 & 0.410 & 0.381 & 0.361 & 0.347 & 0.333 \\
500000 & S1 & MG-Cubed & 0.450 & 0.797 & 0.700 & 0.652 & 0.572 & 0.537 & 0.487 & 0.466 \\
500000 & S2 & MG & 0.175 & 0.602 & 0.546 & 0.514 & 0.433 & 0.369 & 0.295 & 0.271 \\
500000 & S2 & Sine-to-square & 0.314 & 0.454 & 0.414 & 0.393 & 0.374 & 0.367 & 0.361 & 0.354 \\
500000 & S2 & MG-Cubed & 0.457 & 0.832 & 0.773 & 0.718 & 0.607 & 0.555 & 0.504 & 0.482 \\
500000 & S3 & MG & 0.201 & 0.532 & 0.423 & 0.374 & 0.321 & 0.296 & 0.244 & 0.223 \\
500000 & S3 & Sine-to-square & 0.291 & 0.412 & 0.389 & 0.373 & 0.356 & 0.342 & 0.324 & 0.311 \\
500000 & S3 & MG-Cubed & 0.435 & 0.788 & 0.699 & 0.646 & 0.571 & 0.530 & 0.475 & 0.456 \\
500000 & S4 & MG & 0.206 & 0.511 & 0.415 & 0.375 & 0.321 & 0.293 & 0.244 & 0.224 \\
500000 & S4 & Sine-to-square & 0.272 & 0.375 & 0.352 & 0.339 & 0.312 & 0.299 & 0.285 & 0.280 \\
500000 & S4 & MG-Cubed & 0.440 & 0.780 & 0.683 & 0.639 & 0.557 & 0.522 & 0.477 & 0.456 \\
\bottomrule
\end{tabular}
\end{table}

\textbf{Reuse under noise:} \textbf{Table~\ref{tab:si_reuse_noise}} evaluates the noise-aware optimum of each set on the other two tasks at the same receptor number, with the readout retrained on the target task with 20 independent realizations, and compares it with the noise-aware optimum of the target task under the same scheme.

\scriptsize
\begin{longtable}{@{}rlllccc@{}}
\caption{Reuse of the noise-aware optima across tasks at the design receptor number, readout retrained on the target ($L$ and $\lambda$ re-selected), mean $\pm$ standard deviation over 20 realizations, and relative penalty against the noise-aware optimum of the target task under the same scheme.}
\label{tab:si_reuse_noise}\\
\toprule
$N_R$ & \textbf{Set} & \textbf{Tuned for} & \textbf{Target} & \textbf{Reused} & \textbf{Reference} & \textbf{Penalty (\%)} \\
\midrule
\endfirsthead
\toprule
$N_R$ & \textbf{Set} & \textbf{Tuned for} & \textbf{Target} & \textbf{Reused} & \textbf{Reference} & \textbf{Penalty (\%)} \\
\midrule
\endhead
\bottomrule
\endfoot
500 & S1 & Sine-to-square & MG & 0.637 $\pm$ 0.009 & 0.537 $\pm$ 0.008 & 19 \\
500 & S1 & MG-Cubed & MG & 0.672 $\pm$ 0.019 & 0.537 $\pm$ 0.008 & 25 \\
500 & S1 & MG & Sine-to-square & 0.451 $\pm$ 0.003 & 0.421 $\pm$ 0.005 & 7 \\
500 & S1 & MG-Cubed & Sine-to-square & 0.514 $\pm$ 0.008 & 0.421 $\pm$ 0.005 & 22 \\
500 & S1 & MG & MG-Cubed & 0.740 $\pm$ 0.012 & 0.737 $\pm$ 0.017 & 0 \\
500 & S1 & Sine-to-square & MG-Cubed & 0.774 $\pm$ 0.010 & 0.737 $\pm$ 0.017 & 5 \\
500 & S2 & Sine-to-square & MG & 0.665 $\pm$ 0.009 & 0.598 $\pm$ 0.011 & 11 \\
500 & S2 & MG-Cubed & MG & 0.623 $\pm$ 0.015 & 0.598 $\pm$ 0.011 & 4 \\
500 & S2 & MG & Sine-to-square & 0.466 $\pm$ 0.004 & 0.456 $\pm$ 0.013 & 2 \\
500 & S2 & MG-Cubed & Sine-to-square & 0.462 $\pm$ 0.006 & 0.456 $\pm$ 0.013 & 1 \\
500 & S2 & MG & MG-Cubed & 0.795 $\pm$ 0.010 & 0.795 $\pm$ 0.012 & -0 \\
500 & S2 & Sine-to-square & MG-Cubed & 0.824 $\pm$ 0.019 & 0.795 $\pm$ 0.012 & 4 \\
500 & S3 & Sine-to-square & MG & 0.544 $\pm$ 0.006 & 0.512 $\pm$ 0.006 & 6 \\
500 & S3 & MG-Cubed & MG & 0.654 $\pm$ 0.031 & 0.512 $\pm$ 0.006 & 28 \\
500 & S3 & MG & Sine-to-square & 0.446 $\pm$ 0.002 & 0.402 $\pm$ 0.003 & 11 \\
500 & S3 & MG-Cubed & Sine-to-square & 0.516 $\pm$ 0.011 & 0.402 $\pm$ 0.003 & 28 \\
500 & S3 & MG & MG-Cubed & 0.741 $\pm$ 0.007 & 0.727 $\pm$ 0.024 & 2 \\
500 & S3 & Sine-to-square & MG-Cubed & 0.731 $\pm$ 0.015 & 0.727 $\pm$ 0.024 & 1 \\
500 & S4 & Sine-to-square & MG & 0.514 $\pm$ 0.015 & 0.457 $\pm$ 0.007 & 12 \\
500 & S4 & MG-Cubed & MG & 0.641 $\pm$ 0.013 & 0.457 $\pm$ 0.007 & 40 \\
500 & S4 & MG & Sine-to-square & 0.384 $\pm$ 0.007 & 0.377 $\pm$ 0.005 & 2 \\
500 & S4 & MG-Cubed & Sine-to-square & 0.570 $\pm$ 0.009 & 0.377 $\pm$ 0.005 & 51 \\
500 & S4 & MG & MG-Cubed & 0.719 $\pm$ 0.006 & 0.690 $\pm$ 0.010 & 4 \\
500 & S4 & Sine-to-square & MG-Cubed & 0.755 $\pm$ 0.010 & 0.690 $\pm$ 0.010 & 9 \\
5000 & S1 & Sine-to-square & MG & 0.489 $\pm$ 0.011 & 0.431 $\pm$ 0.008 & 13 \\
5000 & S1 & MG-Cubed & MG & 0.442 $\pm$ 0.009 & 0.431 $\pm$ 0.008 & 2 \\
5000 & S1 & MG & Sine-to-square & 0.411 $\pm$ 0.007 & 0.394 $\pm$ 0.006 & 4 \\
5000 & S1 & MG-Cubed & Sine-to-square & 0.396 $\pm$ 0.003 & 0.394 $\pm$ 0.006 & 0 \\
5000 & S1 & MG & MG-Cubed & 0.622 $\pm$ 0.012 & 0.615 $\pm$ 0.015 & 1 \\
5000 & S1 & Sine-to-square & MG-Cubed & 0.630 $\pm$ 0.011 & 0.615 $\pm$ 0.015 & 2 \\
5000 & S2 & Sine-to-square & MG & 0.506 $\pm$ 0.009 & 0.500 $\pm$ 0.014 & 1 \\
5000 & S2 & MG-Cubed & MG & 0.496 $\pm$ 0.011 & 0.500 $\pm$ 0.014 & -1 \\
5000 & S2 & MG & Sine-to-square & 0.410 $\pm$ 0.008 & 0.397 $\pm$ 0.006 & 3 \\
5000 & S2 & MG-Cubed & Sine-to-square & 0.392 $\pm$ 0.007 & 0.397 $\pm$ 0.006 & -1 \\
5000 & S2 & MG & MG-Cubed & 0.673 $\pm$ 0.014 & 0.660 $\pm$ 0.014 & 2 \\
5000 & S2 & Sine-to-square & MG-Cubed & 0.664 $\pm$ 0.016 & 0.660 $\pm$ 0.014 & 1 \\
5000 & S3 & Sine-to-square & MG & 0.486 $\pm$ 0.018 & 0.366 $\pm$ 0.007 & 33 \\
5000 & S3 & MG-Cubed & MG & 0.372 $\pm$ 0.005 & 0.366 $\pm$ 0.007 & 2 \\
5000 & S3 & MG & Sine-to-square & 0.395 $\pm$ 0.008 & 0.373 $\pm$ 0.008 & 6 \\
5000 & S3 & MG-Cubed & Sine-to-square & 0.394 $\pm$ 0.007 & 0.373 $\pm$ 0.008 & 6 \\
5000 & S3 & MG & MG-Cubed & 0.584 $\pm$ 0.011 & 0.582 $\pm$ 0.010 & 0 \\
5000 & S3 & Sine-to-square & MG-Cubed & 0.660 $\pm$ 0.012 & 0.582 $\pm$ 0.010 & 13 \\
5000 & S4 & Sine-to-square & MG & 0.396 $\pm$ 0.013 & 0.362 $\pm$ 0.005 & 9 \\
5000 & S4 & MG-Cubed & MG & 0.385 $\pm$ 0.007 & 0.362 $\pm$ 0.005 & 7 \\
5000 & S4 & MG & Sine-to-square & 0.341 $\pm$ 0.006 & 0.336 $\pm$ 0.006 & 2 \\
5000 & S4 & MG-Cubed & Sine-to-square & 0.392 $\pm$ 0.003 & 0.336 $\pm$ 0.006 & 17 \\
5000 & S4 & MG & MG-Cubed & 0.638 $\pm$ 0.006 & 0.588 $\pm$ 0.010 & 9 \\
5000 & S4 & Sine-to-square & MG-Cubed & 0.662 $\pm$ 0.019 & 0.588 $\pm$ 0.010 & 13 \\
50000 & S1 & Sine-to-square & MG & 0.439 $\pm$ 0.016 & 0.306 $\pm$ 0.005 & 43 \\
50000 & S1 & MG-Cubed & MG & 0.353 $\pm$ 0.007 & 0.306 $\pm$ 0.005 & 15 \\
50000 & S1 & MG & Sine-to-square & 0.397 $\pm$ 0.008 & 0.362 $\pm$ 0.006 & 10 \\
50000 & S1 & MG-Cubed & Sine-to-square & 0.371 $\pm$ 0.007 & 0.362 $\pm$ 0.006 & 2 \\
50000 & S1 & MG & MG-Cubed & 0.539 $\pm$ 0.009 & 0.528 $\pm$ 0.007 & 2 \\
50000 & S1 & Sine-to-square & MG-Cubed & 0.563 $\pm$ 0.014 & 0.528 $\pm$ 0.007 & 7 \\
50000 & S2 & Sine-to-square & MG & 0.373 $\pm$ 0.013 & 0.353 $\pm$ 0.009 & 6 \\
50000 & S2 & MG-Cubed & MG & 0.369 $\pm$ 0.010 & 0.353 $\pm$ 0.009 & 5 \\
50000 & S2 & MG & Sine-to-square & 0.395 $\pm$ 0.009 & 0.369 $\pm$ 0.005 & 7 \\
50000 & S2 & MG-Cubed & Sine-to-square & 0.374 $\pm$ 0.007 & 0.369 $\pm$ 0.005 & 1 \\
50000 & S2 & MG & MG-Cubed & 0.554 $\pm$ 0.013 & 0.546 $\pm$ 0.011 & 2 \\
50000 & S2 & Sine-to-square & MG-Cubed & 0.553 $\pm$ 0.010 & 0.546 $\pm$ 0.011 & 1 \\
50000 & S3 & Sine-to-square & MG & 0.348 $\pm$ 0.010 & 0.279 $\pm$ 0.003 & 25 \\
50000 & S3 & MG-Cubed & MG & 0.334 $\pm$ 0.006 & 0.279 $\pm$ 0.003 & 20 \\
50000 & S3 & MG & Sine-to-square & 0.389 $\pm$ 0.005 & 0.341 $\pm$ 0.005 & 14 \\
50000 & S3 & MG-Cubed & Sine-to-square & 0.352 $\pm$ 0.006 & 0.341 $\pm$ 0.005 & 3 \\
50000 & S3 & MG & MG-Cubed & 0.522 $\pm$ 0.008 & 0.518 $\pm$ 0.010 & 1 \\
50000 & S3 & Sine-to-square & MG-Cubed & 0.532 $\pm$ 0.009 & 0.518 $\pm$ 0.010 & 3 \\
50000 & S4 & Sine-to-square & MG & 0.327 $\pm$ 0.007 & 0.284 $\pm$ 0.003 & 15 \\
50000 & S4 & MG-Cubed & MG & 0.290 $\pm$ 0.004 & 0.284 $\pm$ 0.003 & 2 \\
50000 & S4 & MG & Sine-to-square & 0.389 $\pm$ 0.006 & 0.301 $\pm$ 0.006 & 29 \\
50000 & S4 & MG-Cubed & Sine-to-square & 0.379 $\pm$ 0.008 & 0.301 $\pm$ 0.006 & 26 \\
50000 & S4 & MG & MG-Cubed & 0.525 $\pm$ 0.006 & 0.520 $\pm$ 0.007 & 1 \\
50000 & S4 & Sine-to-square & MG-Cubed & 0.595 $\pm$ 0.006 & 0.520 $\pm$ 0.007 & 14 \\
500000 & S1 & Sine-to-square & MG & 0.312 $\pm$ 0.011 & 0.257 $\pm$ 0.003 & 21 \\
500000 & S1 & MG-Cubed & MG & 0.274 $\pm$ 0.003 & 0.257 $\pm$ 0.003 & 6 \\
500000 & S1 & MG & Sine-to-square & 0.378 $\pm$ 0.009 & 0.333 $\pm$ 0.004 & 13 \\
500000 & S1 & MG-Cubed & Sine-to-square & 0.341 $\pm$ 0.006 & 0.333 $\pm$ 0.004 & 2 \\
500000 & S1 & MG & MG-Cubed & 0.521 $\pm$ 0.006 & 0.469 $\pm$ 0.007 & 11 \\
500000 & S1 & Sine-to-square & MG-Cubed & 0.488 $\pm$ 0.006 & 0.469 $\pm$ 0.007 & 4 \\
500000 & S2 & Sine-to-square & MG & 0.284 $\pm$ 0.007 & 0.273 $\pm$ 0.004 & 4 \\
500000 & S2 & MG-Cubed & MG & 0.293 $\pm$ 0.006 & 0.273 $\pm$ 0.004 & 7 \\
500000 & S2 & MG & Sine-to-square & 0.410 $\pm$ 0.006 & 0.353 $\pm$ 0.004 & 16 \\
500000 & S2 & MG-Cubed & Sine-to-square & 0.365 $\pm$ 0.005 & 0.353 $\pm$ 0.004 & 3 \\
500000 & S2 & MG & MG-Cubed & 0.505 $\pm$ 0.009 & 0.483 $\pm$ 0.007 & 5 \\
500000 & S2 & Sine-to-square & MG-Cubed & 0.491 $\pm$ 0.006 & 0.483 $\pm$ 0.007 & 2 \\
500000 & S3 & Sine-to-square & MG & 0.278 $\pm$ 0.008 & 0.223 $\pm$ 0.003 & 24 \\
500000 & S3 & MG-Cubed & MG & 0.257 $\pm$ 0.005 & 0.223 $\pm$ 0.003 & 15 \\
500000 & S3 & MG & Sine-to-square & 0.328 $\pm$ 0.006 & 0.312 $\pm$ 0.005 & 5 \\
500000 & S3 & MG-Cubed & Sine-to-square & 0.322 $\pm$ 0.005 & 0.312 $\pm$ 0.005 & 3 \\
500000 & S3 & MG & MG-Cubed & 0.479 $\pm$ 0.006 & 0.455 $\pm$ 0.007 & 5 \\
500000 & S3 & Sine-to-square & MG-Cubed & 0.465 $\pm$ 0.006 & 0.455 $\pm$ 0.007 & 2 \\
500000 & S4 & Sine-to-square & MG & 0.277 $\pm$ 0.002 & 0.225 $\pm$ 0.003 & 23 \\
500000 & S4 & MG-Cubed & MG & 0.255 $\pm$ 0.004 & 0.225 $\pm$ 0.003 & 13 \\
500000 & S4 & MG & Sine-to-square & 0.310 $\pm$ 0.006 & 0.280 $\pm$ 0.003 & 11 \\
500000 & S4 & MG-Cubed & Sine-to-square & 0.325 $\pm$ 0.004 & 0.280 $\pm$ 0.003 & 16 \\
500000 & S4 & MG & MG-Cubed & 0.484 $\pm$ 0.007 & 0.457 $\pm$ 0.005 & 6 \\
500000 & S4 & Sine-to-square & MG-Cubed & 0.556 $\pm$ 0.006 & 0.457 $\pm$ 0.005 & 22 \\
\end{longtable}
\normalsize
\setlength{\tabcolsep}{6pt}

\textbf{Readout without memory:} \textbf{Table~\ref{tab:si_lone}} restricts the readout of every mean-field optimum to the current state ($L=1$), with the ridge value re-selected, both in the mean-field model and under noise at $N_R=500$.

\begin{table}[H]
\centering
\caption{Mean-field optima with the readout restricted to the current state ($L=1$). Design value with the tuned $L$, mean-field value with $L=1$, and value under noise at $N_R=500$ with $L=1$ (mean $\pm$ standard deviation over 20 realizations).}
\label{tab:si_lone}
\small
\begin{tabular}{@{}llcccc@{}}
\toprule
\textbf{Set} & \textbf{Task} & \textbf{Tuned} $L$ & \textbf{Design NRMSE} & $L=1$, \textbf{mean field} & $L=1$, $N_R=500$ \\
\midrule
S0 & MG & 5 & 0.122 & 0.362 & 0.801 $\pm$ 0.010 \\
S0 & Sine-to-square & 5 & 0.293 & 0.320 & 0.546 $\pm$ 0.004 \\
S0 & MG-Cubed & 5 & 0.346 & 0.636 & 0.925 $\pm$ 0.010 \\
S1 & MG & 5 & 0.088 & 0.301 & 0.754 $\pm$ 0.013 \\
S1 & Sine-to-square & 5 & 0.256 & 0.273 & 0.539 $\pm$ 0.007 \\
S1 & MG-Cubed & 5 & 0.299 & 0.531 & 0.908 $\pm$ 0.009 \\
S2 & MG & 5 & 0.111 & 0.366 & 0.763 $\pm$ 0.012 \\
S2 & Sine-to-square & 5 & 0.273 & 0.289 & 0.511 $\pm$ 0.005 \\
S2 & MG-Cubed & 5 & 0.339 & 0.633 & 0.907 $\pm$ 0.012 \\
S3 & MG & 5 & 0.088 & 0.300 & 0.743 $\pm$ 0.015 \\
S3 & Sine-to-square & 5 & 0.248 & 0.262 & 0.469 $\pm$ 0.008 \\
S3 & MG-Cubed & 5 & 0.287 & 0.514 & 0.892 $\pm$ 0.011 \\
S4 & MG & 5 & 0.087 & 0.309 & 0.730 $\pm$ 0.010 \\
S4 & Sine-to-square & 5 & 0.231 & 0.244 & 0.390 $\pm$ 0.004 \\
S4 & MG-Cubed & 5 & 0.285 & 0.538 & 0.906 $\pm$ 0.016 \\
\bottomrule
\end{tabular}
\end{table}

\FloatBarrier
\section*{S7. Search-Domain Boundaries}

Several optima reported in the main article lie on a boundary of the search domain. \textbf{Table~\ref{tab:si_wide}} gives the results of the S1 and S4 searches repeated in a wider domain (dissociation rate up to 100~s$^{-1}$, symbol interval up to 5~s, release amplitude up to $10^{5}$ molecules, distance down to 3.5~$\mu$m, and association rate unchanged at its diffusion-limited bound), with five mean-field searches and three noise-aware searches at 50000 receptors per task. The best mean-field errors change by 0 to 2 percent under S1 and by -3 to 1 percent under S4, where positive values indicate improvements, and negative values reflect the larger space searched with the same budget. The operating regimes of the three tasks remain unchanged. Under noise at 50000 receptors, the wider domain lowers the re-evaluated errors of the noise-aware optima by 0 to 9 percent, since the forecasting and MG-Cubed optima shift to dissociation rates above the former bound of 10~s$^{-1}$.

\begin{table}[H]
\centering
\caption{Best configurations in the default and in the wider search domain. $d$ in $\mu$m. NRMSE on the optimization block (mean field), or the re-evaluation with 50 independent realizations for the noise-aware rows. Units are m$^3$~s$^{-1}$ for $k_{\mathrm{on}}$, s$^{-1}$ for $k_{\mathrm{off}}$, s for $T$, $\mu$m for $d$, and m$^2$~s$^{-1}$ for $D$.}
\label{tab:si_wide}
\scriptsize
\setlength{\tabcolsep}{2.5pt}
\begin{tabular}{@{}lllcccccccccccc@{}}
\toprule
\textbf{Set} & \textbf{Task} & \textbf{Domain} & $k_{\mathrm{on}}$ & $k_{\mathrm{off}}$ & $T$ & $d$ & $D$ & $N_{\max}$ & $L$ & \textbf{NRMSE} & $k_{\mathrm{off}}T$ & $c_{\mathrm{peak}}/K_D$ & $\tau_D/T$ \\
\midrule
S1 & MG & default & $8.23\times10^{-18}$ & 8.45 & 1.25 & 7.1 & $3.16\times10^{-11}$ & 10100 & 5 & 0.0883 & 10.6 & 3.4 & 0.21 \\
S1 & MG & wider & $1.08\times10^{-17}$ & 16.25 & 1.25 & 7.1 & $3.16\times10^{-11}$ & 10100 & 5 & 0.0868 & 20.3 & 2.3 & 0.21 \\
S1 & MG & wider, $N_R=50000$ & $1.99\times10^{-17}$ & 78.70 & 1.25 & 7.1 & $3.16\times10^{-11}$ & 10100 & 5 & 0.2783 & 98.4 & 0.87 & 0.21 \\
S1 & Sine-to-square & default & $1.14\times10^{-17}$ & 2.30 & 1.25 & 7.1 & $3.16\times10^{-11}$ & 10100 & 5 & 0.2559 & 2.9 & 18 & 0.21 \\
S1 & Sine-to-square & wider & $1.13\times10^{-17}$ & 2.53 & 1.25 & 7.1 & $3.16\times10^{-11}$ & 10100 & 5 & 0.2561 & 3.2 & 16 & 0.21 \\
S1 & Sine-to-square & wider, $N_R=50000$ & $9.40\times10^{-18}$ & 3.65 & 1.25 & 7.1 & $3.16\times10^{-11}$ & 10100 & 5 & 0.3602 & 4.6 & 9.1 & 0.21 \\
S1 & MG-Cubed & default & $6.75\times10^{-18}$ & 3.69 & 1.25 & 7.1 & $3.16\times10^{-11}$ & 10100 & 5 & 0.2986 & 4.6 & 6.3 & 0.21 \\
S1 & MG-Cubed & wider & $6.47\times10^{-18}$ & 3.30 & 1.25 & 7.1 & $3.16\times10^{-11}$ & 10100 & 5 & 0.2984 & 4.1 & 6.8 & 0.21 \\
S1 & MG-Cubed & wider, $N_R=50000$ & $1.98\times10^{-17}$ & 53.48 & 1.25 & 7.1 & $3.16\times10^{-11}$ & 10100 & 5 & 0.5247 & 66.8 & 1.3 & 0.21 \\
S4 & MG & default & $2.00\times10^{-17}$ & 10.00 & 1.87 & 11.1 & $5.75\times10^{-11}$ & 15597 & 5 & 0.0872 & 18.7 & 2.6 & 0.19 \\
S4 & MG & wider & $1.44\times10^{-18}$ & 3.63 & 5.00 & 8.5 & $2.03\times10^{-11}$ & 44537 & 5 & 0.0902 & 18.2 & 2.7 & 0.12 \\
S4 & MG & wider, $N_R=50000$ & $2.78\times10^{-19}$ & 31.85 & 4.44 & 3.5 & $5.00\times10^{-12}$ & 41353 & 5 & 0.2703 & 141.5 & 0.74 & 0.09 \\
S4 & Sine-to-square & default & $2.00\times10^{-17}$ & 2.37 & 2.00 & 6.0 & $8.80\times10^{-11}$ & 20000 & 5 & 0.2312 & 4.7 & 60 & 0.03 \\
S4 & Sine-to-square & wider & $7.95\times10^{-19}$ & 2.36 & 5.00 & 3.5 & $2.36\times10^{-11}$ & 87277 & 5 & 0.2287 & 11.8 & 51 & 0.02 \\
S4 & Sine-to-square & wider, $N_R=50000$ & $8.99\times10^{-19}$ & 2.06 & 3.79 & 3.5 & $3.83\times10^{-11}$ & 100000 & 3 & 0.2936 & 7.8 & 75 & 0.01 \\
S4 & MG-Cubed & default & $2.00\times10^{-17}$ & 10.00 & 1.52 & 7.6 & $6.68\times10^{-11}$ & 20000 & 5 & 0.2847 & 15.2 & 8 & 0.10 \\
S4 & MG-Cubed & wider & $1.12\times10^{-19}$ & 2.39 & 3.25 & 3.5 & $5.00\times10^{-12}$ & 96477 & 5 & 0.2847 & 7.8 & 10 & 0.13 \\
S4 & MG-Cubed & wider, $N_R=50000$ & $1.91\times10^{-18}$ & 52.41 & 3.58 & 5.0 & $8.56\times10^{-12}$ & 38243 & 5 & 0.5124 & 187.7 & 1.1 & 0.14 \\
\bottomrule
\end{tabular}
\end{table}

\FloatBarrier
\section*{S8. Minimax Configuration}

\textbf{Table~\ref{tab:si_compromise}} gives, for each set, the minimax configuration, i.e., the single configuration that minimizes the largest ratio, over the three benchmark tasks, between its error and the task-specific optimum of the same set, both in the mean-field model and at 50000 receptors, together with the error on every task and the task-specific value in brackets.

\begin{landscape}
\begin{table}[H]
\centering
\caption{Minimax configurations of S1 and S4. $d$ in $\mu$m. Task columns give the NRMSE of the minimax configuration with the readout retrained for that task and, in brackets, the task-specific optimum of the same set, and the last column gives the largest excess in percent. Units are m$^3$~s$^{-1}$ for $k_{\mathrm{on}}$, s$^{-1}$ for $k_{\mathrm{off}}$, s for $T$, $\mu$m for $d$, and m$^2$~s$^{-1}$ for $D$. The ridge column gives the values selected for MG forecasting, sine-to-square, and MG-Cubed in that order, and $k_{\mathrm{off}}T$ and $c_{\mathrm{peak}}/K_D$ are evaluated under the MG input.}
\label{tab:si_compromise}
\footnotesize
\setlength{\tabcolsep}{4pt}
\begin{tabular}{@{}llcccccccccccccc@{}}
\toprule
\textbf{Set} & \textbf{Model} & $k_{\mathrm{on}}$ & $k_{\mathrm{off}}$ & $T$ & $d$ & $D$ & $N_{\max}$ & $L$ & $\lambda$ (MG / sine / MGC) & $k_{\mathrm{off}}T$ & $c_{\mathrm{peak}}/K_D$ & \textbf{MG} & \textbf{Sine-to-square} & \textbf{MG-Cubed} & \textbf{worst (\%)} \\
\midrule
S1 & mean field & $1.10\times10^{-17}$ & 8.12 & 1.25 & 7.1 & $3.16\times10^{-11}$ & 10100 & 5 & $1\times10^{-6}$ / $1\times10^{-6}$ / $1\times10^{-6}$ & 10.2 & 4.7 & 0.091 (0.088) & 0.267 (0.256) & 0.306 (0.299) & 4 \\
S1 & $N_R=50000$ & $3.72\times10^{-18}$ & 6.80 & 1.25 & 7.1 & $3.16\times10^{-11}$ & 10100 & 5 & $1\times10^{-2}$ / $1\times10^{-1}$ / $1\times10^{-2}$ & 8.5 & 1.9 & 0.323 (0.307) & 0.378 (0.361) & 0.535 (0.531) & 5 \\
S4 & mean field & $2.00\times10^{-17}$ & 10.00 & 1.39 & 7.3 & $6.00\times10^{-11}$ & 20000 & 5 & $1\times10^{-6}$ / $1\times10^{-6}$ / $1\times10^{-6}$ & 13.9 & 9.4 & 0.094 (0.087) & 0.249 (0.231) & 0.287 (0.285) & 8 \\
S4 & $N_R=50000$ & $2.00\times10^{-17}$ & 10.00 & 2.00 & 6.0 & $6.31\times10^{-11}$ & 9530 & 5 & $1\times10^{-2}$ / $1\times10^{-2}$ / $1\times10^{-2}$ & 20.0 & 6.9 & 0.297 (0.282) & 0.320 (0.300) & 0.568 (0.522) & 9 \\
\bottomrule
\end{tabular}
\end{table}
\end{landscape}

\FloatBarrier
\section*{S9. Case Study and Task Switching}

\begin{figure}[H]
\centering
\includegraphics[width=\textwidth]{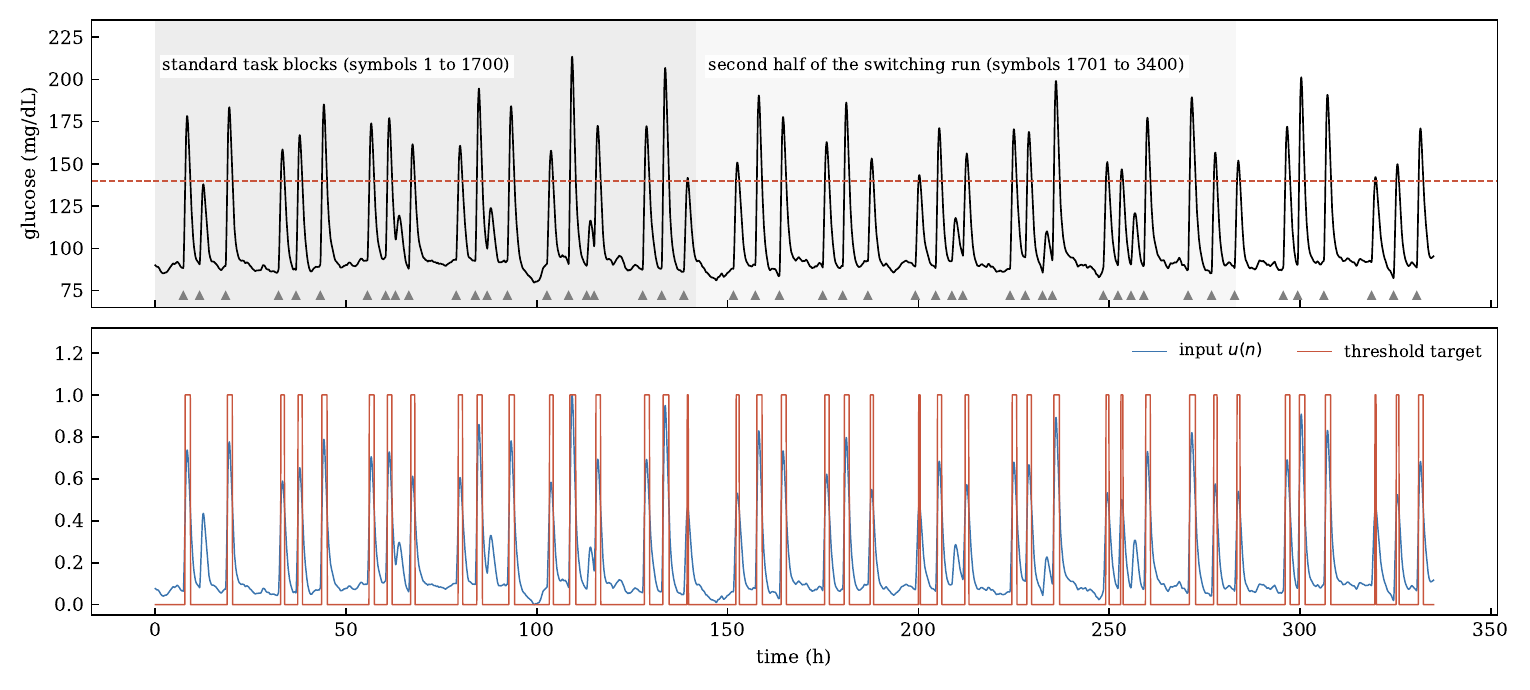}
\caption{Synthetic glucose signal of the case study. Top: glucose level from the minimal model driven by meals (triangles), with the illustrative threshold of 140~mg/dL (dashed). The first 1700 samples carry the standard task blocks, and the first 3400 samples the switching runs, in which the switch occurs after sample 1700, and sample 1701 is the first target-task symbol. Bottom: the normalized input and the threshold-decision target.}
\label{fig:si_signal}
\end{figure}

\textbf{Figure~\ref{fig:si_signal}} shows the complete synthetic glucose signal of the case study, together with the meals and the portions of the signal used for the standard task blocks and for the switching runs, and \textbf{Table~\ref{tab:si_case_params}} lists the parameters of the generator. \textbf{Table~\ref{tab:si_case}} lists the mean-field optima of the two targets with their errors under noise, \textbf{Table~\ref{tab:si_case_taps}} gives the input-tap baselines of the two targets, and \textbf{Table~\ref{tab:si_switch}} gives the switching results of the three task pairs in both directions, i.e., six directed switches.

\begin{table}[H]
\centering
\caption{Parameters of the synthetic biomarker generator (minimal model with meal absorption), integrated at a one-minute step over 14 days and sampled every five minutes, random seed 2028.}
\label{tab:si_case_params}
\small
\begin{tabularx}{\textwidth}{@{}p{4.6cm}p{3.4cm}X@{}}
\toprule
\textbf{Quantity} & \textbf{Value} & \textbf{Role} \\
\midrule
$G_b$, $I_b$ & 90 mg/dL, 10 mU/L & basal glucose and insulin \\
$p_1$, $p_2$ & 0.028 min$^{-1}$, 0.025 min$^{-1}$ & glucose effectiveness and decay of insulin action \\
$p_3$ & $1.3\times10^{-5}$ (mU/L)$^{-1}$ min$^{-2}$ & gain of insulin action \\
Initial state & $G(0)=G_b$, $X(0)=0$, $I(0)=I_b$ & start of the signal \\
$n$, $\sigma$ & 0.14 min$^{-1}$, 0.05 mU/L per mg/dL per min & insulin clearance and secretion \\
$V_G$ & 1.7 dL/kg & glucose distribution volume \\
$A_G$, $t_{\max}$ & 0.8, 40 min & meal absorption fraction and time to peak \\
Meals & 7.5 h, 12.5 h, 19 h with timing SD 0.6 h, sizes $D_G$ of 700, 850, 950 mg/kg times a log-normal factor with standard deviation 0.25 in the logarithm & three meals per day \\
Snack & 16 h with SD 0.5 h, 300 mg/kg times a log-normal factor with standard deviation 0.3 in the logarithm, probability 0.35 per day & optional afternoon snack \\
Insulin-sensitivity modulation & amplitude 0.2, 24 h period & circadian factor $s(t)$ on $p_3$ \\
Basal fluctuation & Ornstein-Uhlenbeck, SD 4 mg/dL, time constant 120 min & slow drift of $G_b$ \\
Threshold, horizon & 140 mg/dL, 6 samples (30 min) & threshold-decision and forecasting targets \\
\bottomrule
\end{tabularx}
\end{table}

\begin{landscape}
\begin{table}[H]
\centering
\caption{Case study (Section 3.7 of the main article). Mean-field optima of S0, S1, and S4 for the two targets on the biomarker signal, their operating point, and their errors in the mean field and under noise. Under noise, the mean-field optimum is evaluated with the readout retrained (r.), and the noise-aware optimum of the same set at that receptor number is evaluated with 50 independent realizations (n.a.). MF is the mean-field value. Optimization block. Units are m$^3$~s$^{-1}$ for $k_{\mathrm{on}}$, s$^{-1}$ for $k_{\mathrm{off}}$, s for $T$, $\mu$m for $d$, and m$^2$~s$^{-1}$ for $D$. n.e. marks conditions that were not evaluated, since no noise-aware S4 search was run at 500 receptors.}
\label{tab:si_case}
\footnotesize
\setlength{\tabcolsep}{4pt}
\begin{tabular}{@{}llccccccccccccccc@{}}
\toprule
\textbf{Target} & \textbf{Set} & $k_{\mathrm{on}}$ & $k_{\mathrm{off}}$ & $T$ & $d$ & $D$ & $N_{\max}$ & $L$ & $\lambda$ & $k_{\mathrm{off}}T$ & $c_{\mathrm{peak}}/K_D$ & \textbf{MF} & \textbf{500 r.} & \textbf{500 n.a.} & \textbf{50000 r.} & \textbf{50000 n.a.} \\
\midrule
Biomarker forecast & S0 & $1.00\times10^{-18}$ & 1.00 & 1.25 & 7.1 & $3.16\times10^{-11}$ & 10100 & 4 & $1\times10^{-6}$ & 1.2 & 3.3 & 0.340 & 0.552 & 0.550 $\pm$ 0.008 & 0.424 & 0.422 $\pm$ 0.004 \\
Biomarker forecast & S1 & $7.85\times10^{-20}$ & 1.90 & 1.25 & 7.1 & $3.16\times10^{-11}$ & 10100 & 4 & $1\times10^{-6}$ & 2.4 & 0.13 & 0.332 & 0.698 & 0.480 $\pm$ 0.005 & 0.469 & 0.383 $\pm$ 0.002 \\
Biomarker forecast & S4 & $2.00\times10^{-17}$ & 10.00 & 1.23 & 35.9 & $2.00\times10^{-10}$ & 20000 & 4 & $1\times10^{-6}$ & 12.3 & 0.23 & 0.329 & 0.643 & n.e. & 0.466 & 0.364 $\pm$ 0.005 \\
Threshold decision & S0 & $1.00\times10^{-18}$ & 1.00 & 1.25 & 7.1 & $3.16\times10^{-11}$ & 10100 & 3 & $1\times10^{-6}$ & 1.2 & 3.3 & 0.418 & 0.608 & 0.608 $\pm$ 0.012 & 0.513 & 0.513 $\pm$ 0.011 \\
Threshold decision & S1 & $2.59\times10^{-18}$ & 1.69 & 1.25 & 7.1 & $3.16\times10^{-11}$ & 10100 & 1 & $1\times10^{-6}$ & 2.1 & 5 & 0.374 & 0.630 & 0.521 $\pm$ 0.008 & 0.481 & 0.478 $\pm$ 0.009 \\
Threshold decision & S4 & $2.00\times10^{-17}$ & 4.05 & 0.50 & 6.0 & $2.00\times10^{-10}$ & 19715 & 1 & $1\times10^{-5}$ & 2.0 & 36 & 0.344 & 0.519 & n.e. & 0.436 & 0.419 $\pm$ 0.011 \\
\bottomrule
\end{tabular}
\end{table}
\end{landscape}

\begin{table}[H]
\centering
\caption{Input-tap baseline for the case-study targets, NRMSE on the optimization block for 1, 2, 3, 5, 10, 20, and 50 taps.}
\label{tab:si_case_taps}
\small
\begin{tabular}{@{}lccccccc@{}}
\toprule
\textbf{Target} & 1 & 2 & 3 & 5 & 10 & 20 & 50 \\
\midrule
Biomarker forecast & 0.675 & 0.410 & 0.340 & 0.335 & 0.333 & 0.332 & 0.332 \\
Threshold decision & 0.513 & 0.518 & 0.520 & 0.512 & 0.499 & 0.501 & 0.500 \\
\bottomrule
\end{tabular}
\end{table}

\begin{table}[H]
\centering
\caption{Online task switching, S1, 50000 receptors, 50 realizations. Steady error of the source task before the switch, steady error of the target configuration, steady error of the source configuration with a readout retrained for the target (reuse), and, for each ramp length, the peak of the windowed error after the switch and the number of symbols until the error of the switched run stays within 10 percent of the target level for 100 symbols. For MG forecasting and MG-Cubed, the S1 optimum is that of the wider search domain (Section S7), which has the lower noise-aware error at 50000 receptors; the other configurations are the default-domain optima.}
\label{tab:si_switch}
\scriptsize
\setlength{\tabcolsep}{3pt}
\begin{tabular}{@{}llccc ccc@{}}
\toprule
\textbf{Source} & \textbf{Target} & \textbf{source level} & \textbf{target level} & \textbf{reuse level} & \textbf{ramp 0, peak / symbols} & \textbf{ramp 10} & \textbf{ramp 50} \\
\midrule
MG & MG-Cubed & 0.278 & 0.497 & 0.498 & 2.0 / 16 & 1.1 / 24 & 1.0 / 52 \\
MG-Cubed & MG & 0.498 & 0.276 & 0.282 & 2.3 / 16 & 0.7 / 17 & 0.4 / 60 \\
Biomarker forecast & Threshold decision & 0.289 & 0.378 & 0.365 & 4.5 / 13 & 4.9 / 21 & 5.6 / 55 \\
Threshold decision & Biomarker forecast & 0.381 & 0.269 & 0.299 & 2.3 / 16 & 2.2 / 20 & 2.6 / 55 \\
Sine-to-square & MG & 0.298 & 0.276 & 0.453 & 14.2 / 17 & 6.6 / 26 & 3.3 / 65 \\
MG & Sine-to-square & 0.278 & 0.295 & 0.339 & 2.8 / 17 & 2.2 / 23 & 2.0 / 63 \\
\bottomrule
\end{tabular}
\end{table}

\FloatBarrier
\section*{S10. Memory and Nonlinear Capacity against Receptor Number}

\begin{figure}[H]
\centering
\includegraphics[width=\textwidth]{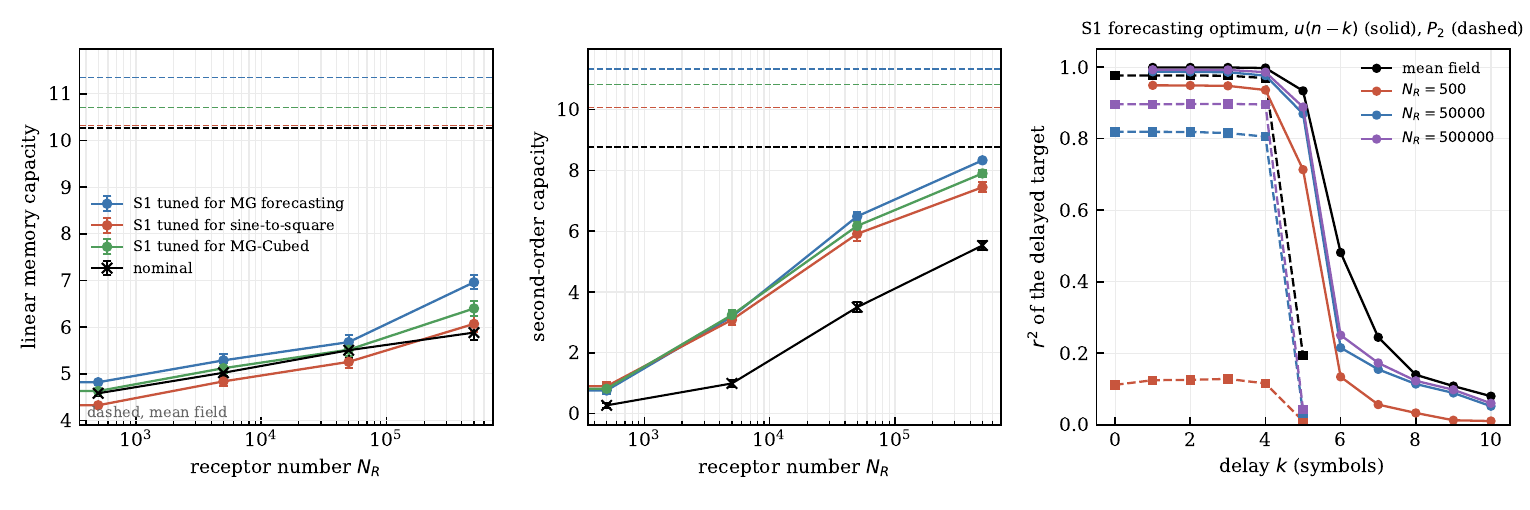}
\caption{Partial linear-memory capacity (left), partial second-order capacity (middle), and delay-resolved reconstruction accuracy for the S1 forecasting configuration (right). The left and middle panels compare the nominal configuration and the task-selected S1 configurations as a function of the receptor number, with the mean and standard deviation over three input sequences and five realizations, and with dashed horizontal lines indicating the mean-field values. In the right panel, solid curves show the delayed-input reconstruction, and dashed curves show the second-order Legendre targets at the indicated receptor numbers. The capacity basis, readout memory, regularization selection, and averaging are specified in Section S10 and in the Experimental Section of the main article.}
\label{fig:si_capacity}
\end{figure}

\textbf{Figure~\ref{fig:si_capacity}} and \textbf{Table~\ref{tab:si_capacity}} give the linear memory capacity and the second-order capacity of the nominal configuration and of the mean-field S1 optima of the three benchmark tasks, both in the mean-field model and under noise, following the construction described in the Experimental Section of the main article. Both capacities drop sharply at 500 receptors and recover slowly as the receptor number increases. The delay profile of the S1 forecasting optimum (Figure~\ref{fig:si_capacity}, right) reveals where the remaining capacity is located. The inputs of the last four symbols, which the readout with $L=5$ stores explicitly, are recovered with $r^2=0.95$ at 500 receptors against 1.00 in the mean-field model, whereas the recovery of the input six symbols back, which is carried only by the memory of the channel itself, decreases from 0.48 to 0.13. The second-order term of the current input decreases from 0.98 to 0.11 at 500 receptors, and recovers to 0.90 at 500000 receptors. Therefore, under noise, the reservoir retains the memory stored by the readout, but loses the fine memory and the weak nonlinearity that the channel itself provides for this configuration, which explains why forecasting, which requires both, suffers the most.

\begin{table}[H]
\centering
\caption{Linear / second-order capacity in the mean field and at 500, 5000, 50000, and 500000 receptors.}
\label{tab:si_capacity}
\small
\begin{tabular}{@{}lccccc@{}}
\toprule
\textbf{Configuration} & \textbf{mean field} & 500 & 5000 & 50000 & 500000 \\
\midrule
nominal & 10.3 / 8.8 & 4.6 / 0.3 & 5.0 / 1.0 & 5.5 / 3.5 & 5.9 / 5.5 \\
S1 tuned for MG & 11.3 / 11.3 & 4.8 / 0.8 & 5.3 / 3.2 & 5.7 / 6.5 & 7.0 / 8.3 \\
S1 tuned for Sine-to-square & 10.3 / 10.1 & 4.3 / 0.9 & 4.8 / 3.1 & 5.3 / 5.9 & 6.1 / 7.4 \\
S1 tuned for MG-Cubed & 10.7 / 10.8 & 4.6 / 0.8 & 5.1 / 3.2 & 5.5 / 6.2 & 6.4 / 7.9 \\
\bottomrule
\end{tabular}
\end{table}

\FloatBarrier
\section*{S11. Additional Tasks}

To test whether the results obtained under noise hold beyond the three primary benchmarks (Section 3.6 of the main article), three additional tasks were studied, i.e., temporal XOR and three-bit pattern detection on an on-off keyed bit stream, and NARMA-10, whose definitions are given in the Experimental Section of the main article. \textbf{Table~\ref{tab:si_bits}} lists the mean-field optima of S0, S1, and S4 with their errors under noise, and \textbf{Table~\ref{tab:si_bits_taps}} gives the input-tap baselines. No linear map of the bit stream can compute XOR, and thus, its baseline is one at every tap count. The pattern target is only partly linear in the bits, and its baseline remains near 0.76. NARMA-10 has a strong linear component, and twenty input taps reach 0.387, which the channel with a five-symbol readout does not outperform in the mean-field model.

\begin{landscape}
\begin{table}[H]
\centering
\caption{Additional tasks. Mean-field optima of S0, S1, and S4, their operating point, and their errors in the mean field and under noise. Under noise, the mean-field optimum is evaluated with the readout retrained (r.), and the noise-aware optimum of the same set at that receptor number is given with its re-evaluation with independent realizations (n.a., mean $\pm$ standard deviation over 50 realizations). MF is the mean-field value. Optimization block. Units are m$^3$~s$^{-1}$ for $k_{\mathrm{on}}$, s$^{-1}$ for $k_{\mathrm{off}}$, s for $T$, $\mu$m for $d$, and m$^2$~s$^{-1}$ for $D$.}
\label{tab:si_bits}
\footnotesize
\setlength{\tabcolsep}{4pt}
\begin{tabular}{@{}llccccccccccccccc@{}}
\toprule
\textbf{Task} & \textbf{Set} & $k_{\mathrm{on}}$ & $k_{\mathrm{off}}$ & $T$ & $d$ & $D$ & $N_{\max}$ & $L$ & $\lambda$ & $k_{\mathrm{off}}T$ & $c_{\mathrm{peak}}/K_D$ & \textbf{MF} & \textbf{500 r.} & \textbf{500 n.a.} & \textbf{50000 r.} & \textbf{50000 n.a.} \\
\midrule
Temporal XOR & S0 & $1.00\times10^{-18}$ & 1.00 & 1.25 & 7.1 & $3.16\times10^{-11}$ & 10100 & 5 & $1\times10^{-6}$ & 1.2 & 3.4 & 0.254 & 0.992 & 0.991 $\pm$ 0.002 & 0.930 & 0.930 $\pm$ 0.013 \\
Temporal XOR & S1 & $2.83\times10^{-18}$ & 8.04 & 1.25 & 7.1 & $3.16\times10^{-11}$ & 10100 & 5 & $1\times10^{-6}$ & 10.0 & 1.2 & 0.159 & 0.997 & 0.889 $\pm$ 0.016 & 0.629 & 0.422 $\pm$ 0.007 \\
Temporal XOR & S4 & $2.00\times10^{-17}$ & 9.71 & 2.00 & 6.0 & $1.48\times10^{-10}$ & 20000 & 5 & $1\times10^{-6}$ & 19.4 & 14 & 0.003 & 0.989 & 0.813 $\pm$ 0.016 & 0.448 & 0.369 $\pm$ 0.008 \\
Bit-pattern detection & S0 & $1.00\times10^{-18}$ & 1.00 & 1.25 & 7.1 & $3.16\times10^{-11}$ & 10100 & 5 & $1\times10^{-6}$ & 1.2 & 3.4 & 0.511 & 0.814 & 0.815 $\pm$ 0.003 & 0.762 & 0.761 $\pm$ 0.008 \\
Bit-pattern detection & S1 & $4.90\times10^{-18}$ & 8.29 & 1.25 & 7.1 & $3.16\times10^{-11}$ & 10100 & 4 & $1\times10^{-6}$ & 10.4 & 2 & 0.467 & 0.787 & 0.775 $\pm$ 0.004 & 0.670 & 0.671 $\pm$ 0.007 \\
Bit-pattern detection & S4 & $2.00\times10^{-17}$ & 4.22 & 1.76 & 6.0 & $1.41\times10^{-10}$ & 20000 & 5 & $1\times10^{-6}$ & 7.4 & 33 & 0.006 & 0.749 & 0.758 $\pm$ 0.005 & 0.336 & 0.388 $\pm$ 0.010 \\
NARMA-10 & S0 & $1.00\times10^{-18}$ & 1.00 & 1.25 & 7.1 & $3.16\times10^{-11}$ & 10100 & 5 & $1\times10^{-6}$ & 1.2 & 3.2 & 0.671 & 0.779 & 0.778 $\pm$ 0.007 & 0.741 & 0.739 $\pm$ 0.009 \\
NARMA-10 & S1 & $5.00\times10^{-20}$ & 0.62 & 1.25 & 7.1 & $3.16\times10^{-11}$ & 10100 & 5 & $1\times10^{-6}$ & 0.8 & 0.26 & 0.609 & 0.835 & 0.751 $\pm$ 0.004 & 0.754 & 0.703 $\pm$ 0.005 \\
NARMA-10 & S4 & $5.38\times10^{-20}$ & 0.16 & 1.48 & 6.9 & $6.59\times10^{-11}$ & 12857 & 4 & $1\times10^{-6}$ & 0.2 & 1.1 & 0.431 & 0.828 & 0.738 $\pm$ 0.005 & 0.750 & 0.682 $\pm$ 0.006 \\
\bottomrule
\end{tabular}
\end{table}
\end{landscape}

\begin{table}[H]
\centering
\caption{Input-tap baseline for the additional tasks, NRMSE on the optimization block for 1, 2, 3, 5, 10, 20, and 50 taps.}
\label{tab:si_bits_taps}
\small
\begin{tabular}{@{}lccccccc@{}}
\toprule
\textbf{Task} & 1 & 2 & 3 & 5 & 10 & 20 & 50 \\
\midrule
Temporal XOR & 1.002 & 1.003 & 1.003 & 1.002 & 1.007 & 1.011 & 1.004 \\
Bit-pattern detection & 0.941 & 0.853 & 0.761 & 0.763 & 0.764 & 0.768 & 0.784 \\
NARMA-10 & 0.999 & 0.859 & 0.822 & 0.801 & 0.819 & 0.387 & 0.394 \\
\bottomrule
\end{tabular}
\end{table}

\FloatBarrier
\section*{S12. Readout Designed for Noise and Joint Search}

\begin{figure}[H]
\centering
\includegraphics[width=\textwidth]{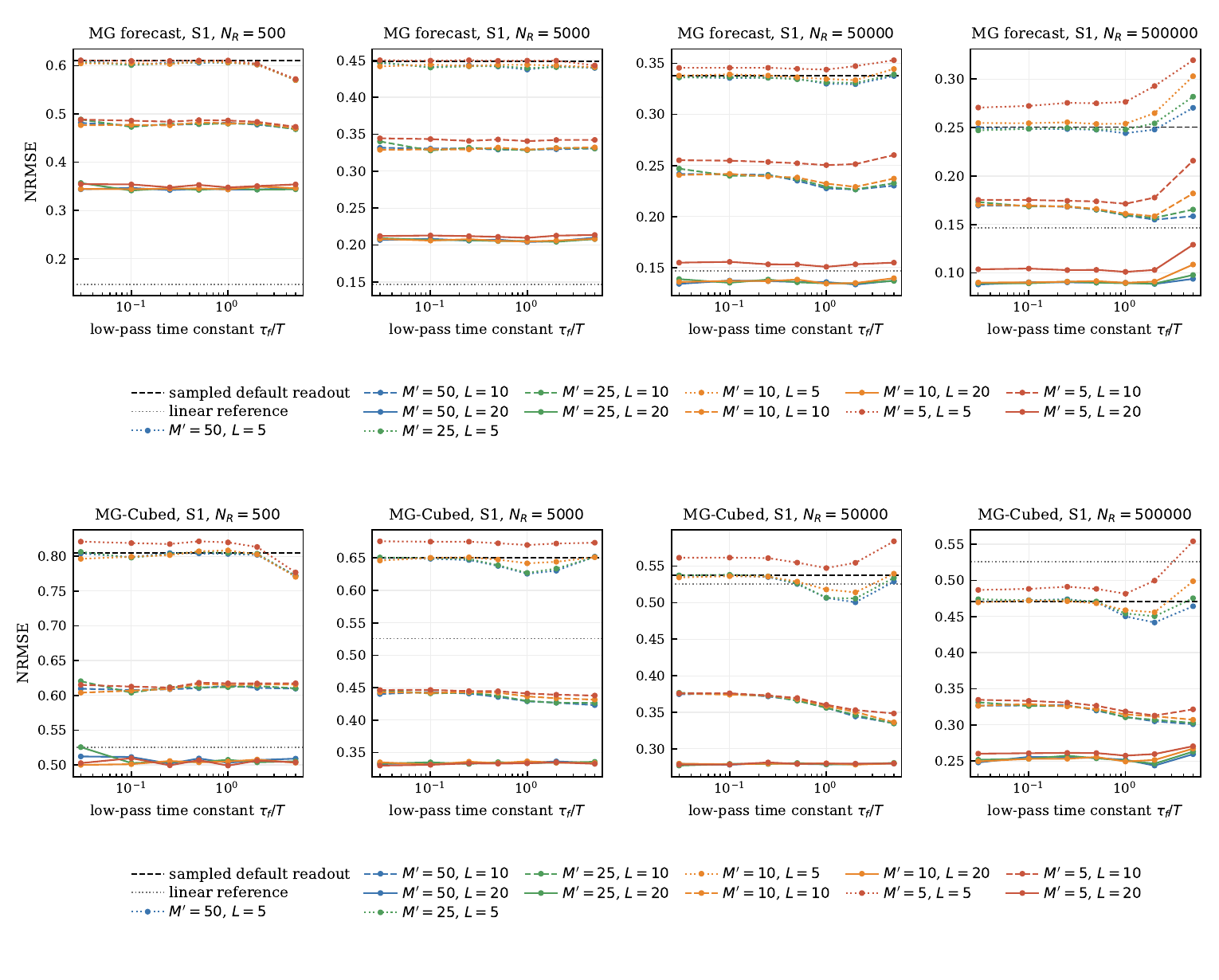}
\caption{Readout grid for the S1 mean-field optima of MG forecasting and MG-Cubed at 500, 5000, 50000, and $5\times10^{5}$ receptors. NRMSE against the low-pass time constant $\tau_f/T$ for every node count $M'$ (color) and readout memory $L$ (line style), with the sampled default readout (dashed) and the linear reference of the task (dotted). The leftmost point of every curve, drawn at 0.03, is the setting without a filter. The legend below each row applies to that row.}
\label{fig:si_remedies}
\end{figure}

\begin{figure}[H]
\centering
\includegraphics[width=\textwidth]{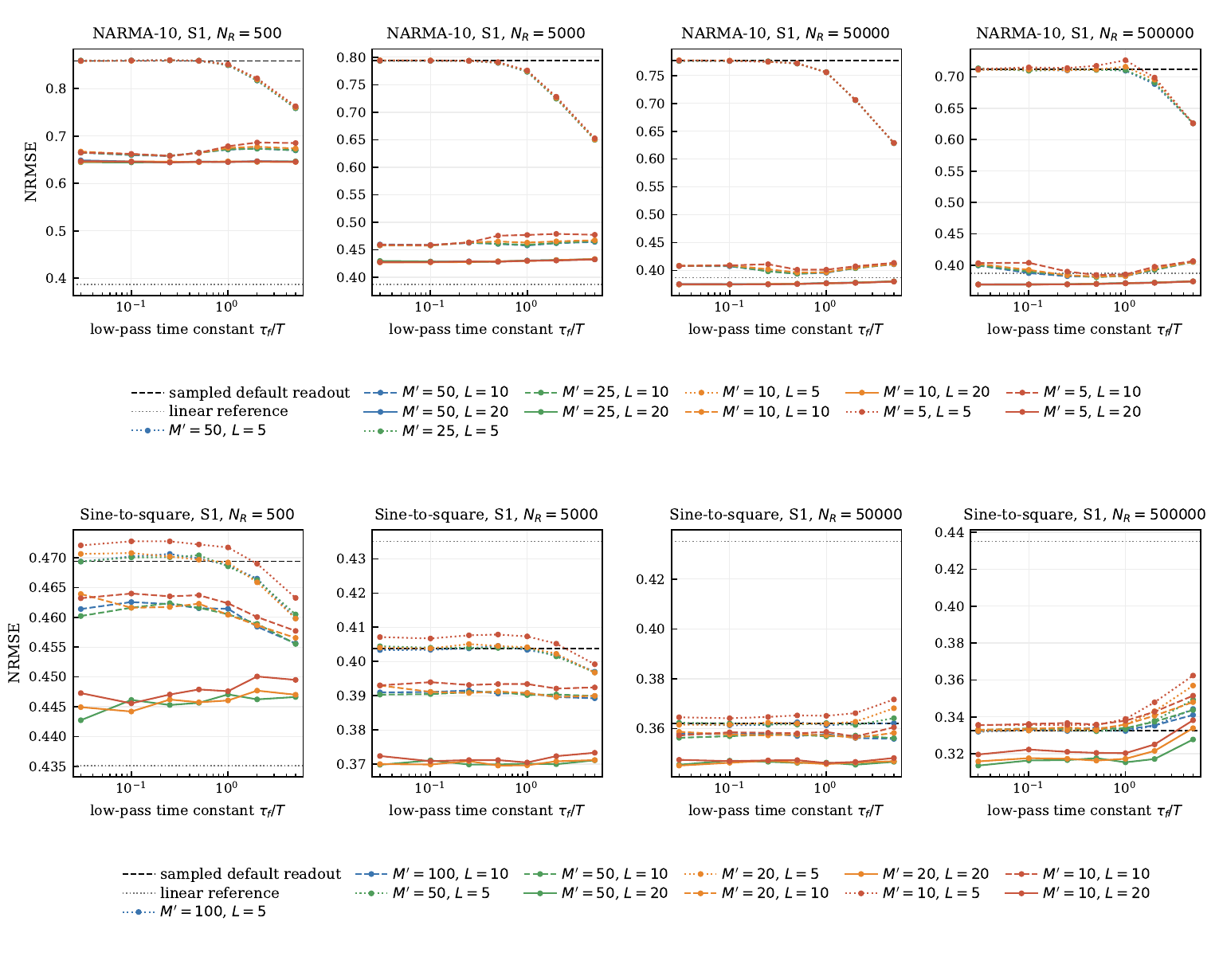}
\caption{Same as Figure~\ref{fig:si_remedies} for NARMA-10 and sine-to-square, whose node counts are $M'\in\{100,50,20,10\}$.}
\label{fig:si_remedies_b}
\end{figure}

\textbf{Table~\ref{tab:si_remedies}} summarizes the readout grid described in the Experimental Section of the main article for the S1 mean-field optima of the eight tasks at four receptor numbers, and \textbf{Table~\ref{tab:si_remedies_sfour}} does the same for the S4 optima, which supply the critical receptor numbers reported in the main article. Each table gives the default sampled readout, the best low-pass filter alone (integrating nodes at the default node count, $L=5$), the best readout memory alone (integrating nodes at the default node count, no filter), the best combination with its settings, and the held-out value of the best combination. \textbf{Table~\ref{tab:si_paired}} gives the crossings of the grid default and of the designed readout on the same grid and realizations, which are not used in the default column of Table 4 of the main article. \textbf{Figures~\ref{fig:si_remedies}} and \textbf{\ref{fig:si_remedies_b}} show the complete grid for the four tasks on which the readout has the largest effect. \textbf{Table~\ref{tab:si_joint}} lists the optima of the joint channel-plus-readout searches, together with their search objective, their re-evaluation with independent realizations, and the mean-field value of the same configuration with the same readout.

\scriptsize
\setlength{\tabcolsep}{3pt}
\begin{longtable}{@{}lrcccccc@{}}
\caption{Readout designed for noise, S1 mean-field optima, 20 independent realizations, optimization block. Default sampled readout with the $L$ of the optimum and a re-selected ridge value, best low-pass filter alone with its $\tau_f/T$ in brackets, best readout memory alone with its $L$ in brackets, best of the grid and the default (Best) with its settings $M'$ / $\tau_f/T$ / $L$ or the word default, and the held-out value of that best. Note that this grid default keeps the $L$ of the optimum, and thus, it differs from the retrained-readout scheme of the receptor-number sweep (Table~\ref{tab:si_nrext}), which also re-selects $L$ from 1 to 5.}
\label{tab:si_remedies}\\
\toprule
\textbf{Task} & $N_R$ & \textbf{Default} & \textbf{Filter alone} & \textbf{Memory alone} & \textbf{Best} & $M'$ / $\tau_f/T$ / $L$ & \textbf{Best, held-out} \\
\midrule
\endfirsthead
\toprule
\textbf{Task} & $N_R$ & \textbf{Default} & \textbf{Filter alone} & \textbf{Memory alone} & \textbf{Best} & $M'$ / $\tau_f/T$ / $L$ & \textbf{Best, held-out} \\
\midrule
\endhead
\bottomrule
\endfoot
MG forecasting & 500 & 0.611 & 0.571 (5) & 0.344 (20) & 0.341 & 25 / 0.1 / 20 & 0.336 \\
MG forecasting & 5000 & 0.449 & 0.437 (1) & 0.207 (20) & 0.204 & 50 / 1 / 20 & 0.207 \\
MG forecasting & 50000 & 0.338 & 0.330 (2) & 0.134 (20) & 0.133 & 50 / 2 / 20 & 0.141 \\
MG forecasting & 500000 & 0.251 & 0.244 (1) & 0.088 (20) & 0.088 & 50 / 0 / 20 & 0.094 \\
Sine-to-square & 500 & 0.469 & 0.460 (5) & 0.461 (10) & 0.443 & 50 / 0 / 20 & 0.467 \\
Sine-to-square & 5000 & 0.404 & 0.397 (5) & 0.391 (10) & 0.370 & 20 / 0.5 / 20 & 0.384 \\
Sine-to-square & 50000 & 0.362 & 0.361 (0.1) & 0.358 (10) & 0.345 & 20 / 0 / 20 & 0.350 \\
Sine-to-square & 500000 & 0.332 & 0.332 (0) & 0.332 (5) & 0.314 & 50 / 0 / 20 & 0.324 \\
MG-Cubed & 500 & 0.805 & 0.772 (5) & 0.512 (20) & 0.499 & 5 / 1 / 20 & 0.538 \\
MG-Cubed & 5000 & 0.651 & 0.626 (1) & 0.332 (20) & 0.330 & 5 / 0 / 20 & 0.394 \\
MG-Cubed & 50000 & 0.537 & 0.500 (2) & 0.280 (20) & 0.277 & 25 / 0 / 20 & 0.339 \\
MG-Cubed & 500000 & 0.471 & 0.442 (2) & 0.248 (20) & 0.244 & 50 / 2 / 20 & 0.291 \\
NARMA-10 & 500 & 0.858 & 0.759 (5) & 0.649 (20) & 0.644 & 5 / 0.25 / 20 & 0.642 \\
NARMA-10 & 5000 & 0.794 & 0.650 (5) & 0.428 (20) & 0.427 & 10 / 0 / 20 & 0.419 \\
NARMA-10 & 50000 & 0.777 & 0.629 (5) & 0.375 (20) & 0.375 & 10 / 0 / 20 & 0.366 \\
NARMA-10 & 500000 & 0.712 & 0.626 (5) & 0.369 (20) & 0.369 & 10 / 0 / 20 & 0.360 \\
Biomarker forecast & 500 & 0.714 & 0.675 (2) & 0.675 (10) & 0.674 & 5 / 5 / 20 & 0.685 \\
Biomarker forecast & 5000 & 0.563 & 0.559 (2) & 0.557 (20) & 0.553 & 5 / 5 / 10 & 0.539 \\
Biomarker forecast & 50000 & 0.474 & 0.469 (0.5) & 0.465 (10) & 0.464 & 5 / 2 / 10 & 0.443 \\
Biomarker forecast & 500000 & 0.405 & 0.401 (2) & 0.402 (10) & 0.401 & 50 / 2 / 5 & 0.384 \\
Threshold decision & 500 & 0.691 & 0.585 (5) & 0.603 (20) & 0.584 & 5 / 5 / 20 & 0.563 \\
Threshold decision & 5000 & 0.594 & 0.552 (0) & 0.552 (5) & 0.550 & 10 / 0 / 5 & 0.492 \\
Threshold decision & 50000 & 0.482 & 0.514 (1) & 0.517 (5) & 0.482 & default & 0.470 \\
Threshold decision & 500000 & 0.461 & 0.480 (0.1) & 0.484 (5) & 0.461 & default & 0.468 \\
Temporal XOR & 500 & 1.002 & 0.998 (0.5) & 1.001 (5) & 0.997 & 5 / 0.5 / 5 & 0.990 \\
Temporal XOR & 5000 & 0.930 & 0.919 (0.5) & 0.927 (5) & 0.913 & 5 / 1 / 5 & 0.907 \\
Temporal XOR & 50000 & 0.629 & 0.608 (5) & 0.618 (5) & 0.605 & 10 / 0.5 / 5 & 0.582 \\
Temporal XOR & 500000 & 0.358 & 0.352 (0.5) & 0.346 (20) & 0.340 & 10 / 0.5 / 20 & 0.328 \\
Bit-pattern detection & 500 & 0.789 & 0.789 (0.1) & 0.792 (5) & 0.789 & 25 / 1 / 5 & 0.789 \\
Bit-pattern detection & 5000 & 0.742 & 0.736 (1) & 0.742 (5) & 0.735 & 25 / 5 / 5 & 0.734 \\
Bit-pattern detection & 50000 & 0.671 & 0.667 (0.5) & 0.671 (5) & 0.666 & 25 / 0.5 / 5 & 0.676 \\
Bit-pattern detection & 500000 & 0.617 & 0.617 (0) & 0.617 (5) & 0.616 & 10 / 0 / 5 & 0.641 \\
\end{longtable}
\normalsize
\setlength{\tabcolsep}{6pt}

\scriptsize
\setlength{\tabcolsep}{3pt}
\begin{longtable}{@{}lrcccccc@{}}
\caption{Same as Table~\ref{tab:si_remedies} for the S4 mean-field optima, which supply the critical receptor numbers of Table 4 and Figure 7 of the main article.}
\label{tab:si_remedies_sfour}\\
\toprule
\textbf{Task} & $N_R$ & \textbf{Default} & \textbf{Filter alone} & \textbf{Memory alone} & \textbf{Best} & $M'$ / $\tau_f/T$ / $L$ & \textbf{Best, held-out} \\
\midrule
\endfirsthead
\toprule
\textbf{Task} & $N_R$ & \textbf{Default} & \textbf{Filter alone} & \textbf{Memory alone} & \textbf{Best} & $M'$ / $\tau_f/T$ / $L$ & \textbf{Best, held-out} \\
\midrule
\endhead
\bottomrule
\endfoot
MG forecasting & 500 & 0.575 & 0.527 (5) & 0.300 (20) & 0.291 & 25 / 0.1 / 20 & 0.300 \\
MG forecasting & 5000 & 0.404 & 0.392 (1) & 0.179 (20) & 0.177 & 50 / 1 / 20 & 0.188 \\
MG forecasting & 50000 & 0.313 & 0.307 (1) & 0.127 (20) & 0.124 & 10 / 0 / 20 & 0.131 \\
MG forecasting & 500000 & 0.241 & 0.232 (1) & 0.089 (20) & 0.089 & 50 / 0 / 20 & 0.095 \\
Sine-to-square & 500 & 0.379 & 0.380 (0) & 0.378 (10) & 0.369 & 50 / 0 / 20 & 0.378 \\
Sine-to-square & 5000 & 0.347 & 0.341 (0) & 0.341 (5) & 0.341 & 100 / 0 / 5 & 0.350 \\
Sine-to-square & 50000 & 0.310 & 0.307 (0) & 0.307 (5) & 0.307 & 100 / 0 / 5 & 0.316 \\
Sine-to-square & 500000 & 0.284 & 0.284 (0) & 0.284 (5) & 0.281 & 50 / 0 / 20 & 0.294 \\
MG-Cubed & 500 & 0.742 & 0.718 (5) & 0.370 (20) & 0.348 & 25 / 5 / 20 & 0.414 \\
MG-Cubed & 5000 & 0.630 & 0.624 (5) & 0.294 (20) & 0.294 & 50 / 0 / 20 & 0.364 \\
MG-Cubed & 50000 & 0.538 & 0.537 (0) & 0.254 (20) & 0.254 & 50 / 0 / 20 & 0.311 \\
MG-Cubed & 500000 & 0.476 & 0.477 (1) & 0.197 (20) & 0.197 & 50 / 0 / 20 & 0.247 \\
NARMA-10 & 500 & 0.839 & 0.745 (5) & 0.641 (20) & 0.641 & 50 / 0 / 20 & 0.625 \\
NARMA-10 & 5000 & 0.780 & 0.630 (5) & 0.423 (20) & 0.422 & 10 / 0 / 20 & 0.419 \\
NARMA-10 & 50000 & 0.770 & 0.608 (5) & 0.377 (20) & 0.377 & 10 / 0 / 20 & 0.370 \\
NARMA-10 & 500000 & 0.706 & 0.601 (5) & 0.372 (20) & 0.372 & 10 / 0 / 20 & 0.365 \\
Biomarker forecast & 500 & 0.643 & 0.634 (5) & 0.637 (5) & 0.626 & 5 / 1 / 20 & 0.617 \\
Biomarker forecast & 5000 & 0.550 & 0.538 (0.5) & 0.541 (10) & 0.528 & 5 / 1 / 20 & 0.519 \\
Biomarker forecast & 50000 & 0.463 & 0.451 (5) & 0.460 (10) & 0.451 & 50 / 5 / 5 & 0.436 \\
Biomarker forecast & 500000 & 0.408 & 0.399 (0.5) & 0.402 (5) & 0.398 & 5 / 0.5 / 5 & 0.377 \\
Threshold decision & 500 & 0.553 & 0.519 (0) & 0.519 (5) & 0.519 & 50 / 0 / 5 & 0.486 \\
Threshold decision & 5000 & 0.480 & 0.467 (0) & 0.467 (5) & 0.460 & 25 / 0 / 5 & 0.396 \\
Threshold decision & 50000 & 0.445 & 0.459 (0) & 0.459 (5) & 0.445 & default & 0.438 \\
Threshold decision & 500000 & 0.453 & 0.457 (0) & 0.457 (5) & 0.446 & 10 / 0 / 5 & 0.388 \\
Temporal XOR & 500 & 0.992 & 0.986 (0.25) & 0.990 (5) & 0.983 & 5 / 1 / 5 & 0.971 \\
Temporal XOR & 5000 & 0.853 & 0.825 (0.25) & 0.835 (5) & 0.818 & 10 / 5 / 5 & 0.809 \\
Temporal XOR & 50000 & 0.457 & 0.439 (0) & 0.439 (5) & 0.439 & 50 / 0 / 5 & 0.418 \\
Temporal XOR & 500000 & 0.163 & 0.155 (0) & 0.155 (5) & 0.155 & 25 / 0 / 5 & 0.154 \\
Bit-pattern detection & 500 & 0.750 & 0.745 (0.1) & 0.748 (5) & 0.743 & 10 / 0 / 5 & 0.748 \\
Bit-pattern detection & 5000 & 0.547 & 0.537 (0) & 0.537 (5) & 0.534 & 25 / 0 / 5 & 0.529 \\
Bit-pattern detection & 50000 & 0.338 & 0.328 (0) & 0.328 (5) & 0.326 & 25 / 0 / 5 & 0.322 \\
Bit-pattern detection & 500000 & 0.208 & 0.193 (0) & 0.188 (10) & 0.188 & 50 / 0 / 10 & 0.185 \\
\end{longtable}
\normalsize
\setlength{\tabcolsep}{6pt}

\begin{table}[H]
\centering
\caption{Paired crossings on the readout grid. For the S1 and S4 mean-field optima, the receptor number at which the sampled default of the grid run (the $L$ of the optimum, ridge value re-selected) and the designed readout (best of the integrating-node grid and that default) cross the linear reference and reach an NRMSE of 0.25, both on the same 20 realizations and the same four receptor numbers, log-interpolated. The entry ${\le}500$ marks a crossing at or below 500 receptors, and n.r. marks no crossing for $N_R\le5\times10^{5}$. The default column of Table 4 of the main article instead uses the denser receptor-number sweep given in Table~\ref{tab:si_nrext}, with $L$ re-selected; hence, differences between the two default schemes cannot be attributed to the readout alone. The interpolation intervals are set by the spacing of the evaluated receptor numbers, and they should not be interpreted as error bounds.}
\label{tab:si_paired}
\small
\begin{tabular}{@{}llcccc@{}}
\toprule
 & & \multicolumn{2}{c}{\textbf{Below the reference}} & \multicolumn{2}{c}{\textbf{Reaches NRMSE 0.25}} \\
\cmidrule(lr){3-4}\cmidrule(lr){5-6}
\textbf{Task} & \textbf{Set} & \textbf{grid default} & \textbf{selected} & \textbf{grid default} & \textbf{selected} \\
\midrule
MG forecasting & S1 & n.r. & $3.2\times10^{4}$ & n.r. & 2300 \\
MG forecasting & S4 & n.r. & $1.9\times10^{4}$ & $3.7\times10^{5}$ & 1100 \\
Sine-to-square & S1 & 1700 & 640 & n.r. & n.r. \\
Sine-to-square & S4 & $\le 500$ & $\le 500$ & n.r. & n.r. \\
MG-Cubed & S1 & $7.5\times10^{4}$ & $\le 500$ & n.r. & $3.2\times10^{5}$ \\
MG-Cubed & S4 & $7.9\times10^{4}$ & $\le 500$ & n.r. & $5.9\times10^{4}$ \\
NARMA-10 & S1 & n.r. & $2.9\times10^{4}$ & n.r. & n.r. \\
NARMA-10 & S4 & n.r. & $3.0\times10^{4}$ & n.r. & n.r. \\
Biomarker forecast & S1 & n.r. & n.r. & n.r. & n.r. \\
Biomarker forecast & S4 & n.r. & n.r. & n.r. & n.r. \\
Threshold decision & S1 & $3.5\times10^{4}$ & $2.8\times10^{4}$ & n.r. & n.r. \\
Threshold decision & S4 & 2800 & 1100 & n.r. & n.r. \\
Temporal XOR & S1 & $\le 500$ & $\le 500$ & n.r. & n.r. \\
Temporal XOR & S4 & $\le 500$ & $\le 500$ & $2.5\times10^{5}$ & $2.3\times10^{5}$ \\
Bit-pattern detection & S1 & 2000 & 1700 & n.r. & n.r. \\
Bit-pattern detection & S4 & $\le 500$ & $\le 500$ & $2.4\times10^{5}$ & $1.8\times10^{5}$ \\
\bottomrule
\end{tabular}
\end{table}

\begin{landscape}
\begin{table}[H]
\centering
\caption{Joint channel-plus-readout optima at 5000 and 50000 receptors under S1 and S3. For S3, the symbol interval was allowed from 0.5 to 5 s, and the feature count was not capped (Table~\ref{tab:si_domain}). Search objective (mean over the ten realizations of the search), re-evaluation with 50 independent realizations (mean $\pm$ standard deviation), its held-out value, and the mean-field error of the same configuration with the same readout. Units are m$^3$~s$^{-1}$ for $k_{\mathrm{on}}$, s$^{-1}$ for $k_{\mathrm{off}}$, s for $T$, $\mu$m for $d$, and m$^2$~s$^{-1}$ for $D$.}
\label{tab:si_joint}
\footnotesize
\setlength{\tabcolsep}{4pt}
\begin{tabular}{@{}llrcccccccccccc@{}}
\toprule
\textbf{Task} & \textbf{Set} & $N_R$ & $k_{\mathrm{on}}$ & $k_{\mathrm{off}}$ & $T$ & $N_{\max}$ & $M'$ & $\tau_f/T$ & $L$ & $\lambda$ & \textbf{search} & \textbf{re-eval.} & \textbf{held} & \textbf{mean field} \\
\midrule
MG forecasting & S1 & 5000 & $6.62\times10^{-18}$ & 9.75 & 1.25 & 10100 & 10 & 1.2 & 20 & $1\times10^{-4}$ & 0.183 & 0.189 $\pm$ 0.012 & 0.194 & 0.071 \\
Sine-to-square & S1 & 5000 & $8.77\times10^{-18}$ & 2.02 & 1.25 & 10100 & 20 & 3.7 & 20 & $1\times10^{-5}$ & 0.367 & 0.371 $\pm$ 0.008 & 0.381 & 0.313 \\
MG-Cubed & S1 & 5000 & $1.43\times10^{-17}$ & 9.87 & 1.25 & 10100 & 50 & 0.39 & 20 & $1\times10^{-2}$ & 0.307 & 0.313 $\pm$ 0.014 & 0.376 & 0.157 \\
NARMA-10 & S1 & 5000 & $1.46\times10^{-18}$ & 10.00 & 1.25 & 10100 & 50 & 5 & 17 & $1\times10^{-3}$ & 0.372 & 0.372 $\pm$ 0.002 & 0.369 & 0.370 \\
MG forecasting & S3 & 5000 & $2.00\times10^{-17}$ & 10.00 & 5.00 & 18431 & 50 & 0.05 & 18 & $1\times10^{-2}$ & 0.117 & 0.123 $\pm$ 0.006 & 0.127 & 0.037 \\
Sine-to-square & S3 & 5000 & $5.51\times10^{-18}$ & 0.41 & 4.58 & 12475 & 10 & 3 & 20 & $1\times10^{-5}$ & 0.346 & 0.350 $\pm$ 0.006 & 0.355 & 0.285 \\
MG-Cubed & S3 & 5000 & $7.05\times10^{-18}$ & 0.56 & 4.18 & 14325 & 50 & 5 & 19 & $1\times10^{-5}$ & 0.265 & 0.272 $\pm$ 0.015 & 0.333 & 0.217 \\
NARMA-10 & S3 & 5000 & $4.93\times10^{-19}$ & 10.00 & 5.00 & 20000 & 50 & 5 & 17 & $1\times10^{-3}$ & 0.369 & 0.368 $\pm$ 0.002 & 0.365 & 0.366 \\
MG forecasting & S1 & 50000 & $4.39\times10^{-18}$ & 10.00 & 1.25 & 10100 & 25 & 5 & 20 & $1\times10^{-6}$ & 0.123 & 0.127 $\pm$ 0.005 & 0.132 & 0.091 \\
Sine-to-square & S1 & 50000 & $8.14\times10^{-18}$ & 0.84 & 1.25 & 10100 & 10 & 0.85 & 20 & $1\times10^{-5}$ & 0.343 & 0.345 $\pm$ 0.006 & 0.351 & 0.286 \\
MG-Cubed & S1 & 50000 & $1.16\times10^{-17}$ & 1.77 & 1.25 & 10100 & 25 & 0.085 & 18 & $1\times10^{-3}$ & 0.264 & 0.271 $\pm$ 0.011 & 0.323 & 0.099 \\
NARMA-10 & S1 & 50000 & $5.85\times10^{-19}$ & 10.00 & 1.25 & 10100 & 50 & 5 & 15 & $1\times10^{-4}$ & 0.367 & 0.367 $\pm$ 0.001 & 0.366 & 0.367 \\
MG forecasting & S3 & 50000 & $2.00\times10^{-17}$ & 10.00 & 5.00 & 17821 & 50 & 5 & 20 & $1\times10^{-6}$ & 0.082 & 0.083 $\pm$ 0.003 & 0.092 & 0.072 \\
Sine-to-square & S3 & 50000 & $4.84\times10^{-18}$ & 0.74 & 5.00 & 15480 & 100 & 0.21 & 20 & $1\times10^{-4}$ & 0.314 & 0.314 $\pm$ 0.008 & 0.324 & 0.233 \\
MG-Cubed & S3 & 50000 & $3.21\times10^{-18}$ & 0.36 & 4.77 & 20000 & 50 & 5 & 20 & $1\times10^{-6}$ & 0.222 & 0.228 $\pm$ 0.010 & 0.279 & 0.201 \\
NARMA-10 & S3 & 50000 & $5.71\times10^{-19}$ & 10.00 & 5.00 & 3923 & 10 & 5 & 17 & $1\times10^{-5}$ & 0.366 & 0.366 $\pm$ 0.001 & 0.362 & 0.366 \\
\bottomrule
\end{tabular}
\end{table}
\end{landscape}

\textbf{Cell-like readout:} \textbf{Table~\ref{tab:si_cell}} restricts the readout of the S1 and S4 optima to one or two integrating windows per symbol, i.e., the occupancy averaged over the whole symbol interval or over each of its halves. The low-pass constants and readout memories are those of the grid above, evaluated at the four receptor numbers and in the mean-field model. The table also gives the receptor numbers at which these readouts outperform the linear reference and reach an NRMSE of 0.25 within the sweep. \textbf{Table~\ref{tab:si_cell_settings}} lists the selected settings.

\begin{table}[H]
\centering
\caption{Cell-like readout with $M'$ integrating windows per symbol, S1 and S4 mean-field optima, best low-pass constant and readout memory at each receptor number, 20 independent realizations, optimization block. The linear reference and the default readout in the mean-field model are given for comparison. The last two columns give the receptor numbers at which this readout outperforms the linear reference and reaches an NRMSE of 0.25, log-interpolated within the sweep up to $5\times10^{5}$ receptors. For temporal XOR, the linear reference is one, and thus, the first criterion is not informative. The entry n.r. marks no crossing for $N_R\le5\times10^{5}$.}
\label{tab:si_cell}
\scriptsize
\setlength{\tabcolsep}{3pt}
\begin{tabular}{@{}llcccccccccc@{}}
\toprule
 & & & & \textbf{Default,} & \multicolumn{5}{c}{\textbf{Cell-like readout}} & & \\
\cmidrule(lr){6-10}
\textbf{Task} & \textbf{Set} & $M'$ & \textbf{Linear ref.} & \textbf{mean field} & \textbf{mean field} & 500 & 5000 & 50000 & 500000 & \textbf{Below ref.} & \textbf{NRMSE 0.25} \\
\midrule
MG forecasting & S1 & 1 & 0.147 & 0.088 & 0.399 & 0.528 & 0.486 & 0.455 & 0.431 & n.r. & n.r. \\
 &  & 2 & 0.147 & 0.088 & 0.104 & 0.435 & 0.242 & 0.161 & 0.135 & $1.8\times10^{5}$ & 4600 \\
 & S4 & 1 & 0.147 & 0.087 & 0.363 & 0.493 & 0.452 & 0.420 & 0.393 & n.r. & n.r. \\
 &  & 2 & 0.147 & 0.087 & 0.096 & 0.344 & 0.194 & 0.143 & 0.123 & $4.3\times10^{4}$ & 2100 \\
Sine-to-square & S1 & 1 & 0.435 & 0.256 & 0.527 & 0.618 & 0.595 & 0.575 & 0.561 & n.r. & n.r. \\
 &  & 2 & 0.435 & 0.256 & 0.336 & 0.521 & 0.400 & 0.357 & 0.348 & 2600 & n.r. \\
 & S4 & 1 & 0.435 & 0.231 & 0.506 & 0.584 & 0.562 & 0.547 & 0.535 & n.r. & n.r. \\
 &  & 2 & 0.435 & 0.231 & 0.354 & 0.415 & 0.398 & 0.380 & 0.372 & $\le 500$ & n.r. \\
MG-Cubed & S1 & 1 & 0.526 & 0.299 & 0.721 & 0.745 & 0.737 & 0.730 & 0.725 & n.r. & n.r. \\
 &  & 2 & 0.526 & 0.299 & 0.264 & 0.665 & 0.433 & 0.309 & 0.279 & 2000 & n.r. \\
 & S4 & 1 & 0.526 & 0.285 & 0.712 & 0.736 & 0.728 & 0.722 & 0.718 & n.r. & n.r. \\
 &  & 2 & 0.526 & 0.285 & 0.301 & 0.350 & 0.313 & 0.304 & 0.302 & $\le 500$ & n.r. \\
Biomarker forecast & S1 & 1 & 0.332 & 0.332 & 0.358 & 0.675 & 0.557 & 0.468 & 0.407 & n.r. & n.r. \\
 &  & 2 & 0.332 & 0.332 & 0.348 & 0.675 & 0.556 & 0.465 & 0.404 & n.r. & n.r. \\
 & S4 & 1 & 0.332 & 0.329 & 0.370 & 0.626 & 0.532 & 0.461 & 0.411 & n.r. & n.r. \\
 &  & 2 & 0.332 & 0.329 & 0.347 & 0.625 & 0.526 & 0.452 & 0.398 & n.r. & n.r. \\
Threshold decision & S1 & 1 & 0.499 & 0.374 & 0.582 & 0.584 & 0.582 & 0.582 & 0.582 & n.r. & n.r. \\
 &  & 2 & 0.499 & 0.374 & 0.485 & 0.584 & 0.575 & 0.521 & 0.491 & $2.7\times10^{5}$ & n.r. \\
 & S4 & 1 & 0.499 & 0.344 & 0.663 & 0.664 & 0.663 & 0.663 & 0.663 & n.r. & n.r. \\
 &  & 2 & 0.499 & 0.344 & 0.473 & 0.556 & 0.491 & 0.474 & 0.473 & 3700 & n.r. \\
Temporal XOR & S1 & 1 & 1.002 & 0.159 & 0.998 & 0.998 & 0.998 & 0.998 & 0.998 & $\le 500$ & n.r. \\
 &  & 2 & 1.002 & 0.159 & 0.292 & 0.997 & 0.930 & 0.639 & 0.365 & $\le 500$ & n.r. \\
 & S4 & 1 & 1.002 & 0.003 & 0.998 & 0.998 & 0.998 & 0.998 & 0.998 & $\le 500$ & n.r. \\
 &  & 2 & 1.002 & 0.003 & 0.551 & 0.979 & 0.852 & 0.650 & 0.567 & $\le 500$ & n.r. \\
Bit-pattern detection & S1 & 1 & 0.761 & 0.467 & 0.792 & 0.792 & 0.792 & 0.792 & 0.792 & n.r. & n.r. \\
 &  & 2 & 0.761 & 0.467 & 0.653 & 0.788 & 0.747 & 0.674 & 0.656 & 2300 & n.r. \\
 & S4 & 1 & 0.761 & 0.006 & 0.786 & 0.786 & 0.786 & 0.786 & 0.786 & n.r. & n.r. \\
 &  & 2 & 0.761 & 0.006 & 0.732 & 0.767 & 0.738 & 0.732 & 0.732 & 840 & n.r. \\
NARMA-10 & S1 & 1 & 0.387 & 0.609 & 0.369 & 0.670 & 0.442 & 0.379 & 0.370 & $3.8\times10^{4}$ & n.r. \\
 &  & 2 & 0.387 & 0.609 & 0.368 & 0.653 & 0.422 & 0.376 & 0.370 & $2.9\times10^{4}$ & n.r. \\
 & S4 & 1 & 0.387 & 0.431 & 0.371 & 0.675 & 0.428 & 0.378 & 0.372 & $3.3\times10^{4}$ & n.r. \\
 &  & 2 & 0.387 & 0.431 & 0.371 & 0.665 & 0.424 & 0.377 & 0.372 & $3.1\times10^{4}$ & n.r. \\
\bottomrule
\end{tabular}
\end{table}

\scriptsize
\begin{longtable}{@{}llcrccccc@{}}
\caption{Settings of the cell-like readout of Table~\ref{tab:si_cell}, both sets. NRMSE with the standard deviation over realizations, low-pass constant, readout memory, ridge value, and held-out value. $^{*}$ marks a sample standard deviation below 0.0005.}
\label{tab:si_cell_settings}\\
\toprule
\textbf{Task} & \textbf{Set} & $M'$ & $N_R$ & \textbf{NRMSE} & $\tau_f/T$ & $L$ & $\lambda$ & \textbf{held} \\
\midrule
\endfirsthead
\toprule
\textbf{Task} & \textbf{Set} & $M'$ & $N_R$ & \textbf{NRMSE} & $\tau_f/T$ & $L$ & $\lambda$ & \textbf{held} \\
\midrule
\endhead
\bottomrule
\endfoot
MG forecasting & S1 & 1 & 500 & 0.528 $\pm$ 0.003 & 2 & 20 & $1\times10^{-4}$ & 0.559 \\
MG forecasting & S1 & 1 & 5000 & 0.486 $\pm$ 0.002 & 2 & 20 & $1\times10^{-5}$ & 0.515 \\
MG forecasting & S1 & 1 & 50000 & 0.455 $\pm$ 0.001 & 2 & 20 & $1\times10^{-6}$ & 0.482 \\
MG forecasting & S1 & 1 & 500000 & 0.431 $\pm$ 0.001 & 1 & 20 & $1\times10^{-6}$ & 0.459 \\
MG forecasting & S1 & 2 & 500 & 0.435 $\pm$ 0.021 & 0 & 20 & $1\times10^{-2}$ & 0.454 \\
MG forecasting & S1 & 2 & 5000 & 0.242 $\pm$ 0.016 & 1 & 20 & $1\times10^{-4}$ & 0.249 \\
MG forecasting & S1 & 2 & 50000 & 0.161 $\pm$ 0.005 & 1 & 20 & $1\times10^{-5}$ & 0.169 \\
MG forecasting & S1 & 2 & 500000 & 0.135 $\pm$ 0.002 & 0.5 & 20 & $1\times10^{-6}$ & 0.146 \\
MG forecasting & S4 & 1 & 500 & 0.493 $\pm$ 0.002 & 2 & 20 & $1\times10^{-4}$ & 0.522 \\
MG forecasting & S4 & 1 & 5000 & 0.452 $\pm$ 0.002 & 2 & 20 & $1\times10^{-5}$ & 0.477 \\
MG forecasting & S4 & 1 & 50000 & 0.420 $\pm$ 0.001 & 2 & 20 & $1\times10^{-6}$ & 0.444 \\
MG forecasting & S4 & 1 & 500000 & 0.393 $\pm$ 0.001 & 1 & 20 & $1\times10^{-6}$ & 0.419 \\
MG forecasting & S4 & 2 & 500 & 0.344 $\pm$ 0.023 & 0.5 & 20 & $1\times10^{-3}$ & 0.335 \\
MG forecasting & S4 & 2 & 5000 & 0.194 $\pm$ 0.011 & 1 & 20 & $1\times10^{-5}$ & 0.197 \\
MG forecasting & S4 & 2 & 50000 & 0.143 $\pm$ 0.004 & 0.5 & 20 & $1\times10^{-5}$ & 0.155 \\
MG forecasting & S4 & 2 & 500000 & 0.123 $\pm$ 0.001 & 0.25 & 20 & $1\times10^{-5}$ & 0.134 \\
Sine-to-square & S1 & 1 & 500 & 0.618 $\pm$ 0.002 & 5 & 20 & $1\times10^{-4}$ & 0.668 \\
Sine-to-square & S1 & 1 & 5000 & 0.595 $\pm$ 0.002 & 2 & 20 & $1\times10^{-4}$ & 0.643 \\
Sine-to-square & S1 & 1 & 50000 & 0.575 $\pm$ 0.002 & 1 & 20 & $1\times10^{-5}$ & 0.622 \\
Sine-to-square & S1 & 1 & 500000 & 0.561 $\pm$ 0.001 & 2 & 20 & $1\times10^{-6}$ & 0.608 \\
Sine-to-square & S1 & 2 & 500 & 0.521 $\pm$ 0.020 & 5 & 20 & $1\times10^{-5}$ & 0.552 \\
Sine-to-square & S1 & 2 & 5000 & 0.400 $\pm$ 0.010 & 0 & 20 & $1\times10^{-3}$ & 0.417 \\
Sine-to-square & S1 & 2 & 50000 & 0.357 $\pm$ 0.004 & 5 & 20 & $1\times10^{-6}$ & 0.366 \\
Sine-to-square & S1 & 2 & 500000 & 0.348 $\pm$ 0.001 & 5 & 20 & $1\times10^{-6}$ & 0.354 \\
Sine-to-square & S4 & 1 & 500 & 0.584 $\pm$ 0.002 & 5 & 20 & $1\times10^{-4}$ & 0.629 \\
Sine-to-square & S4 & 1 & 5000 & 0.562 $\pm$ 0.002 & 2 & 20 & $1\times10^{-4}$ & 0.608 \\
Sine-to-square & S4 & 1 & 50000 & 0.547 $\pm$ 0.001 & 2 & 20 & $1\times10^{-6}$ & 0.593 \\
Sine-to-square & S4 & 1 & 500000 & 0.535 $\pm$ 0.001 & 2 & 20 & $1\times10^{-6}$ & 0.579 \\
Sine-to-square & S4 & 2 & 500 & 0.415 $\pm$ 0.005 & 5 & 20 & $1\times10^{-4}$ & 0.439 \\
Sine-to-square & S4 & 2 & 5000 & 0.398 $\pm$ 0.003 & 5 & 20 & $1\times10^{-5}$ & 0.421 \\
Sine-to-square & S4 & 2 & 50000 & 0.380 $\pm$ 0.002 & 5 & 20 & $1\times10^{-6}$ & 0.398 \\
Sine-to-square & S4 & 2 & 500000 & 0.372 $\pm$ 0.001 & 5 & 20 & $1\times10^{-6}$ & 0.390 \\
MG-Cubed & S1 & 1 & 500 & 0.745 $\pm$ 0.002 & 5 & 20 & $1\times10^{-5}$ & 0.814 \\
MG-Cubed & S1 & 1 & 5000 & 0.737 $\pm$ 0.001 & 5 & 20 & $1\times10^{-6}$ & 0.806 \\
MG-Cubed & S1 & 1 & 50000 & 0.730$^{*}$ & 5 & 20 & $1\times10^{-6}$ & 0.800 \\
MG-Cubed & S1 & 1 & 500000 & 0.725$^{*}$ & 5 & 20 & $1\times10^{-6}$ & 0.796 \\
MG-Cubed & S1 & 2 & 500 & 0.665 $\pm$ 0.055 & 0.25 & 20 & $1\times10^{-2}$ & 0.705 \\
MG-Cubed & S1 & 2 & 5000 & 0.433 $\pm$ 0.025 & 5 & 20 & $1\times10^{-6}$ & 0.468 \\
MG-Cubed & S1 & 2 & 50000 & 0.309 $\pm$ 0.014 & 5 & 20 & $1\times10^{-6}$ & 0.357 \\
MG-Cubed & S1 & 2 & 500000 & 0.279 $\pm$ 0.005 & 0 & 20 & $1\times10^{-5}$ & 0.329 \\
MG-Cubed & S4 & 1 & 500 & 0.736 $\pm$ 0.001 & 5 & 20 & $1\times10^{-5}$ & 0.804 \\
MG-Cubed & S4 & 1 & 5000 & 0.728 $\pm$ 0.001 & 5 & 20 & $1\times10^{-6}$ & 0.797 \\
MG-Cubed & S4 & 1 & 50000 & 0.722$^{*}$ & 5 & 20 & $1\times10^{-6}$ & 0.791 \\
MG-Cubed & S4 & 1 & 500000 & 0.718$^{*}$ & 0.5 & 20 & $1\times10^{-6}$ & 0.782 \\
MG-Cubed & S4 & 2 & 500 & 0.350 $\pm$ 0.020 & 5 & 20 & $1\times10^{-6}$ & 0.408 \\
MG-Cubed & S4 & 2 & 5000 & 0.313 $\pm$ 0.007 & 2 & 20 & $1\times10^{-6}$ & 0.392 \\
MG-Cubed & S4 & 2 & 50000 & 0.304 $\pm$ 0.002 & 2 & 20 & $1\times10^{-6}$ & 0.385 \\
MG-Cubed & S4 & 2 & 500000 & 0.302 $\pm$ 0.001 & 2 & 20 & $1\times10^{-6}$ & 0.382 \\
Biomarker forecast & S1 & 1 & 500 & 0.675 $\pm$ 0.013 & 5 & 20 & $1\times10^{-6}$ & 0.664 \\
Biomarker forecast & S1 & 1 & 5000 & 0.557 $\pm$ 0.007 & 2 & 20 & $1\times10^{-6}$ & 0.545 \\
Biomarker forecast & S1 & 1 & 50000 & 0.468 $\pm$ 0.004 & 0 & 20 & $1\times10^{-6}$ & 0.451 \\
Biomarker forecast & S1 & 1 & 500000 & 0.407 $\pm$ 0.003 & 0 & 20 & $1\times10^{-6}$ & 0.390 \\
Biomarker forecast & S1 & 2 & 500 & 0.675 $\pm$ 0.013 & 5 & 20 & $1\times10^{-5}$ & 0.664 \\
Biomarker forecast & S1 & 2 & 5000 & 0.556 $\pm$ 0.008 & 2 & 20 & $1\times10^{-5}$ & 0.542 \\
Biomarker forecast & S1 & 2 & 50000 & 0.465 $\pm$ 0.005 & 2 & 10 & $1\times10^{-6}$ & 0.447 \\
Biomarker forecast & S1 & 2 & 500000 & 0.404 $\pm$ 0.003 & 0.5 & 10 & $1\times10^{-6}$ & 0.386 \\
Biomarker forecast & S4 & 1 & 500 & 0.626 $\pm$ 0.006 & 2 & 20 & $1\times10^{-6}$ & 0.619 \\
Biomarker forecast & S4 & 1 & 5000 & 0.532 $\pm$ 0.005 & 5 & 10 & $1\times10^{-6}$ & 0.519 \\
Biomarker forecast & S4 & 1 & 50000 & 0.461 $\pm$ 0.003 & 0 & 20 & $1\times10^{-6}$ & 0.448 \\
Biomarker forecast & S4 & 1 & 500000 & 0.411 $\pm$ 0.001 & 0 & 20 & $1\times10^{-6}$ & 0.399 \\
Biomarker forecast & S4 & 2 & 500 & 0.625 $\pm$ 0.008 & 1 & 20 & $1\times10^{-3}$ & 0.618 \\
Biomarker forecast & S4 & 2 & 5000 & 0.526 $\pm$ 0.006 & 5 & 10 & $1\times10^{-6}$ & 0.511 \\
Biomarker forecast & S4 & 2 & 50000 & 0.452 $\pm$ 0.004 & 1 & 10 & $1\times10^{-6}$ & 0.436 \\
Biomarker forecast & S4 & 2 & 500000 & 0.398 $\pm$ 0.003 & 0 & 10 & $1\times10^{-6}$ & 0.380 \\
Threshold decision & S1 & 1 & 500 & 0.584 $\pm$ 0.005 & 5 & 20 & $1\times10^{-2}$ & 0.568 \\
Threshold decision & S1 & 1 & 5000 & 0.582 $\pm$ 0.001 & 5 & 20 & $1\times10^{-2}$ & 0.563 \\
Threshold decision & S1 & 1 & 50000 & 0.582 $\pm$ 0.001 & 5 & 20 & $1\times10^{-2}$ & 0.562 \\
Threshold decision & S1 & 1 & 500000 & 0.582$^{*}$ & 5 & 20 & $1\times10^{-2}$ & 0.562 \\
Threshold decision & S1 & 2 & 500 & 0.584 $\pm$ 0.004 & 5 & 20 & $1\times10^{-1}$ & 0.568 \\
Threshold decision & S1 & 2 & 5000 & 0.575 $\pm$ 0.015 & 2 & 20 & $1\times10^{-6}$ & 0.509 \\
Threshold decision & S1 & 2 & 50000 & 0.521 $\pm$ 0.011 & 0 & 10 & $1\times10^{-6}$ & 0.482 \\
Threshold decision & S1 & 2 & 500000 & 0.491 $\pm$ 0.004 & 2 & 10 & $1\times10^{-6}$ & 0.443 \\
Threshold decision & S4 & 1 & 500 & 0.664 $\pm$ 0.003 & 5 & 10 & $1\times10^{0}$ & 0.710 \\
Threshold decision & S4 & 1 & 5000 & 0.663 $\pm$ 0.001 & 5 & 10 & $1\times10^{0}$ & 0.711 \\
Threshold decision & S4 & 1 & 50000 & 0.663$^{*}$ & 5 & 10 & $1\times10^{0}$ & 0.710 \\
Threshold decision & S4 & 1 & 500000 & 0.663$^{*}$ & 5 & 10 & $1\times10^{0}$ & 0.711 \\
Threshold decision & S4 & 2 & 500 & 0.556 $\pm$ 0.015 & 0.25 & 5 & $1\times10^{-2}$ & 0.512 \\
Threshold decision & S4 & 2 & 5000 & 0.491 $\pm$ 0.010 & 0 & 10 & $1\times10^{-3}$ & 0.463 \\
Threshold decision & S4 & 2 & 50000 & 0.474 $\pm$ 0.004 & 5 & 10 & $1\times10^{-6}$ & 0.464 \\
Threshold decision & S4 & 2 & 500000 & 0.473 $\pm$ 0.001 & 5 & 10 & $1\times10^{-5}$ & 0.427 \\
Temporal XOR & S1 & 1 & 500 & 0.998$^{*}$ & 5 & 20 & $1\times10^{2}$ & 1.000 \\
Temporal XOR & S1 & 1 & 5000 & 0.998$^{*}$ & 5 & 20 & $1\times10^{2}$ & 1.000 \\
Temporal XOR & S1 & 1 & 50000 & 0.998$^{*}$ & 5 & 20 & $1\times10^{2}$ & 1.000 \\
Temporal XOR & S1 & 1 & 500000 & 0.998$^{*}$ & 5 & 20 & $1\times10^{2}$ & 1.000 \\
Temporal XOR & S1 & 2 & 500 & 0.997 $\pm$ 0.006 & 0.25 & 5 & $1\times10^{-2}$ & 0.987 \\
Temporal XOR & S1 & 2 & 5000 & 0.930 $\pm$ 0.018 & 5 & 5 & $1\times10^{-6}$ & 0.917 \\
Temporal XOR & S1 & 2 & 50000 & 0.639 $\pm$ 0.017 & 2 & 10 & $1\times10^{-6}$ & 0.629 \\
Temporal XOR & S1 & 2 & 500000 & 0.365 $\pm$ 0.010 & 0 & 20 & $1\times10^{-6}$ & 0.344 \\
Temporal XOR & S4 & 1 & 500 & 0.998$^{*}$ & 5 & 20 & $1\times10^{2}$ & 1.002 \\
Temporal XOR & S4 & 1 & 5000 & 0.998$^{*}$ & 5 & 20 & $1\times10^{2}$ & 1.002 \\
Temporal XOR & S4 & 1 & 50000 & 0.998$^{*}$ & 5 & 20 & $1\times10^{2}$ & 1.002 \\
Temporal XOR & S4 & 1 & 500000 & 0.998$^{*}$ & 5 & 20 & $1\times10^{2}$ & 1.002 \\
Temporal XOR & S4 & 2 & 500 & 0.979 $\pm$ 0.009 & 2 & 5 & $1\times10^{-4}$ & 0.971 \\
Temporal XOR & S4 & 2 & 5000 & 0.852 $\pm$ 0.010 & 0.5 & 5 & $1\times10^{-4}$ & 0.849 \\
Temporal XOR & S4 & 2 & 50000 & 0.650 $\pm$ 0.023 & 1 & 10 & $1\times10^{-6}$ & 0.633 \\
Temporal XOR & S4 & 2 & 500000 & 0.567 $\pm$ 0.008 & 1 & 10 & $1\times10^{-6}$ & 0.525 \\
Bit-pattern detection & S1 & 1 & 500 & 0.792 $\pm$ 0.001 & 2 & 5 & $1\times10^{-6}$ & 0.791 \\
Bit-pattern detection & S1 & 1 & 5000 & 0.792 $\pm$ 0.001 & 2 & 5 & $1\times10^{-6}$ & 0.790 \\
Bit-pattern detection & S1 & 1 & 50000 & 0.792$^{*}$ & 2 & 5 & $1\times10^{-6}$ & 0.791 \\
Bit-pattern detection & S1 & 1 & 500000 & 0.792$^{*}$ & 2 & 5 & $1\times10^{-6}$ & 0.791 \\
Bit-pattern detection & S1 & 2 & 500 & 0.788 $\pm$ 0.006 & 0.1 & 5 & $1\times10^{-2}$ & 0.784 \\
Bit-pattern detection & S1 & 2 & 5000 & 0.747 $\pm$ 0.010 & 5 & 5 & $1\times10^{-6}$ & 0.746 \\
Bit-pattern detection & S1 & 2 & 50000 & 0.674 $\pm$ 0.007 & 5 & 5 & $1\times10^{-6}$ & 0.680 \\
Bit-pattern detection & S1 & 2 & 500000 & 0.656 $\pm$ 0.002 & 5 & 5 & $1\times10^{-6}$ & 0.658 \\
Bit-pattern detection & S4 & 1 & 500 & 0.786 $\pm$ 0.001 & 2 & 5 & $1\times10^{-2}$ & 0.786 \\
Bit-pattern detection & S4 & 1 & 5000 & 0.786$^{*}$ & 2 & 5 & $1\times10^{-2}$ & 0.785 \\
Bit-pattern detection & S4 & 1 & 50000 & 0.786$^{*}$ & 2 & 5 & $1\times10^{-2}$ & 0.785 \\
Bit-pattern detection & S4 & 1 & 500000 & 0.786$^{*}$ & 2 & 5 & $1\times10^{-2}$ & 0.785 \\
Bit-pattern detection & S4 & 2 & 500 & 0.767 $\pm$ 0.006 & 0.25 & 5 & $1\times10^{-2}$ & 0.770 \\
Bit-pattern detection & S4 & 2 & 5000 & 0.738 $\pm$ 0.005 & 0.25 & 5 & $1\times10^{-6}$ & 0.751 \\
Bit-pattern detection & S4 & 2 & 50000 & 0.732 $\pm$ 0.001 & 0.25 & 5 & $1\times10^{-6}$ & 0.750 \\
Bit-pattern detection & S4 & 2 & 500000 & 0.732$^{*}$ & 0.25 & 5 & $1\times10^{-6}$ & 0.749 \\
NARMA-10 & S1 & 1 & 500 & 0.670 $\pm$ 0.028 & 1 & 20 & $1\times10^{-3}$ & 0.667 \\
NARMA-10 & S1 & 1 & 5000 & 0.442 $\pm$ 0.012 & 0.5 & 20 & $1\times10^{-4}$ & 0.436 \\
NARMA-10 & S1 & 1 & 50000 & 0.379 $\pm$ 0.004 & 0 & 20 & $1\times10^{-4}$ & 0.369 \\
NARMA-10 & S1 & 1 & 500000 & 0.370 $\pm$ 0.001 & 0.5 & 20 & $1\times10^{-5}$ & 0.359 \\
NARMA-10 & S1 & 2 & 500 & 0.653 $\pm$ 0.029 & 1 & 20 & $1\times10^{-3}$ & 0.651 \\
NARMA-10 & S1 & 2 & 5000 & 0.422 $\pm$ 0.011 & 0.25 & 20 & $1\times10^{-3}$ & 0.416 \\
NARMA-10 & S1 & 2 & 50000 & 0.376 $\pm$ 0.004 & 0 & 20 & $1\times10^{-3}$ & 0.367 \\
NARMA-10 & S1 & 2 & 500000 & 0.370 $\pm$ 0.001 & 0 & 20 & $1\times10^{-3}$ & 0.359 \\
NARMA-10 & S4 & 1 & 500 & 0.675 $\pm$ 0.028 & 0.5 & 20 & $1\times10^{-4}$ & 0.655 \\
NARMA-10 & S4 & 1 & 5000 & 0.428 $\pm$ 0.013 & 0 & 20 & $1\times10^{-4}$ & 0.417 \\
NARMA-10 & S4 & 1 & 50000 & 0.378 $\pm$ 0.004 & 0 & 20 & $1\times10^{-4}$ & 0.366 \\
NARMA-10 & S4 & 1 & 500000 & 0.372 $\pm$ 0.001 & 0 & 20 & $1\times10^{-4}$ & 0.362 \\
NARMA-10 & S4 & 2 & 500 & 0.665 $\pm$ 0.027 & 0.5 & 20 & $1\times10^{-3}$ & 0.644 \\
NARMA-10 & S4 & 2 & 5000 & 0.424 $\pm$ 0.013 & 0.1 & 20 & $1\times10^{-3}$ & 0.415 \\
NARMA-10 & S4 & 2 & 50000 & 0.377 $\pm$ 0.004 & 0 & 20 & $1\times10^{-3}$ & 0.369 \\
NARMA-10 & S4 & 2 & 500000 & 0.372 $\pm$ 0.001 & 0 & 20 & $1\times10^{-3}$ & 0.364 \\
\end{longtable}
\normalsize
\setlength{\tabcolsep}{6pt}

\FloatBarrier
\section*{S13. Receptor Sweep to $10^{9}$ Receptors}

\textbf{Table~\ref{tab:si_nrext}} extends the receptor-number sweep presented in Section S6 for the mean-field optima of S0, S1, and S4 of every task to $10^{6}$, $10^{7}$, $10^{8}$, and $10^{9}$ receptors. The readout is retrained under noise, with 50 independent realizations up to $5\times10^{5}$ receptors and 20 realizations beyond. Cells marked n.e. were not evaluated. The errors of every task approach their mean-field values between $10^{8}$ and $10^{9}$ receptors.

\begin{table}[H]
\centering
\caption{Retrained mean-field optima from 500 to $10^{9}$ receptors, NRMSE on the optimization block. The mean-field column is the noise-free value, and n.e. marks receptor numbers that were not evaluated for that configuration.}
\label{tab:si_nrext}
\scriptsize
\setlength{\tabcolsep}{2.2pt}
\begin{tabular}{@{}llcrrrrrrrrrrr@{}}
\toprule
\textbf{Task} & \textbf{Set} & \textbf{MF} & 500 & 2000 & 5000 & 20000 & 50000 & 200000 & 500000 & $10^{6}$ & $10^{7}$ & $10^{8}$ & $10^{9}$ \\
\midrule
MG forecasting & S0 & 0.122 & 0.687 & 0.625 & 0.566 & 0.452 & 0.387 & 0.325 & 0.298 & 0.278 & 0.216 & 0.164 & 0.133 \\
MG forecasting & S1 & 0.088 & 0.615 & 0.509 & 0.449 & 0.373 & 0.338 & 0.279 & 0.249 & 0.229 & 0.173 & 0.118 & 0.093 \\
MG forecasting & S4 & 0.087 & 0.575 & 0.460 & 0.405 & 0.344 & 0.315 & 0.268 & 0.242 & 0.224 & 0.164 & 0.113 & 0.092 \\
Sine-to-square & S0 & 0.293 & 0.531 & 0.491 & 0.460 & 0.423 & 0.409 & 0.388 & 0.378 & 0.370 & 0.346 & 0.323 & 0.301 \\
Sine-to-square & S1 & 0.256 & 0.468 & 0.432 & 0.404 & 0.379 & 0.364 & 0.348 & 0.333 & 0.326 & 0.301 & 0.278 & 0.261 \\
Sine-to-square & S4 & 0.231 & 0.381 & 0.361 & 0.348 & 0.321 & 0.311 & 0.292 & 0.283 & 0.278 & 0.264 & 0.246 & 0.235 \\
MG-Cubed & S0 & 0.346 & 0.843 & 0.790 & 0.740 & 0.626 & 0.572 & 0.511 & 0.491 & 0.487 & 0.451 & 0.392 & 0.355 \\
MG-Cubed & S1 & 0.299 & 0.805 & 0.704 & 0.648 & 0.577 & 0.538 & 0.492 & 0.471 & 0.462 & 0.420 & 0.354 & 0.309 \\
MG-Cubed & S4 & 0.285 & 0.741 & 0.666 & 0.630 & 0.563 & 0.538 & 0.498 & 0.480 & 0.465 & 0.408 & 0.340 & 0.294 \\
Biomarker forecast & S0 & 0.340 & 0.552 & n.e. & n.e. & n.e. & 0.424 & n.e. & n.e. & 0.369 & 0.347 & 0.342 & 0.341 \\
Biomarker forecast & S1 & 0.332 & 0.698 & n.e. & n.e. & n.e. & 0.469 & n.e. & n.e. & 0.387 & 0.352 & 0.336 & 0.332 \\
Biomarker forecast & S4 & 0.329 & 0.643 & n.e. & n.e. & n.e. & 0.466 & n.e. & n.e. & 0.388 & 0.355 & 0.335 & 0.330 \\
Threshold decision & S0 & 0.418 & 0.608 & n.e. & n.e. & n.e. & 0.513 & n.e. & n.e. & 0.488 & 0.469 & 0.427 & 0.420 \\
Threshold decision & S1 & 0.374 & 0.630 & n.e. & n.e. & n.e. & 0.481 & n.e. & n.e. & 0.459 & 0.434 & 0.390 & 0.376 \\
Threshold decision & S4 & 0.344 & 0.519 & n.e. & n.e. & n.e. & 0.436 & n.e. & n.e. & 0.432 & 0.368 & 0.342 & 0.343 \\
Temporal XOR & S0 & 0.254 & 0.992 & 0.989 & 0.985 & 0.964 & 0.930 & 0.812 & 0.705 & 0.610 & 0.395 & 0.319 & 0.268 \\
Temporal XOR & S1 & 0.159 & 0.997 & 0.973 & 0.934 & 0.762 & 0.629 & 0.432 & 0.355 & 0.321 & 0.253 & 0.212 & 0.171 \\
Temporal XOR & S4 & 0.003 & 0.989 & 0.928 & 0.835 & 0.625 & 0.448 & 0.252 & 0.161 & 0.117 & 0.043 & 0.016 & 0.006 \\
Bit-pattern detection & S0 & 0.511 & 0.814 & 0.807 & 0.801 & 0.782 & 0.762 & 0.722 & 0.698 & 0.680 & 0.644 & 0.594 & 0.530 \\
Bit-pattern detection & S1 & 0.467 & 0.787 & 0.768 & 0.742 & 0.702 & 0.670 & 0.637 & 0.612 & 0.601 & 0.570 & 0.536 & 0.480 \\
Bit-pattern detection & S4 & 0.006 & 0.749 & 0.649 & 0.548 & 0.397 & 0.336 & 0.255 & 0.208 & 0.175 & 0.101 & 0.036 & 0.014 \\
NARMA-10 & S0 & 0.671 & 0.779 & 0.768 & 0.762 & 0.751 & 0.741 & 0.723 & 0.713 & 0.708 & 0.692 & 0.679 & 0.673 \\
NARMA-10 & S1 & 0.609 & 0.835 & 0.783 & 0.767 & 0.756 & 0.754 & 0.745 & 0.716 & 0.686 & 0.626 & 0.611 & 0.610 \\
NARMA-10 & S4 & 0.431 & 0.828 & 0.778 & 0.764 & 0.754 & 0.750 & 0.728 & 0.690 & 0.662 & 0.526 & 0.448 & 0.429 \\
\bottomrule
\end{tabular}
\end{table}

\FloatBarrier
\section*{S14. Reference Echo State Network of Equal Size}

\textbf{Table~\ref{tab:si_esn}} gives the error of an echo state network with as many units as the number of features of the default channel readout, tuned with the same optimizer and budget as the channel searches (Experimental Section of the main article), and compares it with the mean-field errors of the S1 and S4 channel optima and the linear reference. Note that the network is a noiseless digital reservoir, and this comparison serves to put the absolute errors of the channel into perspective.

\begin{table}[H]
\centering
\caption{Reference echo state network. Number of units $N$, NRMSE on the optimization block (mean $\pm$ standard deviation over ten independent weight seeds of the best of three searches), its held-out value, the mean-field errors of the S1 and S4 channel optima, and the linear reference. $^{*}$ marks a sample standard deviation below 0.0005, and ${<}0.001$ a mean below that value.}
\label{tab:si_esn}
\small
\begin{tabular}{@{}lcccccc@{}}
\toprule
\textbf{Task} & $N$ & \textbf{ESN} & \textbf{ESN, held} & \textbf{Channel S1} & \textbf{Channel S4} & \textbf{Linear ref.} \\
\midrule
MG forecasting & 250 & 0.003 $\pm$ 0.005 & 0.002 & 0.088 & 0.087 & 0.147 \\
Sine-to-square & 500 & 0.103 $\pm$ 0.007 & 0.091 & 0.256 & 0.231 & 0.435 \\
MG-Cubed & 250 & 0.006 $\pm$ 0.003 & 0.005 & 0.299 & 0.285 & 0.526 \\
Biomarker forecast & 250 & 0.332$^{*}$ & 0.321 & 0.332 & 0.329 & 0.332 \\
Threshold decision & 250 & 0.293 $\pm$ 0.014 & 0.244 & 0.374 & 0.344 & 0.499 \\
Temporal XOR & 250 & ${<}0.001$ & ${<}0.001$ & 0.159 & 0.003 & 1.002 \\
Bit-pattern detection & 250 & ${<}0.001$ & ${<}0.001$ & 0.467 & 0.006 & 0.761 \\
NARMA-10 & 250 & 0.179 $\pm$ 0.018 & 0.177 & 0.609 & 0.431 & 0.387 \\
\bottomrule
\end{tabular}
\end{table}

\FloatBarrier
\section*{S15. Reproducibility}

Every result reported in the main article and in this document was produced by scripts from the simulation result files. A single simulator implements both the mean-field model and the stochastic receptor model, and the optimization scripts run one search per process. The collection, re-evaluation, reuse, error-surface, particle-validation, baseline, robustness, readout, critical-receptor-number, cell-like readout, and echo state network analyses write result tables, from which every number, every table body, and every figure is generated. The scripts run under Python 3.9.6 with NumPy 2.0.2, SciPy 1.13.1, Numba 0.60.0, scikit-optimize 0.10.2, and Matplotlib 3.9.4, and the particle simulations were run with Smoldyn 2.75. The scripts, the optimization logs, one CSV file per search with every evaluation, and the result files are available as described in the Data Availability Statement of the main article.